\documentclass[12pt]{aastex63}
\usepackage{rotating}
\usepackage{longtable}
\usepackage{graphicx}
\usepackage{txfonts}
\usepackage{verbatim}

\usepackage{color}
\usepackage{threeparttablex}

\newcommand{\lsun}{log ($L/L_{\odot})\,$}
\newcommand{\msun}{$M/M_{\odot}\,$}
\newcommand{\rsun}{$R$/$R_{\odot}\,$}

\received{..}
\revised{..}
\accepted{..}
\submitjournal{ApJ}
\shorttitle{An updated theoretical scenario for Classical Cepheids}
\shortauthors{G. De Somma et al. }
\graphicspath{{./}{figures/}}
\begin{document}

\title{An updated metal-dependent theoretical scenario for Classical Cepheids}
\correspondingauthor{Giulia De Somma}
\email{giulia.desomma@inaf.it , gdesomma@na.infn.it}

\author{Giulia De Somma}
\affiliation{ INAF-Osservatorio astronomico di Capodimonte \\
Via Moiariello 16 \\
80131 Napoli, Italy}
\affiliation{Istituto Nazionale di Fisica Nucleare (INFN) Sez. di Napoli\\
Compl. Univ.di Monte S. Angelo, Edificio G, Via Cinthia \\
I-80126, Napoli, Italy}

\author{Marcella Marconi}
\affiliation{ INAF-Osservatorio astronomico di Capodimonte \\
Via Moiariello 16 \\
80131 Napoli, Italy}

\author{Roberto Molinaro}
\affiliation{ INAF-Osservatorio astronomico di Capodimonte \\
Via Moiariello 16 \\
80131 Napoli, Italy}

\author{Vincenzo Ripepi}
\affiliation{ INAF-Osservatorio astronomico di Capodimonte \\
Via Moiariello 16 \\
80131 Napoli, Italy}

\author{Silvio Leccia}
\affiliation{ INAF-Osservatorio astronomico di Capodimonte \\
Via Moiariello 16 \\
80131 Napoli, Italy}

\author{Ilaria Musella}
\affiliation{ INAF-Osservatorio astronomico di Capodimonte \\
Via Moiariello 16 \\
80131 Napoli, Italy}

\begin{abstract}
\noindent
To properly quantify possible residual systematic errors affecting the Classical Cepheid distance scale, a detailed theoretical scenario is recommended.
By extending the set of nonlinear convective pulsation models published for $Z=0.02$ \citep[][]{Desomma2020a} to $Z=0.004$, $Z=0.008$ and $Z=0.03$, we provide a detailed homogeneous nonlinear model grid taking into account simultaneous variations of the mass-luminosity relation, the efficiency of super-adiabatic convection and the chemical composition. The dependence of the inferred Period-Radius, Period-Mass-Radius, and Period-Mass-Luminosity-Temperature relations on the input parameters is discussed for both the Fundamental and First Overtone modes. The trend of the instability strip getting redder as the metallicity increases is confirmed for the additional ML assumptions and mixing length values.
From the obtained multi-filter light curves, we derive mean magnitudes and colors and in turn Period-Luminosity-Color and Period-Wesenheit relations for each assumed chemical composition, mass-luminosity relation and efficiency of super-adiabatic convection. Application to a well-studied sample of Cepheids in the Large Magellanic Cloud allows us to constrain the dependence of the inferred distance modulus on the assumed mass-luminosity relation, and the inclusion of the metallicity term in the derivation of Period-Wesenheit relations allows us, for each assumed mass-luminosity relation, to predict the metallicity dependence of the Cepheid distance scale. The obtained metal-dependent Period-Wesenheit relations are compared with recent results in the literature and applied to a sample of Gaia Early Data Release 3 Galactic Cepheids with known metal abundances to derive individual parallaxes. The comparison of these predictions with Gaia results is finally discussed.

\end{abstract}


\section{Introduction} \label{sec:intro}

The application of Classical Cepheid (CC) Period-Luminosity (PL) and Period-Luminosity-Color (PLC) relations to the calibration of the cosmic distance scale represents one of the most important examples of a connection between stellar and extragalactic astrophysics. 
This stellar route to the extragalactic distance ladder has drawn a renewed interest in the last few years thanks to the results of the Gaia mission providing the most accurate distance determinations obtained so far for more than 1$\%$ of the Milky Way stellar content \citep[see e.g.][]{Brown2021} and the resulting Hubble constant values significantly deviating from early Universe measurements and determinations \citep[see e.g.][and references therein]{Riess2021a,Riess2021b,Verde2019}, despite their reduced uncertainties. \\
On this basis, a detailed investigation of the systematic errors affecting the various rungs of the cosmic distance ladder, in particular the CC distance scale, is needed. Among the possible sources of uncertainty, the dependence of the coefficients of PL and PLC relations on chemical composition has been tested by several authors in previous works from both the observational and the theoretical point of view \citep[see e.g.][and references therein]{Anderson2016, Bono2010, Breuval2021, Fausnaugh2015, Freedman2011,Groenewegen2013, Kennicutt1998, Kodric2013, Macri2006, Pejcha2012, Riess2016, Riess2021arX, Ripepi2019, Ripepi2020, Ripepi2021, Ripepi2022, Romaniello2008, Romaniello2022, Shappee2011, Wiel2017}.\\
As no consensus has yet been reached, despite the contribution provided by the unprecedented accuracy of Gaia astrometric distances, our team is involved in a project devoted to the investigation of the effect of metallicity on the CC distance scale. From the observational point of view, we have started a long term program called “Cepheid–Metallicity in the Leavitt Law” (C-MetaLL) to directly measure the metallicity dependence of CC relations using only Galactic pulsators with accurate metallicities from high-resolution spectroscopy, in combination with optical and Near Infrared (NIR) photometry, and Gaia parallaxes \citep[][]{Ripepi2021}. 
From the theoretical point of view, we are updating the metal-dependent pulsational scenario predicted by \citet[][]{Bono1999, Marconi2005, Marconi2010} by selecting the same chemical compositions of the corresponding evolutionary models presented by \cite{Hidalgo2018} and available at URL:\url{http://basti-iac.oa-abruzzo.inaf.it} and varying simultaneously, for the first time, the chemical composition, the Mass-Luminosity (ML) relation and the efficiency of super-adiabatic convection. This approach aims to provide a self-consistent framework to quantify a variety of possible systematic errors that have not been taken into account so far, in the use of CCs as primary distance indicators as well as young stellar population tracers.

The organization of the paper is as follows. In Section 2, we present the new model sets. The results, including the Instability Strip (IS) boundaries, the period, shape and amplitude of light and radial velocity curves and the Period-Radius (PR) and the Period-Mass-Radius (PMR) relations, are presented in Section 3. The transformation of the bolometric light curves into the photometric filters, the derivation of mean magnitudes and colors, and in turn, of multi-band PLC, Period-Wesenheit (PW) and metal-dependent Period-Wesenheit (PWZ) relations, are presented in Sections 4 and 5, respectively. The application of the derived PWZ relation in the Gaia bands to a sample of Gaia Early Data Release 3 (EDR3) CCs to derive theoretical parallaxes and in turn investigate the Gaia zero-point offset is presented in Section 6. Finally, the conclusions and the discussion of some future developments close the paper.

\section{The extended set of pulsation models}
By adopting the same approach and the same physical and numerical assumptions as in \citet[][]{Desomma2020a} but varying the chemical composition, we built an extended and detailed grid of nonlinear convective pulsation models. We investigated the full amplitude behavior of the computed models in both the Fundamental (F) and the First Overtone (FO) modes. The new adopted chemical compositions are:  $Z=0.004$ $Y=0.25$, $Z=0.008$ $Y=0.25$, $Z=0.03$ $Y=0.28$.\\

For each chemical composition, a wide mass range is adopted (from 3 to 11 $M_{\odot}$ with a step of 1$M_{\odot}$ and for each mass, three luminosity levels were considered in order to predict the effect of the ML relation on CC properties. In particular, a canonical luminosity level, i.e. models computed by neglecting core convective overshooting, rotation, and mass loss, labeled `case A`, and two noncanonical luminosity levels, obtained by increasing the canonical luminosity level by $\Delta\log(L$/$L_\odot)$=$~0.2 \; dex$ and $\Delta\log(L$/$L_\odot)$=$~0.4 \; dex$, labeled `case B` and `case C`, respectively, were adopted. For each combination of chemical composition, mass and luminosity, a wide range of effective temperatures (roughly from $3600$ to $7200K$ with a step of $100K$) and three values of the mixing length parameter describing convective efficiency, namely $\alpha_{ml}=1.5, 1.7, 1.9$ were taken into account. 
All the adopted parameters are reported in the first seven columns of Tables \ref{f_param_model} and \ref{fo_param_model} for F and FO-mode models, respectively.\\ These tables are available in their entirety as supplementary material.
The behavior of the predicted pulsation relation connecting the period to M, L and $T_e$ and of the IS boundaries when varying the chemical composition was investigated in \citet[][]{Desomma2021a}, as well as the new theoretical metal-dependent Period-Age and Period-Age-Color relations and their application to observational data \citep[see][for details]{Desomma2021a}.\\
The combination of these new model sets with the already published framework for $Z=0.02$ $Y=0.28$ \citep[see][]{Desomma2020a, Desomma2020b, DeSomma2021proc, Marconi2020} provides a theoretical scenario that simultaneously takes into account several possible sources of systematic effects in the prediction of CC pulsation properties and distance scale.

\section{The new theoretical results}
As a result of the nonlinear hydrodynamic computations, we obtained the full amplitude behavior of the investigated models in the first three radial modes. We found 696 F-mode, 110 FO-mode and 3 Second Overtone (SO) mode stable Cepheid pulsation models for $Z=0.004$, 639 F-mode, 126 FO-mode and 6 SO-mode stable Cepheid pulsation models for $Z=0.008$, and 127 F-mode and 4 stable FO-mode Cepheid pulsation models for $Z=0.03$. Pulsation in the second overtone mode was found only for a few low mass and metal-poor models so that we decided to focus only on the first two radial modes. The stable limit cycle pulsation period and mean radius (as resulting from an averaging operation of the radius curve), for each model, are reported in the last two columns of Tables \ref{f_param_model} and \ref{fo_param_model}.

\begin{longtable}{ccccccccc}
\caption{\normalsize{\label{f_param_model}}The intrinsic stellar parameters of Z=0.004 Y= 0.25, Z=0.008 Y= 0.25,  and Z=0.03 Y= 0.28 computed F-mode models. This table is available in its entirety in machine readable form.}\\
\hline\hline
Z & Y & \msun & \lsun & $T_{eff}$[K] & $\alpha_{ml}$ & ML & P[d] & $\log(\overline{R}$/$R_{\odot})$ \\
\hline
\endfirsthead
\caption{continued.}\\
\hline
Z & Y & \msun & \lsun & $T_{eff}$[K] & $\alpha_{ml}$ & ML & P[d] & $\log(\overline{R}$/$R_{\odot})$ \\
\hline
\endhead
\hline\hline
0.004 & 0.25 & 3.0 & 2.49 & 5900 & 1.5 & A & 1.46027 & 1.230 \\
0.004 & 0.25 & 3.0 & 2.49 & 6000 & 1.5 & A & 1.38159 & 1.217 \\
0.004 & 0.25 & 3.0 & 2.49 & 6000 & 1.7 & A & 1.38248 & 1.215 \\
...\\
0.008 & 0.25 & 3.0 & 2.39 & 6000 & 1.5 & A & 1.15407 & 1.166 \\
0.008 & 0.25 & 3.0 & 2.59 & 5700 & 1.5 & B & 2.00247 & 1.307 \\
0.008 & 0.25 & 3.0 & 2.59 & 5800 & 1.5 & B & 1.88759 & 1.295 \\
...\\
0.03 & 0.28 & 4.0 & 2.68 & 5400 & 1.5 & A & 2.42133 & 1.398 \\
0.03 & 0.28 & 4.0 & 2.68 & 5500 & 1.5 & A & 2.28760 & 1.383 \\
0.03 & 0.28 & 4.0 & 2.68 & 5600 & 1.5 & A & 2.14887 & 1.367 \\
...\\
\hline\hline
\end{longtable}

\begin{longtable}{ccccccccc}
\caption{\normalsize{\label{fo_param_model}}The intrinsic stellar parameters of Z=0.004 Y= 0.25, Z=0.008 Y= 0.25,  and Z=0.03 Y= 0.28 computed FO-mode models. This table is available in its entirety in machine readable form.}\\
\hline\hline
Z & Y & \msun & \lsun & $T_{eff}$[K] & $\alpha_{ml}$ & ML & P[d] & $\log(\overline{R}$/$R_{\odot})$ \\
\hline
\endfirsthead
\caption{continued.}\\
\hline
Z & Y & \msun & \lsun & $T_{eff}$[K] & $\alpha_{ml}$ & ML & P[d] & $\log(\overline{R}$/$R_{\odot})$ \\
\hline
\endhead
\hline\hline
0.004 & 0.25 & 3.0 & 2.49 & 6100 & 1.5 & A & 0.96086 & 1.202 \\
0.004 & 0.25 & 3.0 & 2.49 & 6200 & 1.5 & A & 0.91345 & 1.190 \\
0.004 & 0.25 & 3.0 & 2.49 & 6300 & 1.5 & A & 0.86753 & 1.176 \\
...\\
0.008 & 0.25 & 3.0 & 2.39 & 6100 & 1.5 & A & 0.79791 & 1.151 \\
0.008 & 0.25 & 3.0 & 2.39 & 6200 & 1.5 & A & 0.75783 & 1.138 \\
0.008 & 0.25 & 3.0 & 2.39 & 6300 & 1.5 & A & 0.72289 & 1.126 \\
...\\
0.03 & 0.28 & 4.0 & 2.68 & 6000 & 1.5 & A & 1.21353 & 1.309 \\
0.03 & 0.28 & 4.0 & 2.68 & 6100 & 1.5 & A & 1.15266 & 1.296 \\
0.03 & 0.28 & 4.0 & 2.68 & 6200 & 1.5 & A & 1.09496 & 1.282 \\
\hline\hline
\end{longtable}

\clearpage

\subsection{The Period-Radius and Period-Mass-Radius relations}
\begin{figure}[th]
\centering
\includegraphics[width=0.6\textwidth]{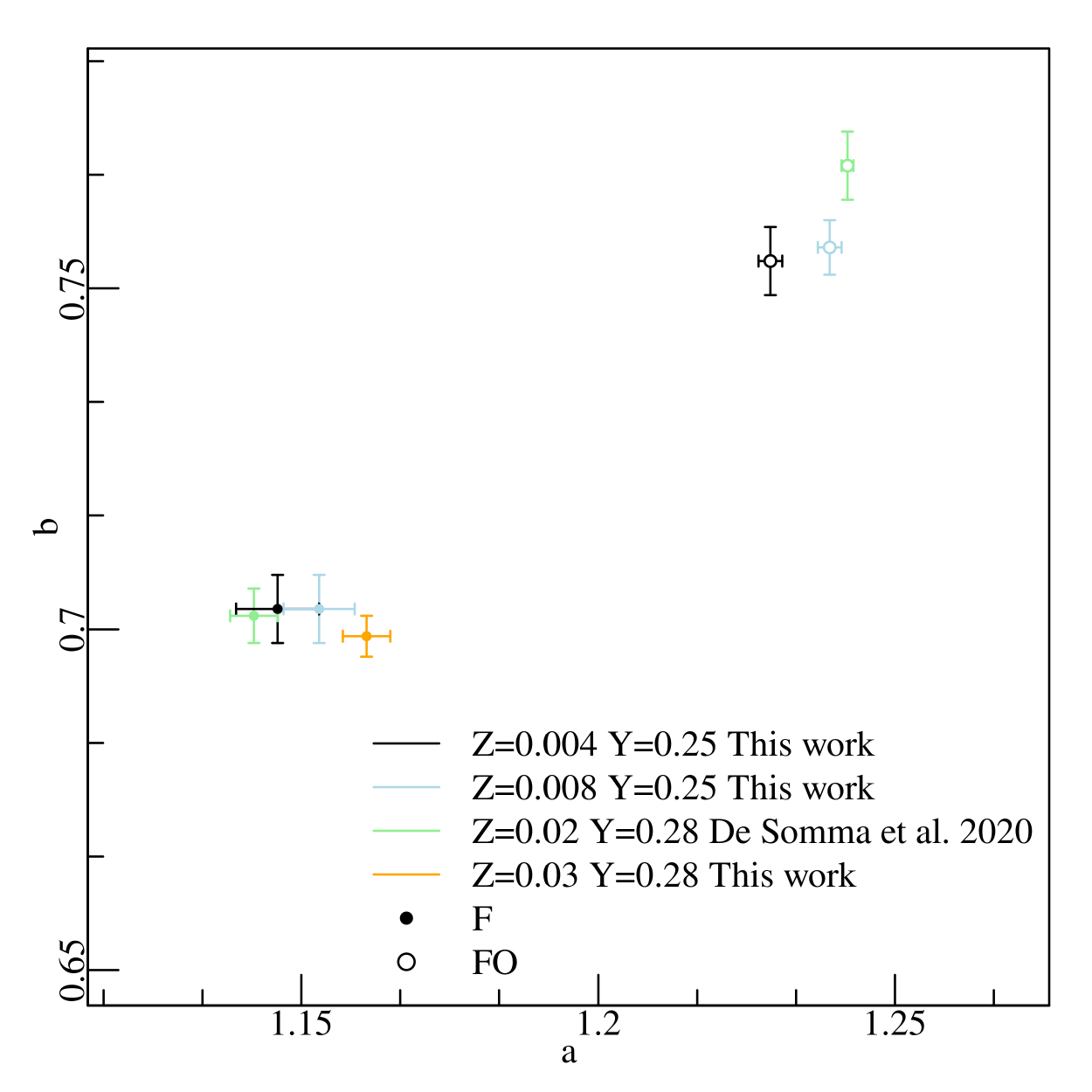}
\caption{The intercept (a) and the slope (b) of the fitted canonical F and FO mode PR relations, at a fixed mixing length parameter $\alpha_{ml}=1.5$ for Z = 0.004 Y = 0.25 (black dots), Z = 0.008 Y = 0.25 (light blue dots), Z = 0.02 Y = 0.28 (green dots) and Z = 0.03 Y = 0.28 (orange dot). The coefficients for FO-mode pulsators are not shown for Z=0.03 because the models were too few to allow an accurate regression.}
\label{Fig:pr_Z_F_FO}
\end{figure}

An important aspect of Cepheid research concerns the use of CCs to infer stellar radii. According to the so-called pulsation relation, the period of a variable at a fixed chemical composition depends on its mean density i.e. on its stellar mass and radius (PMR relation). As such, that independent estimates of both period and mean radius provide an independent evaluation of Cepheid pulsation masses. By neglecting the mass dependence a Period-Radius relation is obtained.
In this work we computed radius curves to derive time-averaged mean radii that were then correlated with the pulsation periods.
As a result, we could obtain updated PR relations for both F and FO models, varying the chemical composition, the assumed ML relation and the efficiency of super-adiabatic convection. The coefficients of the inferred relations are reported in Table \ref{pr_f_fo_all}. In Fig. \ref{Fig:pr_Z_F_FO}, the value of the intercept of our F and FO-mode PR relations (with the associated error) is reported on the x-axis and the value of the slope (with the associated error) is reported on the y axis, for Z = 0.004 (black dots), Z=0.008 (light blue dots), Z=0.02 (green dots taken from \citet[][]{Desomma2020a}) and Z=0.03 (orange dot). This comparison shows that the effect of a variation in the chemical composition is almost negligible, thus suggesting the possibility of deriving a global relation through linear regression of all the periods and mean radii reported in Table \ref{pr_f_fo_togheter}.

\begin{figure}[th]
\centering
\includegraphics[width=0.6\textwidth]{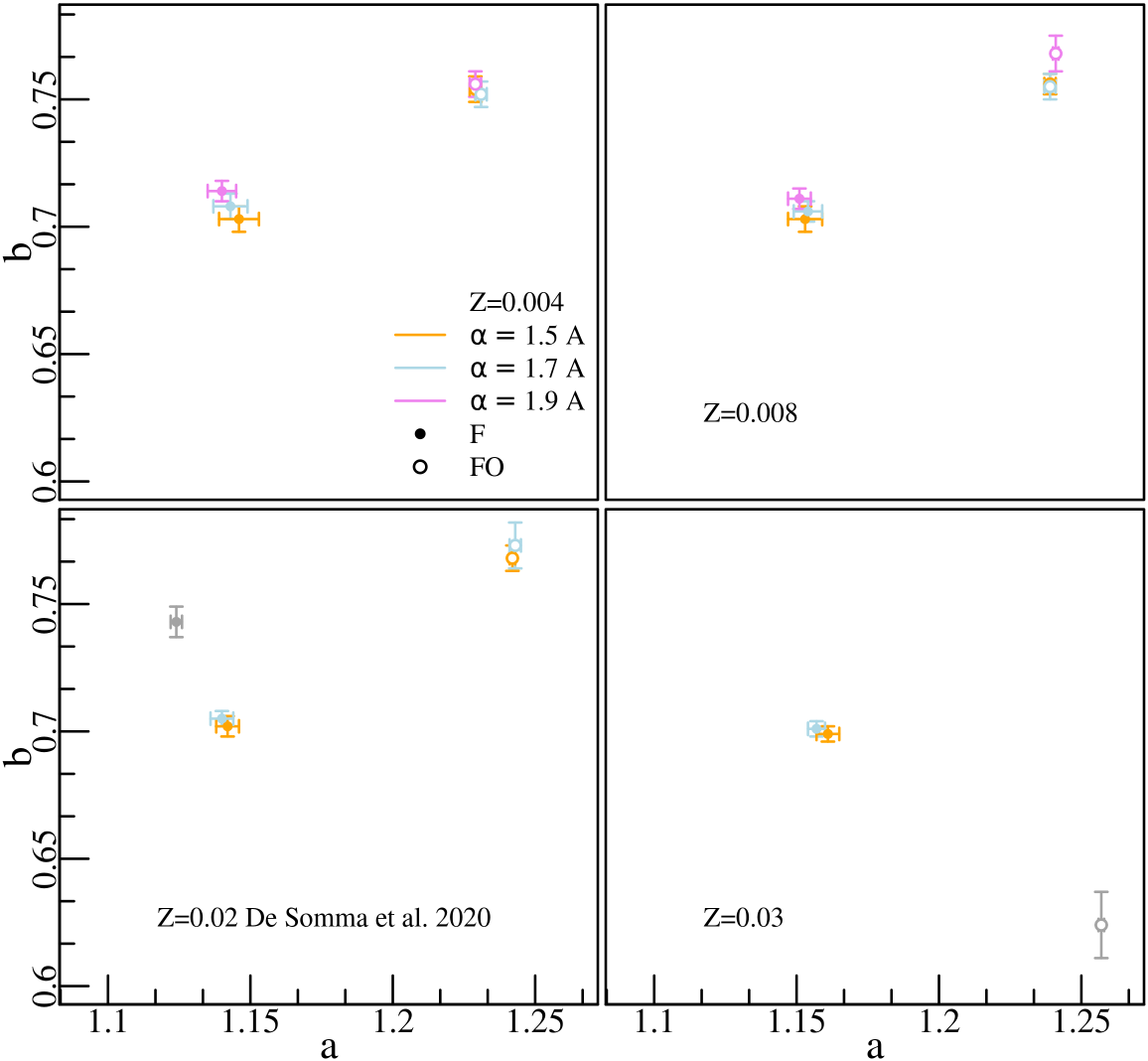}
\caption{The intercept (a) and the slope (b) of the fitted canonical F and FO mode PR relations for Z = 0.004 Y = 0.25, Z = 0.008 Y = 0.25, Z = 0.02 Y = 0.28 (taken from \citet[][]{Desomma2020a}) and Z = 0.03 Y = 0.28 as a function of the mixing length parameter (see text and labels for more details.}
\label{Fig:pr_alfa}
\end{figure}

\begin{figure}[th]
\centering
\includegraphics[width=0.6\textwidth]{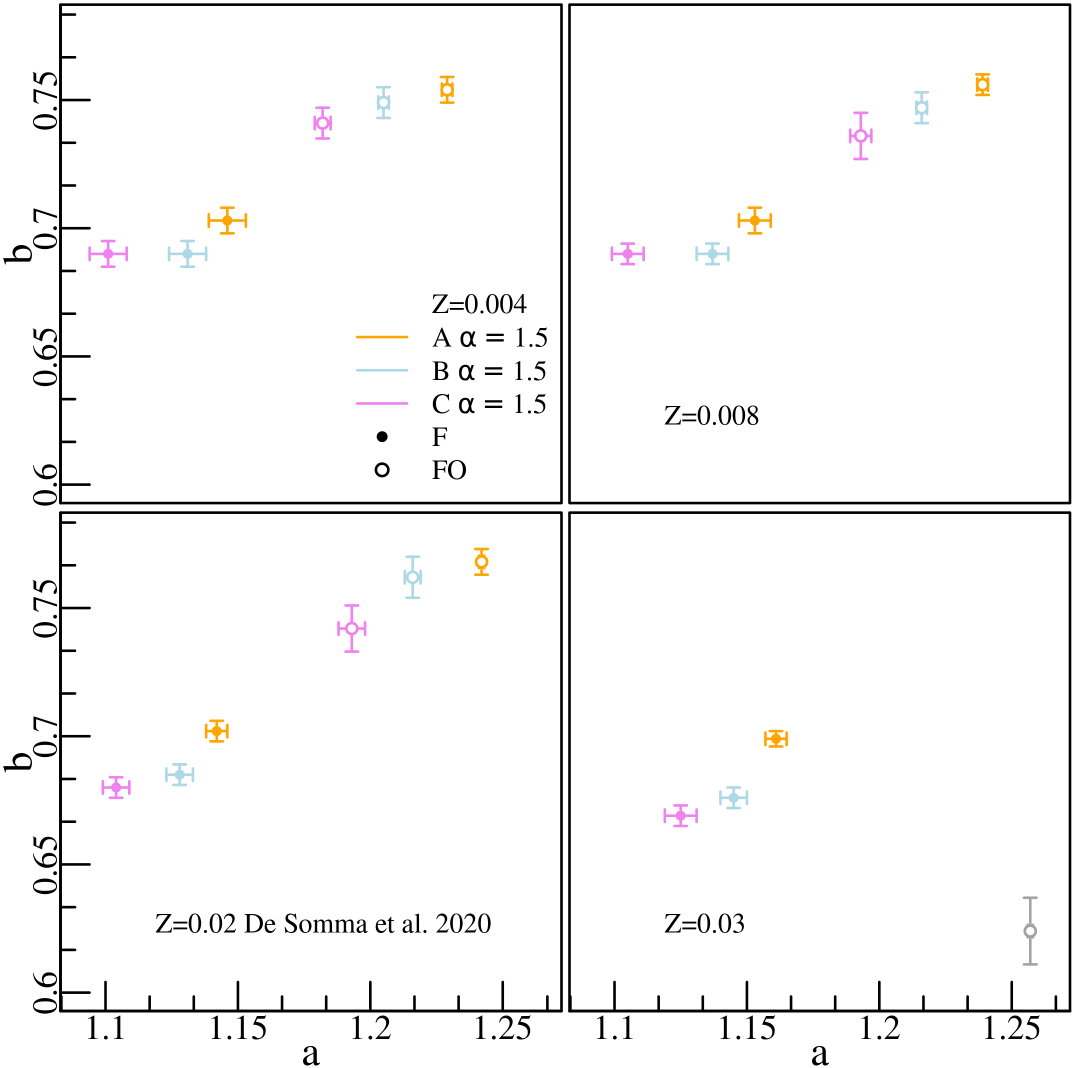}
\caption{The intercept (a) and the slope (b) of the fitted F and FO mode PR relations for Z = 0.004 Y = 0.25, Z = 0.008 Y = 0.25, Z = 0.02 Y = 0.28 (taken from \citet[][]{Desomma2020a}) and Z = 0.03 Y = 0.28, at a fixed mixing length parameter $\alpha_{ml}=1.5$, as a function of the various assumptions about the ML relation (see text and labels for more details.}
\label{Fig:pr_ml}
\end{figure}

We also found (see Fig. \ref{Fig:pr_alfa}) that a variation in the assumed efficiency of      super-adiabatic convection does not significantly affect the slope and the zero point of predicted PR relations. However, a change of the assumed ML relation from case A to case C (see Fig. \ref{Fig:pr_ml}) makes the PR relation flatter, thus predicting a smaller radius at a fixed period. The two grey dots in both Fig. \ref{Fig:pr_alfa} and Fig. \ref{Fig:pr_ml} represent the slope and intercept of the PR relation at Z=0.02 for canonical F mode pulsators with $\alpha_{ml} = 1.9$ and the slope and intercept at Z=0.03 for canonical FO mode pulsators with $\alpha_{ml} = 1.5$. These cases are characterized by a very low number of models and the predicted coefficients are less reliable than the others.

\begin{figure}[th]
\centering
\includegraphics[width=0.9\textwidth]{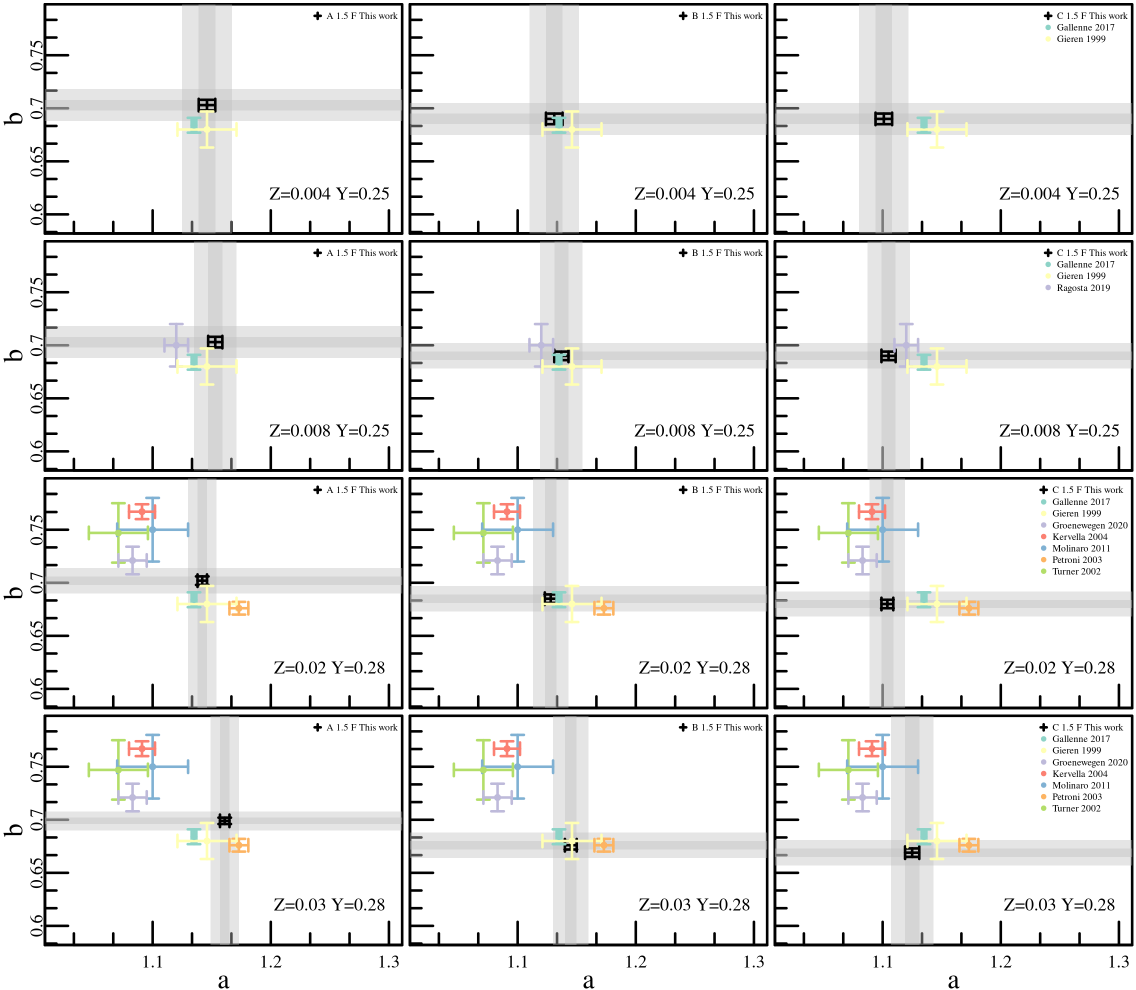}
\caption{The intercept (a) and the slope (b) of the fitted theoretical PR relations for the F-mode pulsators and $\alpha_{ml}=1.5$ are compared to independent results available in the literature. The different panels show the results for different metallicity values, increasing from top (Z=0.004) to bottom (Z=0.03) and different cases for the assumed ML relation: A (left panels), B (middle panels) and C (right panels). In each panel, a (x-axis) and b (y-axis) with their uncertainties for all the considered assumptions are plotted. The theoretical results of this work are indicated with black symbols, while the literature results are plotted using different colors (see labels). The dark and light grey areas indicate the $1\sigma$ error and $3\sigma$ intervals, respectively, of both a and b values of our PR relations.}
\label{Fig:pr_f_comp}
\end{figure}

\begin{figure}[th]
\centering
\includegraphics[width=0.6\textwidth]{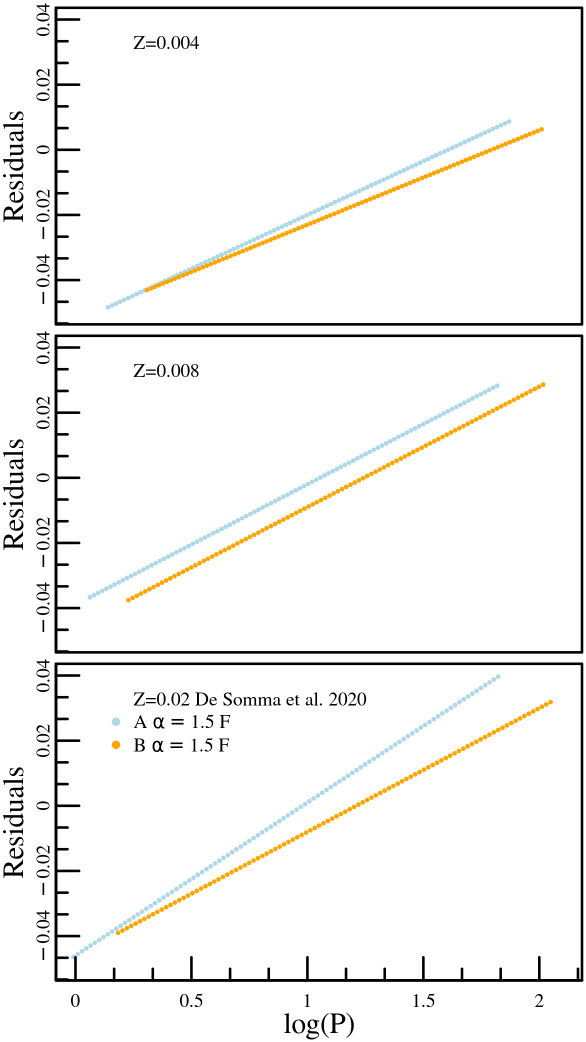}
\caption{This figure shows the residuals, obtained by subtracting the PR relations for F pulsators provided by \citet[][]{Bono1998pr} from the ones derived in this work, chosen as fiducial lines, as a function of the pulsational period. In the top, central and bottom panels, the results for Z = 0.004, Z=0.008, and Z = 0.02 are shown, respectively. In each panel, different ML assumptions are represented with different colors: light blue dots for the canonical case A, and orange dots for the non-canonical case B}
\label{Fig:pr_comp_Bono}
\end{figure}

\begin{figure}[th]
\centering
\includegraphics[width=\textwidth]{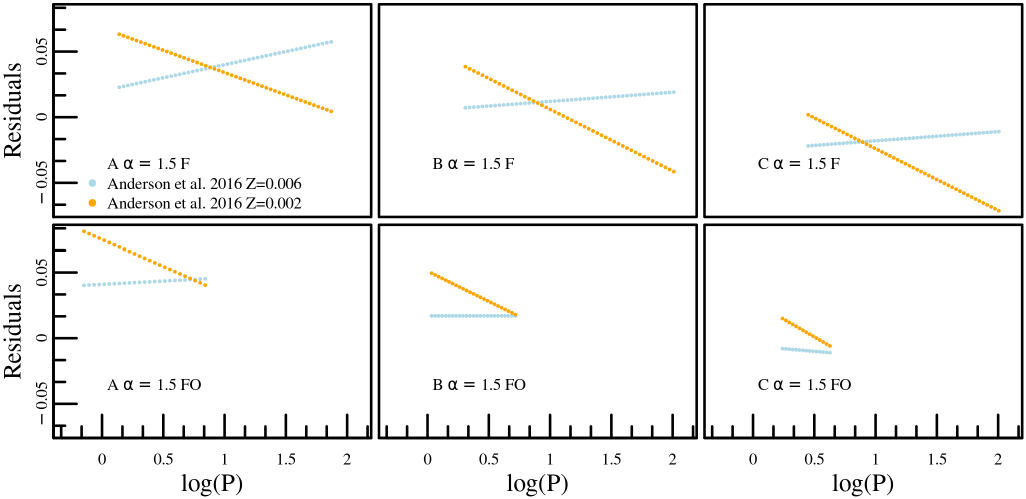}
\caption{As in Fig \ref{Fig:pr_comp_Bono}, but subtracting the PR relations for Z=0.006 and Z=0.002 provided by \citet[][]{Anderson2016}, from the one obtained in this work for Z=0.004 and $\alpha_{ml}=1.5$, as a function of the pulsational period. The top and bottom panels show the results for the F and FO-mode pulsators, respectively. The left panels contain the results for the canonical case A while the central and right panels contain the results for the noncanonical cases B and C, respectively. In each panel, the residuals for Z=0.006 and Z = 0.002 are plotted using light blue and orange dots, respectively.}
\label{Fig:pr_smc_fo_comp_anders}
\end{figure}

The comparison of the PR relations predicted in this paper with independent empirical and semi-empirical relations available in the literature is shown in Figs \ref{Fig:pr_f_comp}, \ref{Fig:pr_comp_Bono} and \ref{Fig:pr_smc_fo_comp_anders}. In fig. \ref{Fig:pr_f_comp} we compared the slope and intercept of our PR relations for $\alpha_{ml}=1.5$, different chemical compositions and ML assumptions, with the slope and intercept with their relative errors of both the empirical and theoretical PR relations taken from the literature (see labels in the plot). The dark grey and light grey areas are the $1\sigma$ error and $3\sigma$ error of the slope and intercept of our theoretical relations, respectively, to investigate the agreement between our relations and ones from the literature. It is shown a better agreement when the A or B cases for $Z=0.004$ and the B case for $Z=0.008$ are adopted. At the highest metal abundances, $Z=0.02$ and $Z=0.03$, it is more difficult to identify the nest ML case as the independent values in the literature are spread on wider ranges, even if the canonical A case seems to be globally less in agreement than B and C.
 
In Fig. \ref{Fig:pr_comp_Bono} the F-mode PR relations for $Z = 0.004$ (upper panels), $Z = 0.008$ (middle panels) and $Z=0.02$ (bottom panels) derived by \citet[][]{Bono1998pr} are subtracted from the ones derived in this paper for the A and B ML relation cases. The agreement is good, as easily expectable considering the adoption of the same hydrodynamical code. We performed this check only to account for the inclusion of a much finer model grid and a slightly different ML relation in the current paper.

In Fig. \ref{Fig:pr_smc_fo_comp_anders} the PR relations for $Z=0.006$ (light blue symbols) and $Z=0.002$ (orange symbols) provided by \citet[][]{Anderson2016}, are subtracted from the ones obtained in this work for $Z=0.004$ and $\alpha_{ml}=1.5$, as a function of the pulsational period, for various assumptions on the ML relation (see labels). 
We noticed that a better agreement is obtained with predictions for $Z=0.006$ than for $Z=0.002$. Moreover, as the \citet[][]{Anderson2016} results are based on brighter models (noncanonical ML relation) which include both overshooting and rotation, we found a general better agreement with the \citet[][]{Anderson2016} relations when considering our noncanonical (cases B and C ML relations) PR relations for both F and FO-modes.

If the width in temperature of the IS is neglected, the pulsation relation connecting the period of a variable, at fixed chemical composition, to the stellar mass, luminosity and effective temperature, can be converted into a period-mass-radius (PMR) relation. This means that independent estimates of both periods and mean radii provide an independent evaluation of Cepheid masses. Theoretical Period-Mass-Radius (PMR) relations are reported in Table \ref{pmr_f_fo_all}. Inspection of this Table suggests that the PMR relation depends on the assumed ML relation and, marginally, on the adopted metal content. In particular, for a short period variable (e.g. $\log P = 0.3$) with $Z=0.004$, the predicted mean radius decreases by 1.4\% from case A to B and by about 5\% from case A to C. For a long period variable (e.g. $\log P = 1.8$) at the same metallicity, the variation in the predicted mean radius decreases by up to 7\% moving from case A to B and by up to 9\% moving from case A to C. On the other hand, for canonical radii, an increase in the metal content from $Z=0.004$ to $Z = 0.008$ and finally to $Z=0.03$ leads to a variation that ranges from 0.5\% to almost 4\%. Similar results on the metallicity dependence were obtained for noncanonical models. Similar to what is shown for the PR relation, we also derived cumulative PMR relations including all the chemical compositions but assumimg various $\alpha_{ml}$ values and ML relations. The coefficients are reported in Table \ref{pmr_f_fo_togheter}.

\clearpage

\begin{longtable}{ccccccc}
\caption{\normalsize{\label{pr_f_fo_all}}The coefficients of the relation $\log$ (\rsun)=a+b $\log P$ for both F and FO Cepheids derived by adopting Z=0.004 Y= 0.25, Z=0.008 Y= 0.25 and Z=0.03 Y= 0.28, as a function of the assumed $\alpha_{ml}$ parameter and  ML relation.}\\
\hline\hline
$\alpha_{ml}$ &ML&a&b& $\sigma_{a}$& $\sigma_{b}$&$R^2$\\
\hline
\endfirsthead
\caption{continued.}\\
\hline\hline
$\alpha_{ml}$ &ML&a&b& $\sigma_{a}$& $\sigma_{b}$&$R^2$\\
\hline
\endhead
\hline\hline
&&&Z=0.004 & Y= 0.25\\
\hline
F\\
\hline
1.5&A&1.146&0.703&0.007&0.005&0.9958\\
1.5&B&1.131&0.690&0.007&0.005&0.9954\\
1.5&C&1.101&0.690&0.007&0.005&0.9955\\
1.7&A&1.143&0.708&0.006&0.005&0.9973\\
1.7&B&1.129&0.692&0.006&0.004&0.9970\\
1.7&C&1.104&0.687&0.006&0.004&0.9970\\
1.9&A&1.140&0.714&0.005&0.004&0.9984\\
1.9&B&1.128&0.694&0.005&0.004&0.9978\\
1.9&C&1.103&0.687&0.006&0.004&0.9978\\
\hline
FO\\
\hline
1.5&A&1.229&0.754&0.002&0.005&0.999\\
1.5&B&1.206&0.725&0.012&0.025&0.9800\\
1.5&C&1.182&0.741&0.003&0.006&0.9997\\
1.7&A&1.231&0.752&0.002&0.005&0.9993\\
1.7&B&1.209&0.713&0.014&0.032&0.9749\\
1.7&C&1.178&0.746&0.007&0.017&0.9989\\
1.9&A&1.229&0.756&0.002&0.005&0.9995\\
1.9&B&1.203&0.752&0.003&0.008&0.9995\\
\hline\hline
&&&Z=0.008 & Y= 0.25\\
\hline
F\\
\hline
1.5&A&1.153&0.703&0.006&0.005&0.9966\\
1.5&B&1.137&0.690&0.006&0.004&0.9961\\
1.5&C&1.105&0.690&0.006&0.004&0.9972\\
1.7&A&1.154&0.706&0.005&0.004&0.9979\\
1.7&B&1.137&0.689&0.006&0.004&0.9973\\
1.7&C&1.111&0.686&0.005&0.004&0.9978\\
1.9&A&1.151&0.711&0.004&0.004&0.9990\\
1.9&B&1.133&0.695&0.005&0.004&0.9986\\
1.9&C&1.115&0.683&0.005&0.003&0.9987\\
\hline
FO\\
\hline
1.5&A&1.239&0.756&0.002&0.004&0.9990\\
1.5&B&1.216&0.747&0.002&0.006&0.9988\\
1.5&C&1.193&0.736&0.004&0.009&0.9987\\
1.7&A&1.239&0.755&0.002&0.005&0.9992\\
1.7&B&1.215&0.747&0.002&0.006&0.9993\\
1.7&C&1.194&0.732&0.003&0.008&0.9993\\
1.9&A&1.241&0.768&0.001&0.007&0.9993\\
1.9&B&1.214&0.748&0.002&0.014&0.9984\\
1.9&C&1.220&0.573&0.001&0.003&1.0000\\
\hline\hline
&&&Z=0.03 & Y= 0.28\\
\hline
F\\
\hline
1.5&A&1.161&0.699&0.004&0.003&0.9988\\
1.5&B&1.145&0.676&0.005&0.004&0.9979\\
1.5&C&1.125&0.669&0.006&0.004&0.9975\\
1.7&A&1.157&0.701&0.003&0.003&0.9997\\
1.7&B&1.140&0.679&0.005&0.004&0.9991\\
1.7&C&1.122&0.670&0.005&0.003&0.9990\\
\hline
FO\\
\hline
1.5&A&1.257&0.624&0.001&0.013&0.9992\\
\hline\hline
\end{longtable}

\clearpage

\begin{table*}
\caption{\label{pr_f_fo_togheter} The coefficients of the relation $\log$ (\rsun)=a+b $\log P$ for both F and FO Cepheids derived by using all the models  as a function of the assumed $\alpha_{ml}$ parameter and ML relation.}
\centering
\begin{tabular}{ccccccc}
\hline\hline
$\alpha_{ml}$&ML&a&b& $\sigma_{a}$& $\sigma_{b}$&$R^2$\\
\hline
F\\
\hline
1.5&A&1.148&0.703&0.003&0.002&0.9972\\
1.5&B&1.134&0.686&0.003&0.002&0.9964\\
1.5&C&1.103&0.687&0.004&0.003&0.9965\\
1.7&A&1.145&0.708&0.003&0.003&0.9978\\
1.7&B&1.132&0.689&0.003&0.002&0.9972\\
1.7&C&1.104&0.687&0.003&0.002&0.9976\\
1.9&A&1.142&0.715&0.003&0.003&0.9986\\
1.9&B&1.127&0.696&0.003&0.003&0.9982\\
1.9&C&1.104&0.688&0.003&0.002&0.9983\\
\hline
FO\\
\hline
1.5&A&1.238&0.740&0.003&0.008&0.9921\\
1.5&B&1.215&0.726&0.004&0.011&0.9903\\
1.5&C&1.193&0.729&0.003&0.007&0.9981\\
1.7&A&1.237&0.754&0.001&0.004&0.9989\\
1.7&B&1.214&0.721&0.005&0.014&0.9887\\
1.7&C&1.190&0.733&0.004&0.011&0.9976\\
1.9&A&1.235&0.754&0.002&0.006&0.9985\\
1.9&B&1.210&0.742&0.002&0.009&0.9982\\
1.9&C&1.212&0.621&0.002&0.012&0.9992\\
\hline\hline
\end{tabular}
\end{table*}

\begin{longtable}{ccccccccc}
\caption{\normalsize{\label{pmr_f_fo_all}}The coefficients of the relation $\log$ \rsun=a+b $\log$ P + c $\log$ M for both F and FO Cepheids derived by adopting Z=0.004 Y= 0.25, Z=0.008 Y= 0.25, and Z=0.03 Y= 0.28 as a function of the assumed $\alpha_{ml}$ parameter and ML relations.}\\
\hline\hline
$\alpha_{ml}$ &ML&a&b&c&$\sigma_{a}$& $\sigma_{b}$&$\sigma_{c}$&$R^2$\\
\hline
\endfirsthead
\caption{continued.}\\
\hline\hline
$\alpha_{ml}$ &ML&a&b&c&$\sigma_{a}$& $\sigma_{b}$&$\sigma_{c}$&$R^2$\\
\hline
\endhead
\hline\hline
&&&&Z=0.004 & Y= 0.25\\
\hline
F\\
\hline
1.5&A&0.893&0.520&0.541&0.009&0.006&0.018&0.9997\\
1.5&B&0.918&0.511&0.536&0.009&0.008&0.022&0.9994\\
1.5&C&0.948&0.515&0.503&0.009&0.010&0.027&0.9992\\
1.7&A&0.893&0.521&0.538&0.009&0.007&0.019&0.9998\\
1.7&B&0.909&0.506&0.550&0.009&0.008&0.022&0.9996\\
1.7&C&0.943&0.515&0.505&0.009&0.009&0.027&0.9995\\
1.9&A&0.884&0.517&0.551&0.011&0.009&0.024&0.9999\\
1.9&B&0.893&0.494&0.585&0.011&0.010&0.028&0.9997\\
1.9&C&0.930&0.506&0.533&0.010&0.010&0.030&0.9996\\
\hline
FO\\
\hline
1.5&A&0.720&0.365&0.988&0.061&0.044&0.117&0.9961\\
1.5&B&0.815&0.418&0.873&0.080&0.065&0.178&0.9920\\
1.5&C&1.072&0.636&0.288&0.022&0.021&0.056&1.0000\\
1.7&A&1.013&0.594&0.418&0.011&0.008&0.021&1.0000\\
1.7&B&0.709&0.335&1.105&0.097&0.076&0.214&0.9922\\
1.7&C&0.973&0.550&0.539&0.031&0.030&0.082&1.0000\\
1.9&A&1.000&0.586&0.441&0.029&0.022&0.056&0.9999\\
1.9&B&1.021&0.595&0.409&0.007&0.006&0.016&1.0000\\
\hline\hline
&&&&Z=0.008 & Y= 0.25\\
\hline
F\\
\hline
1.5&A&0.907&0.542&0.490&0.008&0.005&0.016&0.9997\\
1.5&B&0.913&0.525&0.513&0.010&0.007&0.021&0.9994\\
1.5&C&0.956&0.543&0.437&0.010&0.009&0.027&0.9993\\
1.7&A&0.899&0.537&0.505&0.010&0.006&0.019&0.9999\\
1.7&B&0.881&0.503&0.581&0.010&0.007&0.023&0.9997\\
1.7&C&0.948&0.534&0.460&0.010&0.009&0.028&0.9995\\
1.9&A&0.898&0.538&0.501&0.014&0.010&0.028&0.9999\\
1.9&B&0.865&0.492&0.612&0.016&0.012&0.037&0.9998\\
1.9&C&0.917&0.509&0.539&0.015&0.014&0.042&0.9997\\
\hline
FO\\
\hline
1.5&A&1.012&0.595&0.416&0.009&0.006&0.016&1.0000\\
1.5&B&1.037&0.612&0.370&0.010&0.007&0.020&0.9999\\
1.5&C&1.045&0.617&0.351&0.009&0.007&0.021&1.0000\\
1.7&A&0.996&0.586&0.441&0.016&0.011&0.030&0.9999\\
1.7&B&1.025&0.600&0.397&0.009&0.007&0.018&1.0000\\
1.7&C&1.033&0.603&0.382&0.012&0.010&0.029&1.0000\\
1.9&A&1.020&0.611&0.397&0.019&0.013&0.034&1.0000\\
1.9&B&1.017&0.595&0.410&0.018&0.015&0.038&0.9999\\
\hline\hline
&&&&Z=0.03 & Y= 0.28\\
\hline
F\\
\hline
1.5&A&0.886&0.545&0.500&0.019&0.011&0.035&0.9997\\
1.5&B&0.901&0.522&0.522&0.016&0.010&0.034&0.9995\\
1.5&C&0.960&0.434&0.537&0.012&0.030&0.009&0.9994\\
1.7&A&0.864&0.537&0.531&0.038&0.072&0.023&0.9999\\
1.7&B&0.853&0.617&0.494&0.028&0.061&0.018&0.9998\\
1.7&C&0.962&0.419&0.542&0.020&0.053&0.016&0.9996\\
\hline\hline
\end{longtable}

\begin{table*}
\caption{\label{pmr_f_fo_togheter} The coefficients of the relation $\log$ \rsun=a+b $\log$ P + c $\log$ M for both F and FO Cepheids derived by using all the models as a function of the assumed $\alpha_{ml}$ parameter and ML relation.}
\centering
\begin{tabular}{ccccccccc}
\hline\hline
$\alpha_{ml}$ &ML&a&b&c&$\sigma_{a}$& $\sigma_{b}$&$\sigma_{c}$&$R^2$\\
\hline
F\\
\hline
1.5&A&0.916&0.554&0.460&0.008&0.005&0.015&0.9994\\
1.5&B&0.915&0.527&0.506&0.007&0.005&0.015&0.9991\\
1.5&C&0.954&0.539&0.446&0.006&0.006&0.018&0.9990\\
1.7&A&0.916&0.552&0.465&0.011&0.007&0.022&0.9994\\
1.7&B&0.896&0.512&0.550&0.009&0.007&0.020&0.9994\\
1.7&C&0.949&0.538&0.451&0.007&0.007&0.021&0.9993\\
1.9&A&0.930&0.564&0.435&0.012&0.008&0.024&0.9997\\
1.9&B&0.909&0.524&0.518&0.011&0.009&0.027&0.9995\\
1.9&C&0.941&0.534&0.468&0.010&0.009&0.028&0.9995\\
\hline
FO\\
\hline
1.5&A&0.924&0.531&0.577&0.026&0.018&0.047&0.9974\\
1.5&B&0.957&0.549&0.539&0.030&0.022&0.063&0.9963\\
1.5&C&1.057&0.626&0.322&0.006&0.005&0.014&0.9999\\
1.7&A&1.048&0.623&0.348&0.009&0.007&0.017&0.9999\\
1.7&B&0.909&0.512&0.639&0.048&0.034&0.100&0.9953\\
1.7&C&1.041&0.611&0.361&0.007&0.006&0.017&0.9999\\
1.9&A&1.031&0.613&0.379&0.010&0.007&0.018&0.9999\\
1.9&B&1.055&0.626&0.331&0.010&0.007&0.020&0.9999\\
\hline\hline
\end{tabular}
\end{table*}

\subsection{The Period-Luminosity-Mass-Temperature relations}

The pulsation relation derived from the combination of the definition of the pulsation period as a function of the stellar mean density (period-density relation) and the Stefan-Boltzmann law, is a Period-Luminosity-Mass-effective Temperature (hereinafter PLMT) relation, crucially connecting the characteristic oscillation time to structural parameters resulting from stellar evolution. We combined all the parameters reported in Tables \ref{f_param_model} and \ref{fo_param_model} to derive new PMLT relations (see Table \ref{pmlt_f_fo}) for each assumed chemical composition and both pulsation modes. As a result, we found that fundamental periods show a small correlation with the assumed metal content at a fixed mass, luminosity and effective temperature, whereas no significant dependence is found for first overtone models. On this basis, we also derived metal-dependent PLMT relations (hereinafter called PLMTZ relations) for F-mode models only. The coefficients and the intrinsic dispersion of the PLMTZ relations derived by varying the efficiency of the super-adiabatic convection are reported in Table \ref{pmlt_f_Z}. We noticed that the effect of a variation in the mixing length parameter is negligible within the error, thus confirming previous results obtained for solar chemical composition models \citep[see][]{Desomma2020a}.

\begin{table*}
\caption{\label{pmlt_f_fo} The coefficients of the PMLT relations $\log P$ = a +b$\log T_{eff}$ + c $\log$ (\msun) + d \lsun for both F and FO pulsators as a function of the assumed $\alpha_{ml}$ parameter for Z=0.004 Y= 0.25, Z=0.008 Y= 0.25 and Z=0.03 Y= 0.28.}
\centering
\begin{tabular}{cccccccccc}
\hline\hline
$\alpha_{ml}$&a&b&c&d&$\sigma_{a}$&$\sigma_{b}$&$\sigma_{c}$&$\sigma_{d}$&$R^2$\\
\hline\hline
&&&&Z=0.004 & Y= 0.25\\
\hline
F\\
\hline
1.5&10.711&-3.315&-0.776&0.918&0.109&0.028&0.017&0.005&0.9990\\
1.7&10.699&-3.312&-0.788&0.921&0.104&0.026&0.014&0.005&0.9995\\
1.9&10.764&-3.327&-0.791&0.919&0.124&0.031&0.014&0.005&0.9996\\
\hline
FO\\
\hline
1.5&12.042&-3.636&-0.574&0.799&0.271&0.325&0.112&0.037&0.9867\\
1.7&11.590&-3.524&-0.677&0.828&0.385&0.352&0.102&0.036&0.9929\\
1.9&11.182&-3.410&-0.591&0.801&0.434&0.111&0.023&0.009&0.9999\\
\hline\hline
&&&&Z=0.008 & Y= 0.25\\
\hline
F\\
\hline
1.5&10.482&-3.254&-0.773&0.920&0.103&0.026&0.017&0.005&0.9991\\
1.7&10.588&-3.283&-0.777&0.922&0.110&0.028&0.015&0.005&0.9995\\
1.9&11.077&-3.405&-0.770&0.911&0.184&0.046&0.017&0.007&0.9996\\
\hline
FO\\
\hline
1.5&10.880&-3.337&-0.622&0.816&0.122&0.031&0.009&0.003&0.9999\\
1.7&10.927&-3.348&-0.622&0.813&0.223&0.057&0.014&0.005&0.9998\\
1.9&10.774&-3.300&-0.618&0.802&0.302&0.077&0.014&0.006&0.9998\\
\hline\hline
&&&&Z=0.03 & Y= 0.28\\
\hline
F\\
\hline
1.5&10.414&-3.227&-0.765&0.918&0.119&0.029&0.023&0.008&0.9998\\
\hline\hline
\end{tabular}
\end{table*}

\begin{table*}
\caption{\label{pmlt_f_Z} The coefficients of the PMLTZ relations $\log P$ = a +b$\log T_{eff}$ + c $\log$ (\msun) + d \lsun + e [Fe/H] for both F and FO pulsators as a function of the assumed $\alpha_{ml}$ parameter obtained by using all the computed models for Z = 0.004 Y= 0.25, Z = 0.008 Y = 0.25, Z= 0.02 Y = 0.25 and Z = 0.03 Y = 0.28.}
\centering
\begin{tabular}{cccccccccccc}
\hline\hline
$\alpha_{ml}$&a&b&c&d&e&$\sigma_{a}$&$\sigma_{b}$&$\sigma_{c}$&$\sigma_{d}$&$\sigma_{e}$&$R^2$\\
\hline
F\\
\hline
1.5&10.513&-3.255&-0.772&0.919&0.046&0.054&0.014&0.009&0.003&0.002&0.9993\\
1.7&10.663&-3.295&-0.780&0.919&0.043&0.071&0.018&0.009&0.003&0.002&0.9995\\
1.9&10.891&-3.351&-0.783&0.914&0.036&0.107&0.027&0.011&0.004&0.003&0.9996\\
\hline\hline
\end{tabular}
\end{table*}

\subsection{The new predicted instability strip}
As a result of the integration of the nonlinear system of hydrodynamic equations, the limit cycle stability of each selected pulsation model was obtained and the topology of the instability strips for the aforementioned chemical compositions was derived. The effective temperatures of the inferred F and FO boundaries are reported in Tables \ref{boundaries_smc}, \ref{boundaries_lmc} and \ref{boundaries_m31}, for $Z=0.004$, $Z=0.008$ and $Z=0.03$, respectively, in Section \ref{sec:boundaries} of the Appendix. In each table, columns from 1 to 8 report the mass, the luminosity level, the adopted mixing length parameter, the ML label and the effective temperatures for the First Overtone Blue Edge (FOBE), the Fundamental Blue Edge (FBE), the First Overtone Red Edge (FORE) and the Fundamental Red Edge (FRE).

A linear regression of the inferred boundary effective temperature values, as a function of the luminosity level, for different assumptions of the $\alpha_{ml}$ parameter, provides the relations reported in Tables \ref{tl_lin_smc}, \ref{tl_lin_lmc}, \ref{tl_lin_m31}, for the F and FO-mode respectively and varying the metal abundance from $Z=0.004$ to $Z=0.03$. The quadratic fit of the F-mode boundaries provides the relations reported in Tables \ref{tl_quad_smc}, \ref{tl_quad_lmc} and \ref{tl_quad_m31}.

\begin{figure}[th]
\centering
\includegraphics[width=0.8\textwidth]{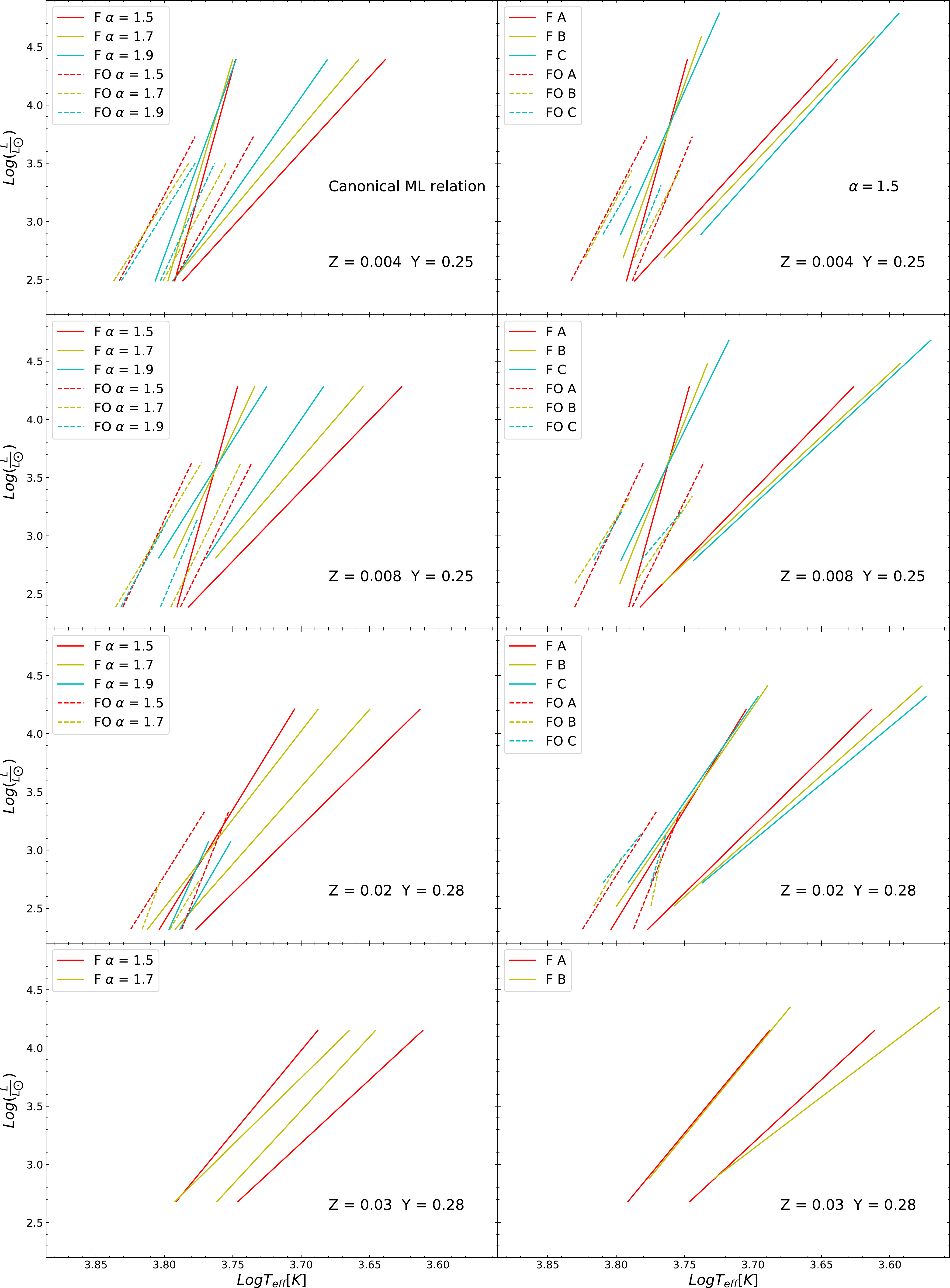}
\caption{The F and FO instability strips for Z = 0.004 Y = 0.25, Z = 0.008 Y = 0.25, Z = 0.02 Y = 0.28 and Z = 0.03 Y = 0.28 pulsators at a fixed case A ML relation for the various assumptions about the superadiabatic convective efficiency (left panels) and at a fixed mixing length parameter $\alpha_{ml}=1.5$ for the assumed A, B, C ML relations  (right panels).}
\label{Fig:strips}
\end{figure}

Fig. \ref{Fig:strips} shows the linear relations for the predicted boundaries, at the labeled chemical compositions, varying both the super-adiabatic convective efficiency (left panels) and the ML assumption (right panels). These plots show that, in agreement with previous investigations \citep[][]{Bono1997, Bono1999} based on an earlier version of the same hydrodynamical code and an older version of the ML relation, a change in the chemical composition does significantly affect the topology of the instability strip.
Indeed, as the metallicity increases from $Z=0.004$ to $Z=0.03$, at each fixed mixing length parameter and ML relation, the strip gets redder. The width of the instability strip decreases as the efficiency of super-adiabatic convection increases. This is particularly evident for the FO-mode that becomes less and less efficient till it disappears at super-solar metallicity ($Z=0.03$) for $\alpha_{ml}$=1.9. This occurrence is related to both the significant decrease of the Hydrogen abundance that reduces the efficiency of the more external FO-mode pulsation, and to the increased opacity related to the iron peak that enhances the convective efficiency and its damping effect on pulsation.

\begin{figure}[th]
\centering
\includegraphics[width=\textwidth]{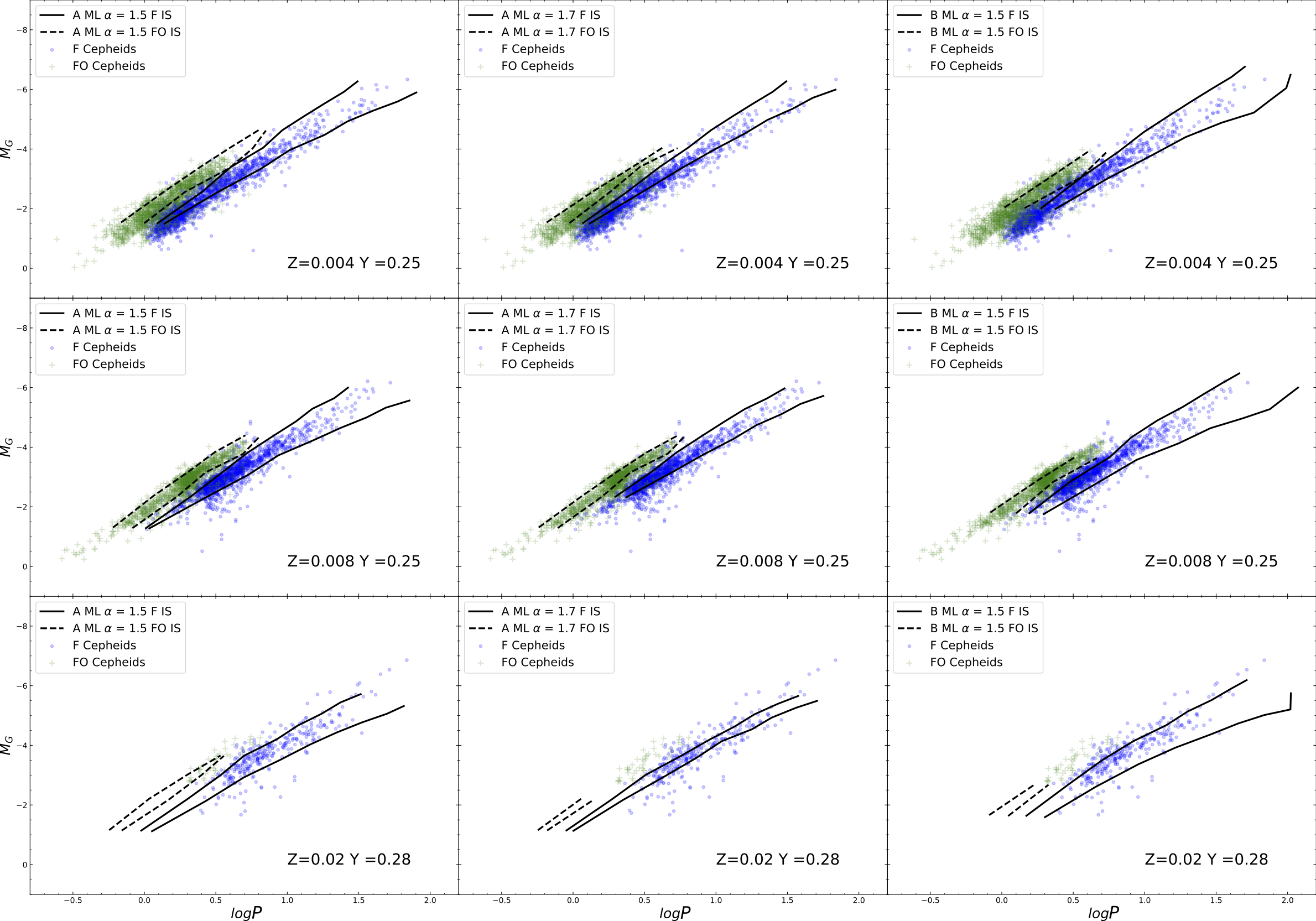}
\caption{The location of F and FO-mode pulsators (green plus and blue points, respectively) for $Z=0.004$ (upper panels), $Z=0.008$ (middle panels) and $Z=0.02$ (bottom panels) in the $logP$-G plane. The solid and dashed black lines are the theoretical F and FO-mode instability strip boundaries, respectively, for the labeled ML and $\alpha_{ml}$ assumptions.}
\label{Fig:strip_oss}
\end{figure}

However, other authors \citep[see e.g.][]{Anderson2016} found a lower dependence of the instability strip morphology on the chemical composition based on linear nonadiabatic pulsation models. To test the accuracy of our predictions, in Fig. \ref{Fig:strip_oss} we compared our theoretical boundaries for $Z = 0.004$, $Z=0.008$ and $Z= 0.02$ with available cepheid data with Gaia parallaxes. In particular, we adopted Gaia DR2 for SMC and LMC assuming a mean distance modulus of $\mu=18.977$ mag \citep[][]{Grac_smc_2020} and $\mu=18.48$ mag \citep[][]{Pietr_lmc_2019}, respectively. On the other hand, the boundaries for $Z=0.02$ were compared with Cepheids with Gaia EDR3 individual distances based on the parallaxes with relative errors better than 10\%. We noted that the agreement is generally good with a slightly better match for ML case A with $\alpha_{ml}=1.7$ for the Magellanic Clouds, at least for the assumed distance moduli. For the MW the situation is more confused but the best agreement seems to be obtained for the canonical ML relation with $\alpha_{ml} =1.5$ for the blue boundary and $\alpha_{ml}= 1.7$ for the red boundary. The possibility that the assumed $\alpha_{ml}$ value varies across the instability strip was already discussed in previous papers \citep[][]{Dicrisci2004, Fiorentino2007}.

\clearpage

\begin{table*}
\caption{\label{tl_lin_smc} The coefficients of the linear relation $\log T_{eff}$ = a +b \lsun for the boundaries of the F and FO-mode IS for Z = 0.004 Y= 0.25 varying both the ML relation and the mixing length parameter.}
\centering
\begin{tabular}{ccccccccc}
\hline\hline
$\alpha_{ml}$&ML&a&b&$\sigma_{a}$&$\sigma_{b}$&$R^2$\\
\hline
FOBE\\
\hline
1.5&A&3.941&-0.044&0.012&0.004&0.9796\\
1.5&B&3.945&-0.045&0.005&0.002&0.9988\\
1.7&A&3.971&-0.054&0.007&0.002&0.9963\\
1.7&B&3.970&-0.054&0.014&0.005&0.9932\\
1.9&A&3.961&-0.052&0.017&0.005&0.9791\\
1.9&B&3.966&-0.055&0.014&0.005&0.9931\\
\hline
FBE\\
\hline
1.5&A&3.825&-0.016&0.012&0.003&0.8717\\
1.5&B&3.849&-0.022&0.011&0.003&0.8879\\
1.5&C&3.889&-0.033&0.010&0.003&0.9589\\
1.7&A&3.842&-0.020&0.009&0.002&0.9075\\
1.7&B&3.876&-0.029&0.013&0.003&0.9188\\
1.7&C&3.911&-0.038&0.008&0.002&0.9803\\
1.9&A&3.873&-0.028&0.013&0.004&0.9009\\
1.9&B&3.895&-0.034&0.011&0.003&0.9565\\
1.9&C&3.924&-0.042&0.005&0.001&0.9939\\
\hline
FORE\\
\hline
1.5&A&3.844&-0.026&0.043&0.013&0.5588\\
1.5&B&3.888&-0.042&0.099&0.032&0.6320\\
1.7&A&3.870&-0.031&0.046&0.015&0.6807\\
1.7&B&3.894&-0.041&0.097&0.031&0.6321\\
1.9&A&3.882&-0.033&0.031&0.010&0.8443\\
1.9&B&3.867&-0.028&0.058&0.019&0.6879\\
\hline
FRE\\
\hline
1.5&A&3.958&-0.074&0.014&0.004&0.9814\\
1.5&B&3.937&-0.069&0.033&0.009&0.9031\\
1.5&C&3.762&-0.021&0.073&0.018&0.7445\\
1.7&A&3.941&-0.064&0.011&0.003&0.9840\\
1.7&B&3.949&-0.069&0.014&0.004&0.9819\\
1.7&C&3.825&-0.035&0.069&0.017&0.3785\\
1.9&A&3.920&-0.055&0.007&0.002&0.9918\\
1.9&B&3.950&-0.065&0.015&0.004&0.9758\\
1.9&C&3.860&-0.042&0.061&0.015&0.5236\\
\hline\hline
\end{tabular}
\end{table*}

\begin{table*}
\caption{\label{tl_quad_smc} The coefficients of the quadratic relation $\log T_{eff}$ = a + b \lsun +c (\lsun)$^2$ for the boundaries of the F-mode IS for Z = 0.004 Y= 0.25 varying both the ML relation and the mixing length parameter.}
\centering
\begin{tabular}{ccccccccc}
\hline\hline
$\alpha_{ml}$&ML&a&b&c&$\sigma_{a}$&$\sigma_{b}$&$\sigma_{c}$&$R^2$\\
\hline
FBE\\
\hline
1.5&A&3.708&0.054&-0.010&0.057&0.034&0.005&0.9003\\
1.5&B&3.681&0.072&-0.013&0.033&0.018&0.003&0.9791\\
1.5&C&3.740&0.047&-0.010&0.040&0.021&0.003&0.9878\\
1.7&A&3.734&0.045&-0.009&0.033&0.019&0.003&0.9676\\
1.7&B&3.674&0.084&-0.015&0.010&0.005&0.001&0.9989\\
1.7&C&3.847&-0.004&-0.004&0.053&0.028&0.004&0.9842\\
1.9&A&3.702&0.074&-0.015&0.039&0.023&0.003&0.9766\\
1.9&B&3.788&0.026&-0.008&0.055&0.031&0.004&0.9736\\
1.9&C&3.885&-0.021&-0.003&0.032&0.017&0.002&0.9951\\
\hline
FRE\\
\hline
1.5&A&3.776&0.035&-0.016&0.046&0.027&0.004&0.9950\\
1.5&B&4.157&-0.193&0.017&0.202&0.112&0.015&0.9195\\
1.5&C&4.917&-0.633&0.079&0.196&0.103&0.013&0.9199\\
1.7&A&3.789&0.026&-0.013&0.034&0.020&0.003&0.9964\\
1.7&B&3.880&-0.030&-0.005&0.087&0.048&0.007&0.9836\\
1.7&C&4.728&-0.515&0.062&0.316&0.166&0.021&0.7406\\
1.9&A&3.851&-0.014&-0.006&0.033&0.019&0.003&0.9953\\
1.9&B&3.738&0.053&-0.016&0.050&0.028&0.004&0.9940\\
1.9&C&4.520&-0.392&0.045&0.339&0.178&0.023&0.7108\\
\hline\hline
\end{tabular}
\end{table*}

\begin{table*}
\caption{\label{tl_lin_lmc} The coefficients of the linear relation $\log T_{eff}$ = a +b \lsun for the boundaries of the F and FO-mode IS for Z = 0.008 Y= 0.25 varying both the ML relation and the mixing length parameter.}
\centering
\begin{tabular}{ccccccccc}
\hline\hline
$\alpha_{ml}$&ML&a&b&$\sigma_{a}$&$\sigma_{b}$&$R^2$\\
\hline
FOBE\\
\hline
1.5&A&3.916&-0.037&0.021&0.007&0.9073\\
1.5&B&3.969&-0.054&0.013&0.004&0.9932\\
1.7&A&3.949&-0.048&0.020&0.006&0.9506\\
1.7&B&3.965&-0.054&0.014&0.005&0.9932\\
1.9&A&3.936&-0.044&0.024&0.009&0.9626\\
\hline
FBE\\
\hline
1.5&A&3.833&-0.020&0.012&0.003&0.8362\\
1.5&B&3.865&-0.029&0.017&0.005&0.8444\\
1.5&C&3.905&-0.040&0.013&0.003&0.9527\\
1.7&A&3.888&-0.035&0.020&0.005&0.8745\\
1.7&B&3.910&-0.041&0.024&0.006&0.8601\\
1.7&C&3.921&-0.044&0.012&0.003&0.9660\\
1.9&A&3.934&-0.048&0.025&0.007&0.8960\\
1.9&B&3.942&-0.051&0.020&0.005&0.9288\\
1.9&C&3.967&-0.058&0.012&0.003&0.9802\\
\hline
FORE\\
\hline
1.5&A&3.864&-0.034&0.029&0.009&0.8122\\
1.5&B&3.879&-0.039&0.078&0.026&0.6879\\
1.7&A&3.870&-0.034&0.029&0.009&0.8123\\
1.7&B&3.885&-0.038&0.076&0.025&0.6879\\
1.9&A&3.869&-0.029&0.040&0.014&0.8073\\
\hline
FRE\\
\hline
1.5&A&3.975&-0.081&0.014&0.004&0.9839\\
1.5&B&3.987&-0.087&0.022&0.006&0.9692\\
1.5&C&3.857&-0.050&0.058&0.015&0.6218\\
1.7&A&3.965&-0.072&0.011&0.003&0.9899\\
1.7&B&3.993&-0.084&0.026&0.007&0.9540\\
1.7&C&3.900&-0.058&0.050&0.013&0.7475\\
1.9&A&3.930&-0.058&0.009&0.002&0.9899\\
1.9&B&3.962&-0.069&0.015&0.004&0.9778\\
1.9&C&3.943&-0.066&0.038&0.010&0.8726\\
\hline\hline
\end{tabular}
\end{table*}

\begin{table*}
\caption{\label{tl_quad_lmc} The coefficients of the quadratic relation $\log T_{eff}$ = a + b \lsun +c (\lsun)$^2$ for the boundaries of the F-mode IS for Z = 0.008 Y= 0.25 varying both the ML relation and the mixing length parameter.}
\centering
\begin{tabular}{ccccccccc}
\hline\hline
$\alpha_{ml}$&ML&a&b&c&$\sigma_{a}$&$\sigma_{b}$&$\sigma_{c}$&$R^2$\\
\hline
FBE\\
\hline
1.5&A&3.696&0.065&-0.013&0.044&0.026&0.004&0.9391\\
1.5&B&3.633&0.106&-0.019&0.061&0.035&0.005&0.9549\\
1.5&C&3.711&0.066&-0.014&0.046&0.025&0.003&0.9885\\
1.7&A&3.528&0.171&-0.029&0.066&0.038&0.005&0.9820\\
1.7&B&3.551&0.166&-0.029&0.044&0.025&0.004&0.9885\\
1.7&C&3.765&0.041&-0.011&0.056&0.031&0.004&0.9852\\
1.9&A&3.483&0.211&-0.036&0.069&0.040&0.006&0.9892\\
1.9&B&3.633&0.128&-0.025&0.030&0.017&0.002&0.9963\\
1.9&C&3.812&0.027&-0.011&0.056&0.031&0.004&0.9913\\
\hline
FRE\\
\hline
1.5&A&3.806&0.023&-0.015&0.049&0.030&0.004&0.9947\\
1.5&B&3.891&-0.031&-0.008&0.139&0.080&0.011&0.9715\\
1.5&C&4.646&-0.481&0.057&0.237&0.128&0.017&0.8691\\
1.7&A&3.820&0.011&-0.012&0.071&0.040&0.006&0.9946\\
1.7&B&3.607&0.139&-0.031&0.062&0.036&0.005&0.9939\\
1.7&C&4.482&-0.376&0.042&0.253&0.137&0.018&0.8674\\
1.9&A&3.936&-0.061&0.000&0.076&0.043&0.006&0.9899\\
1.9&B&3.768&0.043&-0.016&0.055&0.031&0.004&0.9929\\
1.9&C&4.159&-0.184&0.016&0.247&0.133&0.018&0.8873\\
\hline\hline
\end{tabular}
\end{table*}

\begin{table*}
\caption{\label{tl_lin_m31} The coefficients of the linear relation $\log T_{eff}$ = a +b \lsun for the boundaries of the F-mode IS for Z = 0.03 Y= 0.28 varying both the ML relation and the mixing length parameter.}
\centering
\begin{tabular}{ccccccccc}
\hline\hline
$\alpha_{ml}$&ML&a&b&$\sigma_{a}$&$\sigma_{b}$&$R^2$\\
\hline
FBE\\
\hline
1.5&A&3.979&-0.070&0.007&0.002&0.9953\\
1.5&B&3.976&-0.070&0.009&0.002&0.9937\\
\hline
FRE\\
\hline
1.5&A&4.009&-0.096&0.015&0.004&0.9892\\
1.5&B&4.048&-0.111&0.013&0.004&0.9940\\
\hline\hline
\end{tabular}
\end{table*}

\begin{table*}
\caption{\label{tl_quad_m31} The coefficients of the quadratic relation $\log T_{eff}$ = a + b \lsun +c (\lsun)$^2$ for the boundaries of the F-mode IS for Z = 0.03 Y= 0.28 varying both the ML relation and the mixing length parameter.}
\centering
\begin{tabular}{ccccccccc}
\hline\hline
$\alpha_{ml}$&ML&a&b&c&$\sigma_{a}$&$\sigma_{b}$&$\sigma_{c}$&$R^2$\\
\hline
FBE\\
\hline
1.5&A&3.878&-0.010&-0.009&0.037&0.022&0.003&0.9981\\
1.5&B&3.957&-0.059&-0.001&0.076&0.042&0.006&0.9937\\
\hline
FRE\\
\hline
1.5&A&3.764&0.049&-0.021&0.052&0.031&0.004&0.9980\\
1.5&B&3.822&0.016&-0.017&0.059&0.033&0.005&0.9985\\
\hline\hline
\end{tabular}
\end{table*}

\subsection{The new atlas of light and radial velocity curves}

\begin{figure}[ht!]
\centering
\textbf{\msun=3.0\,\,\,\,\lsun=2.39\,\,\,\,$\alpha_{ml} = 1.5$ }\par\medskip
\includegraphics[trim=20 20 0 55,clip,width=0.6\textwidth]{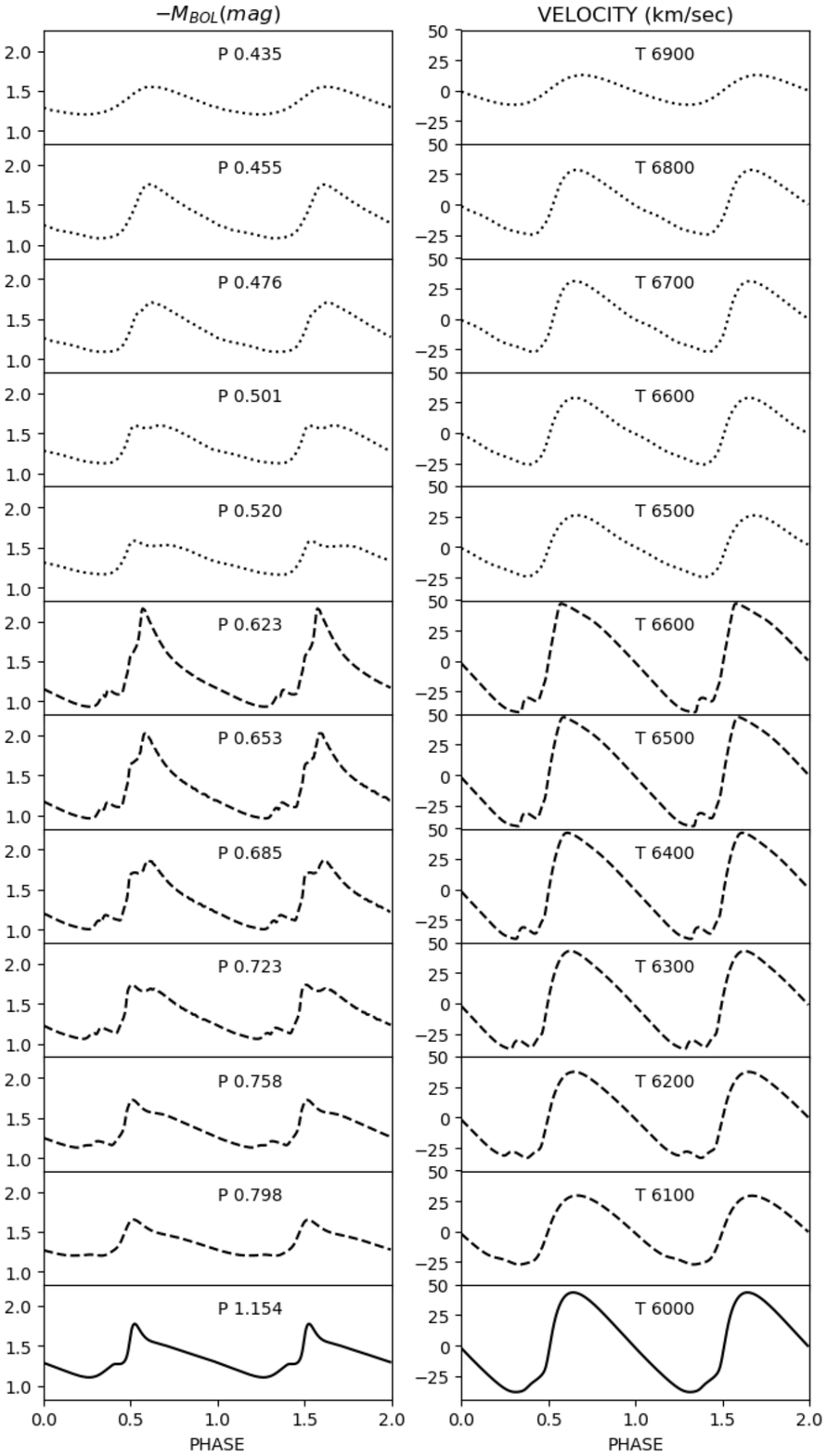}
\caption{Bolometric light curves (left panel) and radial velocity curves (right panel) for a sequence of nonlinear F (solid line), FO (dashed lines) and if any SO (dotted lines) mode models (dashed lines) derived for Z = 0.008 Y = 0.25 at a fixed mass, luminosity, $\alpha_{ml}$ parameter (see labeled values on the top of the plot) adopting the canonical ML relation. The complete figure set (9 images) is available in the online journal.}
\label{Fig:lc_vr_2.39_3.0_1.5_A_lmc}
\end{figure}

\begin{figure}[ht!]
\centering
\textbf{\msun=4.0\,\,\,\,\lsun=2.68\,\,\,\,$\alpha_{ml} = 1.5$ }\par\medskip
\includegraphics[trim=20 20 0 75,clip,width=0.6\textwidth]{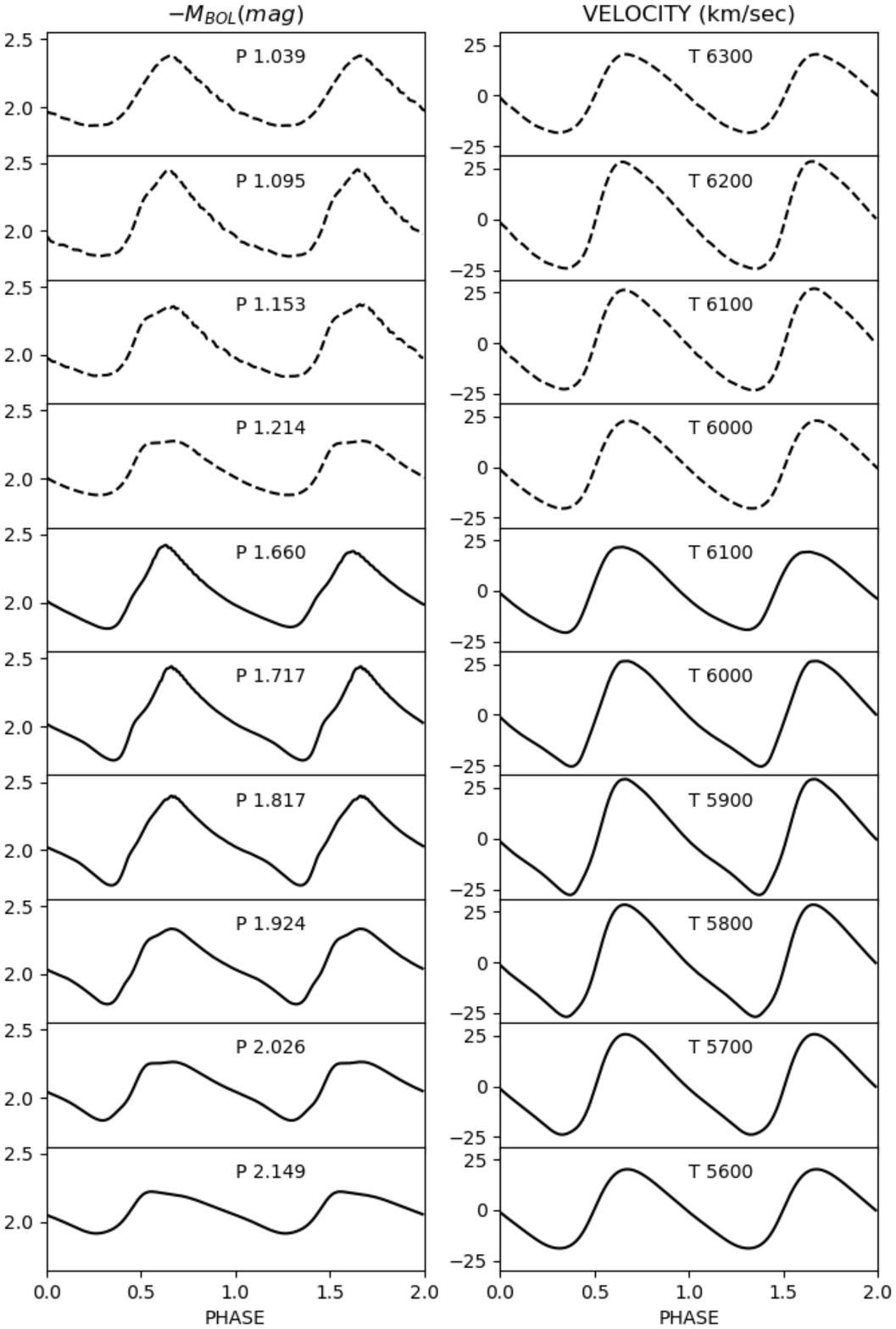}
\caption{Bolometric light curves (left panel) and radial velocity curves (right panel) for a sequence of non linear F (solid line) and FO (dashed lines) models derived for Z = 0.03 Y = 0.28, at fixed mass, luminosity and $\alpha_{ml}$ parameter (see labeled values on the top of the plot) adopting the canonical ML relation. The complete figure set (8 images) is available in the online journal.}
\label{Fig:lc_vr_2.68_4.0_1.5_A_m31}
\end{figure}

The computation of nonlinear pulsation models allowed us to predict variations of all relevant quantities, namely the luminosity, radius, radial velocity, surface temperature and gravity along a model pulsation cycle. The bolometric light curves for a sequence of canonical models listed in Tables \ref{boundaries_smc} to \ref{boundaries_m31} shown in Section \ref{sec:boundaries} of the Appendix are shown in the left panels of Figs \ref{Fig:lc_vr_2.49_3.0_1.5_A_smc} to \ref{Fig:lc_vr_2.68_4.0_1.5_A_m31}, for each labeled mass, luminosity and mixing length parameter (labels at the top of each plot). The radial velocity curves of these models are represented in the right panels of the same plots. In these figures dotted lines refer to the few inferred SO-mode models, dashed lines correspond to FO-mode models, whereas solid lines represent F-mode models. The model period, in days, and the effective temperature, in kelvin, are also reported in the left and right panels, respectively. The entire atlas of light and radial velocity curves is available as supplementary material. We noticed that in agreement with previous investigations \citet[see e.g.][]{BonoCM2000}, the morphology and amplitude of both light and radial velocity curves are affected by a variation in the metal and helium abundance, as well as by the location inside the IS. We also notice a general decrease in the pulsation amplitudes as the metallicity increases, due to the increased contribution of opacity to convection. 
Quite interestingly, within the period range of 6 \textless P \textless 16 days, CCs show the Hertzsprung progression phenomenon i.e. the evolution of the position in phase of the secondary maximum (bump) in both light and radial velocity curves, as a function of the pulsation period \citep[][]{Hertzprung1926, BonoHP2000}. As expected, based on previous theoretical studies for 0.004 \textless Z \textless 0.02 \citep[see e.g.][]{BonoHP2000}, an analysis of Figs \ref{Fig:lc_vr_2.49_3.0_1.5_A_smc}, \ref{Fig:lc_vr_2.39_3.0_1.5_A_lmc} and \ref{Fig:lc_vr_2.68_4.0_1.5_A_m31} suggests that an increase in the metal content causes a shift of the HP center (defined as the period at which the secondary bump reaches the same brightness as the primary one during its evolution from the decreasing to the rising branch of the curve) towards shorter periods. For example, passing from $Z = 0.004$ to $Z = 0.008$ and finally to $Z = 0.02$ for a \msun = 7 model, the period corresponding to the HP center, in the case of canonical models, moves from about 11.4 to about 10.3 and finally decreases to about 7.5 days. A detailed investigation of the dependence on metallicity of the period corresponding to the HP center will be the subject of a future work (Marconi et al., in preparation).

\clearpage
\figsetstart
\figsetnum{9}
\figsettitle{Bolometric light and radial velocity curves for Z = 0.004 Y = 0.25}
\figsetgrpstart
\figsetgrpnum{figurenumber.1}
\figsetgrptitle{3 solar mass}
\figsetplot{lc_vr_2p49_3p0_1p5_A_smc.pdf}
\figsetgrpnote{Bolometric light curves (left panel) and radial velocity curves (right panel) for a sequence of nonlinear F (solid line), FO (dashed lines) and if any SO (dotted lines) mode models (dashed lines) derived for Z = 0.004 Y = 0.25 at a fixed mass, luminosity, $\alpha_{ml}$ parameter (see labeled values on the top of the plot) adopting the canonical ML relation.}
\figsetgrpend
\figsetgrpstart
\figsetgrpnum{figurenumber.2}
\figsetgrptitle{4 solar mass}
\figsetplot{lc_vr_2p91_4p0_1p5_A_smc.pdf}
\figsetgrpnote{Bolometric light curves (left panel) and radial velocity curves (right panel) for a sequence of nonlinear F (solid line), FO (dashed lines) and if any SO (dotted lines) mode models (dashed lines) derived for Z = 0.004 Y = 0.25 at a fixed mass, luminosity, $\alpha_{ml}$ parameter (see labeled values on the top of the plot) adopting the canonical ML relation.}
\figsetgrpend
\figsetgrpstart
\figsetgrpnum{figurenumber.3}
\figsetgrptitle{5 solar mass}
\figsetplot{lc_vr_3p24_5p0_1p5_A_smc.pdf}
\figsetgrpnote{Bolometric light curves (left panel) and radial velocity curves (right panel) for a sequence of nonlinear F (solid line), FO (dashed lines) and if any SO (dotted lines) mode models (dashed lines) derived for Z = 0.004 Y = 0.25 at a fixed mass, luminosity, $\alpha_{ml}$ parameter (see labeled values on the top of the plot) adopting the canonical ML relation.}
\figsetgrpend
\figsetgrpstart
\figsetgrpnum{figurenumber.4}
\figsetgrptitle{6 solar mass}
\figsetplot{lc_vr_3p50_6p0_1p5_A_smc.pdf}
\figsetgrpnote{Bolometric light curves (left panel) and radial velocity curves (right panel) for a sequence of nonlinear F (solid line), FO (dashed lines) and if any SO (dotted lines) mode models (dashed lines) derived for Z = 0.004 Y = 0.25 at a fixed mass, luminosity, $\alpha_{ml}$ parameter (see labeled values on the top of the plot) adopting the canonical ML relation.}
\figsetgrpend
\figsetgrpstart
\figsetgrpnum{figurenumber.5}
\figsetgrptitle{7 solar mass}
\figsetplot{lc_vr_3p73_7p0_1p5_A_smc.pdf}
\figsetgrpnote{Bolometric light curves (left panel) and radial velocity curves (right panel) for a sequence of nonlinear F (solid line), FO (dashed lines) and if any SO (dotted lines) mode models (dashed lines) derived for Z = 0.004 Y = 0.25 at a fixed mass, luminosity, $\alpha_{ml}$ parameter (see labeled values on the top of the plot) adopting the canonical ML relation.}
\figsetgrpend
\figsetgrpstart
\figsetgrpnum{figurenumber.6}
\figsetgrptitle{8 solar mass}
\figsetplot{lc_vr_3p92_8p0_1p5_A_smc.pdf}
\figsetgrpnote{Bolometric light curves (left panel) and radial velocity curves (right panel) for a sequence of nonlinear F (solid line), FO (dashed lines) and if any SO (dotted lines) mode models (dashed lines) derived for Z = 0.004 Y = 0.25 at a fixed mass, luminosity, $\alpha_{ml}$ parameter (see labeled values on the top of the plot) adopting the canonical ML relation.}
\figsetgrpend
\figsetgrpstart
\figsetgrpnum{figurenumber.7}
\figsetgrptitle{9 solar mass}
\figsetplot{lc_vr_4p09_9p0_1p5_A_smc.pdf}
\figsetgrpnote{Bolometric light curves (left panel) and radial velocity curves (right panel) for a sequence of nonlinear F (solid line), FO (dashed lines) and if any SO (dotted lines) mode models (dashed lines) derived for Z = 0.004 Y = 0.25 at a fixed mass, luminosity, $\alpha_{ml}$ parameter (see labeled values on the top of the plot) adopting the canonical ML relation.}
\figsetgrpend
\figsetgrpstart
\figsetgrpnum{figurenumber.8}
\figsetgrptitle{10 solar mass}
\figsetplot{lc_vr_4p25_10p0_1p5_A_smc.pdf}
\figsetgrpnote{Bolometric light curves (left panel) and radial velocity curves (right panel) for a sequence of nonlinear F (solid line), FO (dashed lines) and if any SO (dotted lines) mode models (dashed lines) derived for Z = 0.004 Y = 0.25 at a fixed mass, luminosity, $\alpha_{ml}$ parameter (see labeled values on the top of the plot) adopting the canonical ML relation.}
\figsetgrpend
\figsetgrpstart
\figsetgrpnum{figurenumber.9}
\figsetgrptitle{11 solar mass}
\figsetplot{lc_vr_4p39_11p0_1p5_A_smc.pdf}
\figsetgrpnote{Bolometric light curves (left panel) and radial velocity curves (right panel) for a sequence of nonlinear F (solid line), FO (dashed lines) and if any SO (dotted lines) mode models (dashed lines) derived for Z = 0.004 Y = 0.25 at a fixed mass, luminosity, $\alpha_{ml}$ parameter (see labeled values on the top of the plot) adopting the canonical ML relation.}
\figsetgrpend
\figsetend

\begin{figure}[ht!]
\centering
\textbf{\msun=3.0\,\,\,\,\lsun=2.39\,\,\,\,$\alpha_{ml} = 1.5$ }\par\medskip
\includegraphics[trim=20 20 0 55,clip,width=0.6\textwidth]{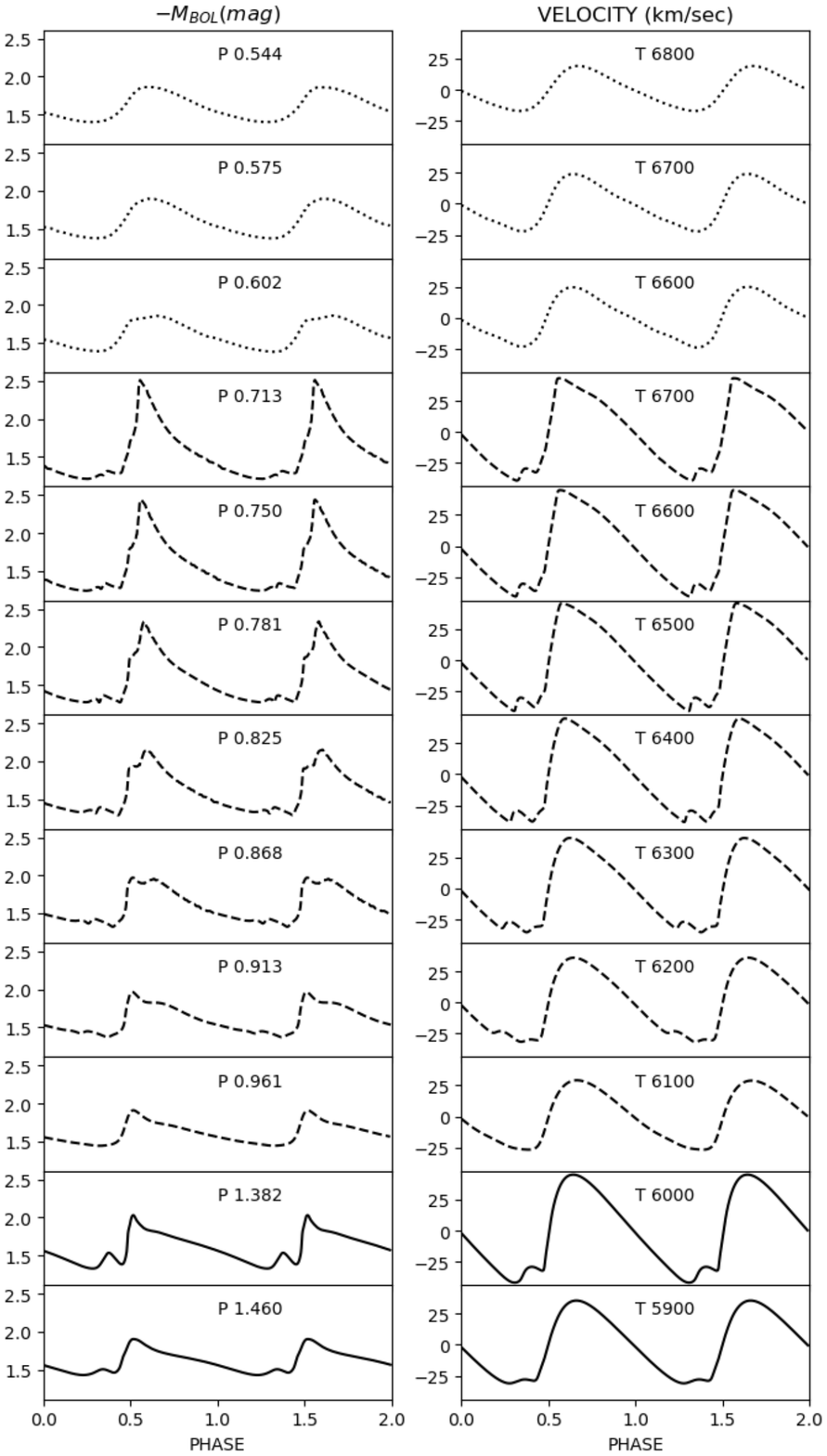}
\caption{Bolometric light curves (left panel) and radial velocity curves (right panel) for a sequence of nonlinear F (solid line), FO (dashed lines) and if any SO (dotted lines) mode models (dashed lines) derived for Z = 0.004 Y = 0.25 at a fixed mass, luminosity, $\alpha_{ml}$ parameter (see labeled values on the top of the plot) adopting the canonical ML relation. The complete figure set (9 images) is available in the online journal.}
\label{Fig:lc_vr_2.49_3.0_1.5_A_smc}
\end{figure}
\clearpage

\figsetstart
\figsetnum{9}
\figsettitle{Bolometric light and radial velocity curves for Z = 0.008 Y = 0.25}
\figsetgrpstart
\figsetgrpnum{figurenumber.1}
\figsetgrptitle{3 solar mass}
\figsetplot{lc_vr_2p39_3p0_1p5_A_lmc.pdf}
\figsetgrpnote{Bolometric light curves (left panel) and radial velocity curves (right panel) for a sequence of nonlinear F (solid line), FO (dashed lines) and if any SO (dotted lines) mode models (dashed lines) derived for Z = 0.008 Y = 0.25 at a fixed mass, luminosity, $\alpha_{ml}$ parameter (see labeled values on the top of the plot) adopting the canonical ML relation.}
\figsetgrpend
\figsetgrpstart
\figsetgrpnum{figurenumber.2}
\figsetgrptitle{4 solar mass}
\figsetplot{lc_vr_2p81_4p0_1p5_A_lmc.pdf}
\figsetgrpnote{Bolometric light curves (left panel) and radial velocity curves (right panel) for a sequence of nonlinear F (solid line), FO (dashed lines) and if any SO (dotted lines) mode models (dashed lines) derived for Z = 0.008 Y = 0.25 at a fixed mass, luminosity, $\alpha_{ml}$ parameter (see labeled values on the top of the plot) adopting the canonical ML relation.}
\figsetgrpend
\figsetgrpstart
\figsetgrpnum{figurenumber.3}
\figsetgrptitle{5 solar mass}
\figsetplot{lc_vr_3p14_5p0_1p5_A_lmc.pdf}
\figsetgrpnote{Bolometric light curves (left panel) and radial velocity curves (right panel) for a sequence of nonlinear F (solid line), FO (dashed lines) and if any SO (dotted lines) mode models (dashed lines) derived for Z = 0.008 Y = 0.25 at a fixed mass, luminosity, $\alpha_{ml}$ parameter (see labeled values on the top of the plot) adopting the canonical ML relation.}
\figsetgrpend
\figsetgrpstart
\figsetgrpnum{figurenumber.4}
\figsetgrptitle{6 solar mass}
\figsetplot{lc_vr_3p40_6p0_1p5_A_lmc.pdf}
\figsetgrpnote{Bolometric light curves (left panel) and radial velocity curves (right panel) for a sequence of nonlinear F (solid line), FO (dashed lines) and if any SO (dotted lines) mode models (dashed lines) derived for Z = 0.008 Y = 0.25 at a fixed mass, luminosity, $\alpha_{ml}$ parameter (see labeled values on the top of the plot) adopting the canonical ML relation.}
\figsetgrpend
\figsetgrpstart
\figsetgrpnum{figurenumber.5}
\figsetgrptitle{7 solar mass}
\figsetplot{lc_vr_3p63_7p0_1p5_A_lmc.pdf}
\figsetgrpnote{Bolometric light curves (left panel) and radial velocity curves (right panel) for a sequence of nonlinear F (solid line), FO (dashed lines) and if any SO (dotted lines) mode models (dashed lines) derived for Z = 0.008 Y = 0.25 at a fixed mass, luminosity, $\alpha_{ml}$ parameter (see labeled values on the top of the plot) adopting the canonical ML relation.}
\figsetgrpend
\figsetgrpstart
\figsetgrpnum{figurenumber.6}
\figsetgrptitle{8 solar mass}
\figsetplot{lc_vr_3p82_8p0_1p5_A_lmc.pdf}
\figsetgrpnote{Bolometric light curves (left panel) and radial velocity curves (right panel) for a sequence of nonlinear F (solid line), FO (dashed lines) and if any SO (dotted lines) mode models (dashed lines) derived for Z = 0.008 Y = 0.25 at a fixed mass, luminosity, $\alpha_{ml}$ parameter (see labeled values on the top of the plot) adopting the canonical ML relation.}
\figsetgrpend
\figsetgrpstart
\figsetgrpnum{figurenumber.7}
\figsetgrptitle{9 solar mass}
\figsetplot{lc_vr_3p99_9p0_1p5_A_lmc.pdf}
\figsetgrpnote{Bolometric light curves (left panel) and radial velocity curves (right panel) for a sequence of nonlinear F (solid line), FO (dashed lines) and if any SO (dotted lines) mode models (dashed lines) derived for Z = 0.008 Y = 0.25 at a fixed mass, luminosity, $\alpha_{ml}$ parameter (see labeled values on the top of the plot) adopting the canonical ML relation.}
\figsetgrpend
\figsetgrpstart
\figsetgrpnum{figurenumber.8}
\figsetgrptitle{10 solar mass}
\figsetplot{lc_vr_4p14_10p0_1p5_A_lmc.pdf}
\figsetgrpnote{Bolometric light curves (left panel) and radial velocity curves (right panel) for a sequence of nonlinear F (solid line), FO (dashed lines) and if any SO (dotted lines) mode models (dashed lines) derived for Z = 0.008 Y = 0.25 at a fixed mass, luminosity, $\alpha_{ml}$ parameter (see labeled values on the top of the plot) adopting the canonical ML relation.}
\figsetgrpend
\figsetgrpstart
\figsetgrpnum{figurenumber.9}
\figsetgrptitle{11 solar mass}
\figsetplot{lc_vr_4p28_11p0_1p5_A_lmc.pdf}
\figsetgrpnote{Bolometric light curves (left panel) and radial velocity curves (right panel) for a sequence of nonlinear F (solid line), FO (dashed lines) and if any SO (dotted lines) mode models (dashed lines) derived for Z = 0.008 Y = 0.25 at a fixed mass, luminosity, $\alpha_{ml}$ parameter (see labeled values on the top of the plot) adopting the canonical ML relation.}
\figsetgrpend
\figsetend

\clearpage

\figsetstart
\figsetnum{8}
\figsettitle{Bolometric light and radial velocity curves for Z = 0.03 Y = 0.28}
\figsetgrpstart
\figsetgrpnum{figurenumber.1}
\figsetgrptitle{4 solar mass}
\figsetplot{lc_vr_2p68_4p0_1p5_A_m31.pdf}
\figsetgrpnote{Bolometric light curves (left panel) and radial velocity curves (right panel) for a sequence of non linear F (solid line) and FO (dashed lines) models derived for Z = 0.03 Y = 0.28, at fixed mass, luminosity and $\alpha_{ml}$ parameter (see labeled values on the top of the plot) adopting the canonical ML relation.}
\figsetgrpend
\figsetgrpstart
\figsetgrpnum{figurenumber.2}
\figsetgrptitle{5 solar mass}
\figsetplot{lc_vr_3p01_5p0_1p5_A_m31.pdf}
\figsetgrpnote{Bolometric light curves (left panel) and radial velocity curves (right panel) for a sequence of non linear F (solid line) and FO (dashed lines) models derived for Z = 0.03 Y = 0.28, at fixed mass, luminosity and $\alpha_{ml}$ parameter (see labeled values on the top of the plot) adopting the canonical ML relation.}
\figsetgrpend
\figsetgrpstart
\figsetgrpnum{figurenumber.3}
\figsetgrptitle{6 solar mass}
\figsetplot{lc_vr_3p27_6p0_1p5_A_m31.pdf}
\figsetgrpnote{Bolometric light curves (left panel) and radial velocity curves (right panel) for a sequence of non linear F (solid line) and FO (dashed lines) models derived for Z = 0.03 Y = 0.28, at fixed mass, luminosity and $\alpha_{ml}$ parameter (see labeled values on the top of the plot) adopting the canonical ML relation.}
\figsetgrpend
\figsetgrpstart
\figsetgrpnum{figurenumber.4}
\figsetgrptitle{7 solar mass}
\figsetplot{lc_vr_3p50_7p0_1p5_A_m31.pdf}
\figsetgrpnote{Bolometric light curves (left panel) and radial velocity curves (right panel) for a sequence of non linear F (solid line) and FO (dashed lines) models derived for Z = 0.03 Y = 0.28, at fixed mass, luminosity and $\alpha_{ml}$ parameter (see labeled values on the top of the plot) adopting the canonical ML relation.}
\figsetgrpend
\figsetgrpstart
\figsetgrpnum{figurenumber.5}
\figsetgrptitle{8 solar mass}
\figsetplot{lc_vr_3p69_8p0_1p5_A_m31.pdf}
\figsetgrpnote{Bolometric light curves (left panel) and radial velocity curves (right panel) for a sequence of non linear F (solid line) and FO (dashed lines) models derived for Z = 0.03 Y = 0.28, at fixed mass, luminosity and $\alpha_{ml}$ parameter (see labeled values on the top of the plot) adopting the canonical ML relation.}
\figsetgrpend
\figsetgrpstart
\figsetgrpnum{figurenumber.6}
\figsetgrptitle{9 solar mass}
\figsetplot{lc_vr_3p86_9p0_1p5_A_m31.pdf}
\figsetgrpnote{Bolometric light curves (left panel) and radial velocity curves (right panel) for a sequence of non linear F (solid line) and FO (dashed lines) models derived for Z = 0.03 Y = 0.28, at fixed mass, luminosity and $\alpha_{ml}$ parameter (see labeled values on the top of the plot) adopting the canonical ML relation.}
\figsetgrpend
\figsetgrpstart
\figsetgrpnum{figurenumber.7}
\figsetgrptitle{10 solar mass}
\figsetplot{lc_vr_4p02_10p0_1p5_A_m31.pdf}
\figsetgrpnote{Bolometric light curves (left panel) and radial velocity curves (right panel) for a sequence of non linear F (solid line) and FO (dashed lines) models derived for Z = 0.03 Y = 0.28, at fixed mass, luminosity and $\alpha_{ml}$ parameter (see labeled values on the top of the plot) adopting the canonical ML relation.}
\figsetgrpend
\figsetgrpstart
\figsetgrpnum{figurenumber.8}
\figsetgrptitle{11 solar mass}
\figsetplot{lc_vr_4p15_11p0_1p5_A_m31.pdf}
\figsetgrpnote{Bolometric light curves (left panel) and radial velocity curves (right panel) for a sequence of non linear F (solid line) and FO (dashed lines) models derived for Z = 0.03 Y = 0.28, at fixed mass, luminosity and $\alpha_{ml}$ parameter (see labeled values on the top of the plot) adopting the canonical ML relation.}
\figsetgrpend
\figsetend

\clearpage

\section{The predicted multifilters light curves}
To facilitate observational comparisons, the bolometric light curves obtained from the nonlinear model computations have been transformed into various photometric filters, namely the three Gaia bands $G$, $G_{BP}$ and $G_{RP}$, the HST WFC3 filters F110W,  F160W,  F814W,  F125W,  F475W,  F606W and F555W and the Johnson-Cousins (JC) bands U, B, V, R, J, H, K by adopting PHOENIX model atmospheres \citep[][]{Chen2019}. The complete multi-filter atlas, obtained by varying the chemical composition, the ML relation and the efficiency of super-adiabatic convection, is available as supplementary material.

\subsection{Mean magnitudes, colors and pulsation amplitudes}

From the transformed light curves, we derived the intensity weighted mean magnitudes and the pulsation amplitudes reported in Tables \ref{dr3_F_amp} to \ref{wfc3_FO_amp} for the various filter combinations and the two pulsation modes. Mean colors are obtained as differences between couples of mean magnitudes and used to derive PLC and PW relations (see next section). In Fig. \ref{Fig:ampl_F_FO_Z} shown in Section \ref{sec:amplitudes} of the appendix, the predicted multi-filter amplitudes are plotted as a function of the pulsation period for F (right panels) and FO (left panels), models for the labeled metal contents, assuming the case A (canonical) ML relation and $\alpha_{ml}$=1.5. We notice that, in agreement with previous theoretical results \citep[see e.g.][]{BonoCM2000}, the pulsation amplitudes tend to decrease as the metallicity increases. Moreover, F-mode amplitudes show a linear trend with the pulsation periods at the lowest luminosity levels. On the other hand, at higher luminosities, in a range where the FO-mode pulsation is no more efficient, the F-mode amplitudes show the same bell-shape displayed by FO models at lower luminosity levels.
Figures from \ref{Fig:ampl_F_SMC} to \ref{Fig:ampl_F_M31} shown in Section \ref{sec:amplitudes} of the appendix display the effect of a variation of $\alpha_{ml}$ (left panels) and the assumed ML relation (right panels) for all the chemical compositions and both F and FO-mode models, in the same filters as Fig.\ref{Fig:ampl_F_FO_Z} shown in Section \ref{sec:amplitudes} of the appendix. We notice that independently of the assumed abundances, the predicted pulsation amplitudes decrease as $\alpha_{ml}$ increases, down to the total quenching of pulsation for the more metallic FO sets at $\alpha=1.9$ and F amplitude values close to zero for $Z=0.03$ at $\alpha_{ml}=1.7$. On the other hand, the main effect of a variation of the ML relation is a shift of the covered period range, in both pulsation modes, to longer values, as the ML relation moves from case A to C.
To test these theoretical predictions we used a sample of Magellanic Cepheids in the Gaia DR2 database \citep[][]{Brown2018} for which amplitudes in the Gaia $G_{BP}$ and $G_{RP}$ bands are available.
In Figures \ref{Fig:amp_comp_gbp_smc} to \ref{Fig:amp_comp_grp_lmc} shown in Section \ref{sec:amplitudes} of the appendix, we compared the predicted period-amplitude diagram in the Gaia $G_{BP}$ and $G_{RP}$ filters at Z=0.008 and Z=0.004 to Gaia LMC and SMC results, respectively. In each figure, the upper and middle panels show the comparison between observed and theoretical amplitudes obtained for the canonical ML relation with $\alpha_{ml} = 1.5$ and $1.7$, respectively, while the bottom panel shows the comparison for the B-case ML relation with $\alpha_{ml} = 1.5$ . 
In general, we noticed that the period range of F and FO mode pulsators is well reproduced by canonical models both in the LMC and the SMC, apart from the lack of observed Cepheids at the longest predicted periods. Moreover, in the case of Z=0.008, we notice that the combination of a canonical ML relation with $\alpha_{ml} = 1.5$ tends to predict higher amplitudes than observed, in particular at the shortest and longest periods, whereas the assumption of $\alpha_{ml} = 1.7$ seems to better reproduce the data. The adoption of a brighter ML relation (case B) provides a satisfactory agreement for what concerns the amplitudes but seems to predict a period range shifted towards longer values with respect to the observations. Similar conclusions can be drawn in the case of Z=0.004.

\begin{table*}
\centering
\caption{\label{dr3_F_amp} Mean magnitudes and theoretical amplitudes in the Gaia DR3 filters for the computed F-mode models with Z=0.004 Y= 0.25, Z=0.008 Y= 0.25, Z=0.02 Y= 0.28 and Z=0.03 Y= 0.28. This table is available in its entirety in machine readable form.}
\begin{tabular}{ccccccccccccccccccccc}
\hline\hline
Z & Y & \msun & \lsun & $T_{eff}$[K] & $\alpha_{ml}$ & ML & G.m & G.amp & $G_{BP}.m$ & $G_{BP}.amp$ & $G_{RP}.m$ & $G_{RP}.amp$ \\
\hline\hline
0.004 & 0.25 & 3.0 & 2.49 & 5900 & 1.5 & A & -1.700 & 0.521 & -1.457 & 0.640 & -2.099 & 0.365 \\
0.004 & 0.25 & 3.0 & 2.49 & 6000 & 1.5 & A & -1.703 & 0.751 & -1.472 & 0.910 & -2.087 & 0.536 \\
0.004 & 0.25 & 3.0 & 2.49 & 6000 & 1.7 & A & -1.705 & 0.408 & -1.475 & 0.500 & -2.087 & 0.285 \\
0.004 & 0.25 & 3.0 & 2.49 & 6100 & 1.7 & A & -1.707 & 0.647 & -1.488 & 0.783 & -2.075 & 0.459 \\
...\\
0.008 & 0.25 & 3.0 & 2.39 & 6000 & 1.5 & A & -1.468 & 0.718 & -1.228 & 0.876 & -1.859 & 0.507 \\
0.008 & 0.25 & 3.0 & 2.59 & 5700 & 1.5 & B & -1.958 & 0.350 & -1.678 & 0.436 & -2.399 & 0.249 \\
0.008 & 0.25 & 3.0 & 2.59 & 5800 & 1.5 & B & -1.963 & 0.611 & -1.696 & 0.745 & -2.387 & 0.442 \\
0.008 & 0.25 & 3.0 & 2.59 & 5900 & 1.5 & B & -1.967 & 0.776 & -1.714 & 0.936 & -2.375 & 0.570 \\
...\\
0.02 & 0.28 & 3.0 & 2.32 & 5900 & 1.5 & A & -1.322 & 0.109 & -1.054 & 0.137 & -1.744 & 0.077 \\
0.02 & 0.28 & 3.0 & 2.32 & 6000 & 1.5 & A & -1.326 & 0.321 & -1.071 & 0.392 & -1.731 & 0.233 \\
0.02 & 0.28 & 3.0 & 2.32 & 6100 & 1.5 & A & -1.330 & 0.428 & -1.090 & 0.520 & -1.716 & 0.330 \\
0.02 & 0.28 & 3.0 & 2.32 & 6100 & 1.7 & A & -1.331 & 0.166 & -1.092 & 0.204 & -1.718 & 0.120 \\
...\\
0.03 & 0.28 & 4.0 & 2.68 & 5400 & 1.5 & A & -2.186 & 0.039 & -1.822 & 0.050 & -2.712 & 0.029 \\
0.03 & 0.28 & 4.0 & 2.68 & 5500 & 1.5 & A & -2.196 & 0.086 & -1.849 & 0.109 & -2.704 & 0.064 \\
0.03 & 0.28 & 4.0 & 2.68 & 5600 & 1.5 & A & -2.198 & 0.357 & -1.870 & 0.445 & -2.686 & 0.260 \\
0.03 & 0.28 & 4.0 & 2.68 & 5700 & 1.5 & A & -2.206 & 0.486 & -1.896 & 0.591 & -2.675 & 0.373 \\
...\\
\hline\hline
\end{tabular}
\end{table*}

\begin{table*}
\centering
\caption{\label{dr3_FO_amp} Mean magnitudes and theoretical amplitudes in the Gaia DR3 filters for the computed FO-mode models with Z=0.004 Y= 0.25, Z=0.008 Y= 0.25, Z=0.02 Y= 0.28 and Z=0.03 Y= 0.28. This table is available in its entirety in machine readable form.}
\begin{tabular}{ccccccccccccccccccccc}
\hline\hline
\hline\hline
Z & Y & \msun & \lsun & $T_{eff}$[K] & $\alpha_{ml}$ & ML & G.m & G.amp & $G_{BP}.m$ & $G_{BP}.amp$ & $G_{RP}.m$ & $G_{RP}.amp$ \\
\hline\hline
0.004 & 0.25 & 3.0 & 2.49 & 6100 & 1.5 & A & -1.708 & 0.495 & -1.490 & 0.588 & -2.076 & 0.367 \\
0.004 & 0.25 & 3.0 & 2.49 & 6200 & 1.5 & A & -1.711 & 0.626 & -1.504 & 0.740 & -2.062 & 0.464 \\
0.004 & 0.25 & 3.0 & 2.49 & 6300 & 1.5 & A & -1.713 & 0.681 & -1.519 & 0.800 & -2.046 & 0.506 \\
0.004 & 0.25 & 3.0 & 2.49 & 6400 & 1.5 & A & -1.713 & 0.872 & -1.533 & 1.018 & -2.027 & 0.650 \\
...\\
0.008 & 0.25 & 3.0 & 2.39 & 6100 & 1.5 & A & -1.474 & 0.477 & -1.248 & 0.578 & -1.847 & 0.349 \\
0.008 & 0.25 & 3.0 & 2.39 & 6200 & 1.5 & A & -1.476 & 0.622 & -1.263 & 0.752 & -1.832 & 0.442 \\
0.008 & 0.25 & 3.0 & 2.39 & 6300 & 1.5 & A & -1.477 & 0.713 & -1.277 & 0.856 & -1.816 & 0.510 \\
0.008 & 0.25 & 3.0 & 2.39 & 6400 & 1.5 & A & -1.478 & 0.881 & -1.292 & 1.047 & -1.798 & 0.640 \\
...\\
0.02 & 0.28 & 3.0 & 2.32 & 6200 & 1.5 & A & -1.333 & 0.453 & -1.107 & 0.555 & -1.703 & 0.319 \\
0.02 & 0.28 & 3.0 & 2.32 & 6300 & 1.5 & A & -1.335 & 0.564 & -1.123 & 0.689 & -1.687 & 0.409 \\
0.02 & 0.28 & 3.0 & 2.32 & 6400 & 1.5 & A & -1.336 & 0.736 & -1.138 & 0.884 & -1.668 & 0.531 \\
0.02 & 0.28 & 3.0 & 2.32 & 6500 & 1.5 & A & -1.335 & 0.884 & -1.152 & 1.063 & -1.646 & 0.624 \\
...\\
0.03 & 0.28 & 4.0 & 2.68 & 6000 & 1.5 & A & -2.227 & 0.432 & -1.964 & 0.522 & -2.642 & 0.324 \\
0.03 & 0.28 & 4.0 & 2.68 & 6100 & 1.5 & A & -2.233 & 0.549 & -1.985 & 0.663 & -2.627 & 0.410 \\
0.03 & 0.28 & 4.0 & 2.68 & 6200 & 1.5 & A & -2.236 & 0.663 & -2.005 & 0.799 & -2.610 & 0.485 \\
0.03 & 0.28 & 4.0 & 2.68 & 6300 & 1.5 & A & -2.241 & 0.533 & -2.022 & 0.643 & -2.598 & 0.387 \\
\hline\hline
\end{tabular}
\end{table*}

\begin{table*}
\centering
\caption{\label{wfc3_F_amp} Mean magnitudes and theoretical amplitudes in the HST-WFC3 filters for the computed F-mode models with Z=0.004 Y= 0.25, Z=0.008 Y= 0.25, Z=0.02 Y= 0.28 and Z=0.03 Y= 0.28.  This table is available in its entirety in machine readable form.}
\resizebox{\textwidth}{!}{%
\begin{tabular}{ccccccccccccccccccccc}
\hline\hline
Z & Y & \msun & \lsun & $T_{eff}$[K] & $\alpha_{ml}$ & ML & F110W.m & F110W.amp & F160W.m & F160W.amp & F814W.m & F814W.amp & F125W.m & F125W.amp & F475W.m & F475W.amp & F606W.m & F606W.amp & F555W.m & F555W.amp  \\
\hline\hline
0.004 &  0.25 &   3.0 &  2.49 &  5900 &  1.5 &  A & -2.435 &  0.245 & -2.765 &  0.188 & -2.152 &  0.342 & -2.543 &  0.208 & -1.315 &  0.713 & -1.707 &  0.535 & -1.520 &  0.616 \\
0.004 &  0.25 &   3.0 &  2.49 &  6000 &  1.5 &  A & -2.409 &  0.372 & -2.724 &  0.266 & -2.137 &  0.505 & -2.512 &  0.319 & -1.338 &  1.009 & -1.711 &  0.769 & -1.532 &  0.880 \\
0.004 &  0.25 &   3.0 &  2.49 &  6000 &  1.7 &  A & -2.404 &  0.170 & -2.712 &  0.140 & -2.137 &  0.268 & -2.506 &  0.137 & -1.342 &  0.557 & -1.713 &  0.419 & -1.535 &  0.483 \\
0.004 &  0.25 &   3.0 &  2.49 &  6100 &  1.7 &  A & -2.380 &  0.293 & -2.674 &  0.215 & -2.123 &  0.432 & -2.478 &  0.245 & -1.362 &  0.868 & -1.717 &  0.661 & -1.547 &  0.758 \\
...\\
0.008 &  0.25 &   3.0 &  2.39 &  6000 &  1.5 &  A & -2.174 &  0.319 & -2.482 &  0.193 & -1.908 &  0.478 & -2.276 &  0.260 & -1.090 &  0.976 & -1.481 &  0.732 & -1.298 &  0.842 \\
0.008 &  0.25 &   3.0 &  2.59 &  5700 &  1.5 &  B & -2.764 &  0.149 & -3.122 &  0.118 & -2.456 &  0.235 & -2.880 &  0.122 & -1.515 &  0.491 & -1.968 &  0.359 & -1.758 &  0.415 \\
0.008 &  0.25 &   3.0 &  2.59 &  5800 &  1.5 &  B & -2.737 &  0.276 & -3.079 &  0.207 & -2.441 &  0.419 & -2.848 &  0.224 & -1.543 &  0.829 & -1.974 &  0.625 & -1.774 &  0.714 \\
0.008 &  0.25 &   3.0 &  2.59 &  5900 &  1.5 &  B & -2.709 &  0.373 & -3.034 &  0.257 & -2.426 &  0.541 & -2.815 &  0.318 & -1.570 &  1.036 & -1.980 &  0.792 & -1.789 &  0.900 \\
...\\
0.02 &  0.28 &   3.0 &  2.32 &  5900 &  1.5 &  A & -2.064 &  0.049 & -2.372 &  0.037 & -1.795 &  0.073 & -2.168 &  0.043 & -0.896 &  0.155 & -1.344 &  0.112 & -1.141 &  0.130 \\
0.02 &  0.28 &   3.0 &  2.32 &  6000 &  1.5 &  A & -2.036 &  0.163 & -2.327 &  0.125 & -1.779 &  0.221 & -2.136 &  0.150 & -0.922 &  0.439 & -1.349 &  0.326 & -1.155 &  0.374 \\
0.02 &  0.28 &   3.0 &  2.32 &  6100 &  1.5 &  A & -2.005 &  0.249 & -2.278 &  0.172 & -1.762 &  0.318 & -2.099 &  0.225 & -0.950 &  0.579 & -1.354 &  0.435 & -1.170 &  0.498 \\
0.02 &  0.28 &   3.0 &  2.32 &  6100 &  1.7 &  A & -2.004 &  0.087 & -2.274 &  0.065 & -1.764 &  0.115 & -2.098 &  0.078 & -0.952 &  0.229 & -1.356 &  0.169 & -1.171 &  0.195 \\
...\\
0.03 &  0.28 &   4.0 &  2.68 &  5400 &  1.5 &  A & -3.128 &  0.018 & -3.528 &  0.011 & -2.774 &  0.028 & -3.258 &  0.015 & -1.598 &  0.058 & -2.201 &  0.041 & -1.934 &  0.048 \\
0.03 &  0.28 &   4.0 &  2.68 &  5500 &  1.5 &  A & -3.102 &  0.041 & -3.482 &  0.028 & -2.765 &  0.061 & -3.226 &  0.036 & -1.638 &  0.125 & -2.214 &  0.090 & -1.958 &  0.105 \\
0.03 &  0.28 &   4.0 &  2.68 &  5600 &  1.5 &  A & -3.064 &  0.173 & -3.426 &  0.122 & -2.744 &  0.248 & -3.184 &  0.155 & -1.672 &  0.508 & -2.218 &  0.369 & -1.974 &  0.428 \\
0.03 &  0.28 &   4.0 &  2.68 &  5700 &  1.5 &  A & -3.034 &  0.260 & -3.376 &  0.162 & -2.730 &  0.358 & -3.148 &  0.230 & -1.711 &  0.669 & -2.229 &  0.500 & -1.997 &  0.569 \\
...\\
\hline\hline
\end{tabular}}
\end{table*}

\begin{table*}
\centering
\caption{\label{wfc3_FO_amp} Mean magnitudes and theoretical amplitudes in the HST-WFC3 filters for the computed FO-mode models with Z=0.004 Y= 0.25, Z=0.008 Y= 0.25, Z=0.02 Y= 0.28 and Z=0.03 Y= 0.28. This table is available in its entirety in machine readable form.}
\resizebox{\textwidth}{!}{%
\begin{tabular}{ccccccccccccccccccccc}
\hline\hline
Z & Y & \msun & \lsun & $T_{eff}$[K] & $\alpha_{ml}$ & ML & F110W.m & F110W.amp & F160W.m & F160W.amp & F814W.m & F814W.amp & F125W.m & F125W.amp & F475W.m & F475W.amp & F606W.m & F606W.amp & F555W.m & F555W.amp  \\
\hline\hline
0.004 &  0.25 &  3.0 &  2.49 &  6100 &  1.5 &  A & -2.378 &  0.251 & -2.670 &  0.119 & -2.124 &  0.348 & -2.476 &  0.212 & -1.363 &  0.646 & -1.718 &  0.506 & -1.548 &  0.572 \\
0.004 &  0.25 &  3.0 &  2.49 &  6200 &  1.5 &  A & -2.350 &  0.323 & -2.625 &  0.163 & -2.108 &  0.440 & -2.443 &  0.273 & -1.385 &  0.811 & -1.722 &  0.638 & -1.560 &  0.720 \\
0.004 &  0.25 &  3.0 &  2.49 &  6300 &  1.5 &  A & -2.321 &  0.373 & -2.580 &  0.215 & -2.090 &  0.481 & -2.409 &  0.327 & -1.408 &  0.876 & -1.725 &  0.692 & -1.572 &  0.780 \\
0.004 &  0.25 &  3.0 &  2.49 &  6400 &  1.5 &  A & -2.288 &  0.480 & -2.531 &  0.264 & -2.068 &  0.619 & -2.370 &  0.413 & -1.431 &  1.111 & -1.725 &  0.880 & -1.583 &  0.990 \\
...\\
0.008 &  0.25 &  3.0 &  2.39 &  6100 &  1.5 &  A & -2.142 &  0.237 & -2.425 &  0.107 & -1.894 &  0.331 & -2.238 &  0.198 & -1.117 &  0.641 & -1.489 &  0.486 & -1.314 &  0.557 \\
0.008 &  0.25 &  3.0 &  2.39 &  6200 &  1.5 &  A & -2.113 &  0.306 & -2.380 &  0.155 & -1.877 &  0.416 & -2.204 &  0.260 & -1.140 &  0.833 & -1.492 &  0.633 & -1.326 &  0.725 \\
0.008 &  0.25 &  3.0 &  2.39 &  6300 &  1.5 &  A & -2.083 &  0.339 & -2.335 &  0.198 & -1.858 &  0.481 & -2.169 &  0.292 & -1.163 &  0.948 & -1.494 &  0.723 & -1.338 &  0.827 \\
0.008 &  0.25 &  3.0 &  2.39 &  6400 &  1.5 &  A & -2.051 &  0.449 & -2.287 &  0.241 & -1.838 &  0.607 & -2.132 &  0.378 & -1.187 &  1.151 & -1.496 &  0.890 & -1.350 &  1.011 \\
...\\
0.02 &  0.28 &  3.0 &  2.32 &  6200 &  1.5 &  A & -1.976 &  0.208 & -2.232 &  0.124 & -1.747 &  0.300 & -2.066 &  0.174 & -0.975 &  0.620 & -1.359 &  0.461 & -1.183 &  0.531 \\
0.02 &  0.28 &  3.0 &  2.32 &  6300 &  1.5 &  A & -1.945 &  0.281 & -2.185 &  0.168 & -1.727 &  0.388 & -2.029 &  0.243 & -1.001 &  0.769 & -1.361 &  0.572 & -1.196 &  0.660 \\
0.02 &  0.28 &  3.0 &  2.32 &  6400 &  1.5 &  A & -1.912 &  0.367 & -2.136 &  0.198 & -1.706 &  0.502 & -1.991 &  0.306 & -1.026 &  0.978 & -1.363 &  0.743 & -1.208 &  0.849 \\
0.02 &  0.28 &  3.0 &  2.32 &  6500 &  1.5 &  A & -1.876 &  0.433 & -2.084 &  0.223 & -1.682 &  0.590 & -1.950 &  0.359 & -1.049 &  1.176 & -1.362 &  0.885 & -1.218 &  1.017 \\
...\\
0.03 &  0.28 &  4.0 &  2.68 &  6000 &  1.5 &  A & -2.946 &  0.227 & -3.229 &  0.125 & -2.689 &  0.310 & -3.044 &  0.196 & -1.810 &  0.582 & -2.255 &  0.441 & -2.054 &  0.502 \\
0.03 &  0.28 &  4.0 &  2.68 &  6100 &  1.5 &  A & -2.915 &  0.290 & -3.180 &  0.155 & -2.672 &  0.391 & -3.008 &  0.249 & -1.841 &  0.738 & -2.261 &  0.559 & -2.072 &  0.636 \\
0.03 &  0.28 &  4.0 &  2.68 &  6200 &  1.5 &  A & -2.882 &  0.343 & -3.130 &  0.179 & -2.653 &  0.461 & -2.969 &  0.292 & -1.872 &  0.889 & -2.266 &  0.671 & -2.088 &  0.768 \\
0.03 &  0.28 &  4.0 &  2.68 &  6300 &  1.5 &  A & -2.853 &  0.273 & -3.081 &  0.138 & -2.638 &  0.367 & -2.935 &  0.230 & -1.896 &  0.715 & -2.271 &  0.539 & -2.101 &  0.618 \\
\hline\hline
\end{tabular}}
\end{table*}
%

\clearpage

\section{The updated Period-Luminosity-Color and the Period-Wesenheit relations in the optical and near-infrared filters}\label{sec_pwz}

The obtained intensity-weighted mean magnitudes and colors are used to derive updated theoretical PLC and PW relations for different filter combinations varying the pulsation mode, the chemical composition, the ML relation and the $\alpha_{ml}$ parameter. PL relations are being derived in an accompanying paper (Musella et al. in preparation) investigating their dependence on the Cepheid distribution inside the predicted instability strip boundaries, in a variety of photometric bands. Tables \ref{plc_gaia} and \ref{plc_hst} in Section \ref{sec:plc_coeff} of the Appendix provide the PLC coefficients for F and FO-mode in the Gaia EDR3 filters ($<G>$=a+b$\log P$ +c($<G_{BP}>$ - $<G_{RP}>$) and HST-WFC filters ($<F_{160W}>$=a+b$\log P$ +c($<F_{555W}>$ - $<F_{814W}>$), respectively, while Tables \ref{plc_johI} and \ref{plc_johK} reported in the same section of the Appendix provide the PLC coefficients for two different Johnson-Cousins filter combinations ($<V>$ = a + b$\log P$ +c($<V>$ - $<I>$); $<V>$ = a + b$\log P$ +c($<V>$ - $<K>$)). The coefficients of the PW relations in the Gaia EDR3, HST-WFC3 and Johnson-Cousins filters are instead reported in Tables \ref{pw_gaia} to \ref{pw_joh_K} in Section \ref{sec:pw_coeff} of the Appendix. In each of the aforementioned tables, apart from the chemical composition, we report the $\alpha_{ml}$ parameter, the selected ML case, the coefficients of the relations with their relative errors, the root-mean-square deviation ($\sigma$) and the R-squared ($R^2$) coefficients. 
To take into account the non-negligible metallicity effect on the predicted pulsation properties, we also derived the first theoretical PWZ relations for each different assumption on the $\alpha_{ml}$ parameter and the ML relation, for the four adopted filter combinations and the two pulsation modes. The coefficients are reported in Tables \ref{pwz_gaia} to \ref{pwz_johK} (See Table captions for more details). Inspection of these tables suggests that a small metallicity effect is predicted for the adopted filter combinations, thus confirming the advantage of using PW relations instead of simple PL relations to infer the extragalactic distance scale \citep[see][for details]{Fiorentino2007}. Indeed, for an uncertainty of $\sim 0.1$ dex on the metal abundance of the investigated pulsators, a distance modulus error depending on the filter, ML and $\alpha_{ml}$ combination, but generally smaller than $0.02$-$0.03$ mag, is expected. 

\subsection{The comparison with published PWZ relations}
The inferred theoretical PWZ relations can be tested against current empirical determinations in the literature. Some of the most recent published PWZ relations are listed in Table \ref{pwz_lit}. We notice that the metallicity coefficient of the PWZ relations predicted in this paper and reported in Tables \ref{pwz_gaia} and \ref{pwz_johK} is significantly smaller than in the corresponding empirical relations derived by \citet[][]{Ripepi2021} and \citet[][]{Ripepi2022} for F and fundamentalized FO-mode Galactic Cepheids, in the Gaia and the  V and K filters, respectively. A better agreement is found between our PWZ relation in the JC V, I filters (see Table\ref{pwz_johI}) and the relation provided by \citet[][]{Breuval2021} for a sample of MW, LMC and SMC CCs.
The reason for the discrepancy between the predicted metallicity effect and the result by Ripepi et al. could be related to the different metallicity range of the selected observational and model samples and deserves further investigation.

\clearpage

\begin{ThreePartTable}
\begin{longtable}{cccccccccc}
\caption{\label{pwz_gaia} PWZ coefficients in the Gaia EDR3 filters (
W ($G$, $G_{BP}$-$G_{RP}$)=a + b ($\log P$ -1) +c [Fe/H]) for F and FO CCs derived by adopting A, B, C ML relations and $\alpha_{ml}$ = 1.5, 1.7 and 1.9.}\\
\hline\hline
$\alpha_{ml}$&ML&a&b&c&$\sigma_{a}$&$\sigma_{b}$&$\sigma_{c}$&$\sigma$&$R^2$\\
\hline
F\\
\hline
\endfirsthead
\caption{continued.}\\
\hline\hline
$\alpha_{ml}$&ML&a&b&c&$\sigma_{a}$&$\sigma_{b}$&$\sigma_{c}$&$\sigma$&$R^2$\\
\hline
\endhead
\hline
\endfoot
1.5&A&-6.018&-3.314&-0.189&0.009&0.016&0.021&0.118&0.993\\
1.7&A&-6.072&-3.379&-0.129&0.010&0.016&0.021&0.090&0.996\\
1.9&A&-6.170&-3.472&-0.245&0.023&0.018&0.040&0.072&0.998\\
1.5&B&-5.853&-3.234&-0.190&0.011&0.016&0.022&0.139&0.991\\
1.7&B&-5.871&-3.262&-0.260&0.012&0.015&0.023&0.118&0.995\\
1.9&B&-5.968&-3.370&-0.189&0.026&0.017&0.047&0.092&0.997\\
1.5&C&-5.694&-3.270&-0.105&0.012&0.017&0.023&0.141&0.991\\
1.7&C&-5.722&-3.274&-0.140&0.012&0.015&0.022&0.116&0.994\\
1.9&C&-5.800&-3.327&-0.167&0.023&0.016&0.043&0.094&0.997\\
\hline
FO\\
\hline
1.5&A&-6.676&-3.450&-0.221&0.051&0.048&0.059&0.145&0.985\\
1.7&A&-6.818&-3.627&-0.243&0.040&0.034&0.049&0.073&0.996\\
1.9&A&-6.933&-3.688&-0.349&0.045&0.030&0.052&0.034&0.999\\
1.5&B&-6.634&-3.566&-0.304&0.063&0.063&0.062&0.097&0.988\\
1.7&B&-6.616&-3.533&-0.303&0.095&0.083&0.095&0.103&0.987\\
1.9&B&-6.719&-3.627&-0.304&0.066&0.050&0.068&0.030&0.998\\
1.5&C&-6.473&-3.510&-0.235&0.043&0.051&0.038&0.038&0.996\\
1.7&C&-6.486&-3.506&-0.261&0.049&0.056&0.051&0.030&0.998\\
\hline
\end{longtable}
\end{ThreePartTable}

\clearpage

\begin{ThreePartTable}
\begin{longtable}{cccccccccc}
\caption{\label{pwz_hst} PWZ coefficients in the HST-WFC3 filters (W (F160W, F555W-F814W)=a + b ($\log P$ -1) +c [Fe/H]) for F and FO CCs derived by adopting A, B, C ML relations and $\alpha_{ml}$ = 1.5, 1.7 and 1.9.}\\
\hline\hline
$\alpha_{ml}$&ML&a&b&c&$\sigma_{a}$&$\sigma_{b}$&$\sigma_{c}$&$\sigma$&$R^2$\\
\hline
F\\
\hline
\endfirsthead
\caption{continued.}\\
\hline\hline
$\alpha_{ml}$&ML&a&b&c&$\sigma_{a}$&$\sigma_{b}$&$\sigma_{c}$&$\sigma$&$R^2$\\
\hline
\endhead
\hline
\endfoot
1.5&A&-6.023&-3.340&-0.161&0.008&0.015&0.019&0.109&0.994\\
1.7&A&-6.060&-3.393&-0.135&0.010&0.015&0.020&0.085&0.997\\
1.9&A&-6.155&-3.484&-0.195&0.022&0.017&0.038&0.067&0.998\\
1.5&B&-5.858&-3.254&-0.125&0.010&0.015&0.021&0.129&0.993\\
1.7&B&-5.865&-3.280&-0.121&0.011&0.014&0.021&0.110&0.995\\
1.9&B&-5.952&-3.376&-0.136&0.025&0.016&0.045&0.088&0.997\\
1.5&C&-5.702&-3.274&-0.145&0.011&0.016&0.021&0.127&0.993\\
1.7&C&-5.717&-3.281&-0.130&0.011&0.014&0.020&0.104&0.995\\
1.9&C&-5.782&-3.329&-0.110&0.022&0.015&0.040&0.088&0.997\\
\hline
FO\\
\hline
1.5&A&-6.684&-3.477&-0.105&0.048&0.045&0.056&0.137&0.987\\
1.7&A&-6.821&-3.651&-0.131&0.037&0.031&0.046&0.067&0.997\\
1.9&A&-6.930&-3.710&-0.240&0.038&0.025&0.045&0.029&0.999\\
1.5&B&-6.643&-3.595&-0.186&0.059&0.059&0.058&0.090&0.990\\
1.7&B&-6.627&-3.568&-0.192&0.091&0.080&0.091&0.099&0.989\\
1.9&B&-6.723&-3.653&-0.204&0.061&0.046&0.062&0.027&0.999\\
1.5&C&-6.499&-3.573&-0.125&0.034&0.041&0.030&0.030&0.998\\
1.7&C&-6.518&-3.576&-0.162&0.041&0.046&0.042&0.025&0.998\\
\hline
\end{longtable}
\end{ThreePartTable}

\clearpage

\begin{ThreePartTable}
\begin{longtable}{cccccccccc}
\caption{\label{pwz_johI} PWZ coefficients in the Johnson-Cousins filters (W(V,I)=a + b ($\log P$ -1) +c [Fe/H]) for F and FO CCs derived by adopting A, B, C ML relations and $\alpha_{ml}$ = 1.5, 1.7 and 1.9.}\\
\hline\hline
$\alpha_{ml}$&ML&a&b&c&$\sigma_{a}$&$\sigma_{b}$&$\sigma_{c}$&$\sigma$&$R^2$\\
\hline
F\\
\hline
\endfirsthead
\caption{continued.}\\
\hline\hline
$\alpha_{ml}$&ML&a&b&c&$\sigma_{a}$&$\sigma_{b}$&$\sigma_{c}$&$\sigma$&$R^2$\\
\hline
\endhead
\hline
\endfoot
1.5&A&-6.118&-3.341&-0.150&0.008&0.015&0.019&0.108&0.994\\
1.7&A&-6.162&-3.396&-0.175&0.010&0.015&0.020&0.085&0.997\\
1.9&A&-6.248&-3.483&-0.175&0.022&0.017&0.038&0.068&0.998\\
1.5&B&-5.957&-3.269&-0.101&0.010&0.014&0.020&0.125&0.993\\
1.7&B&-5.969&-3.289&-0.104&0.011&0.014&0.021&0.108&0.995\\
1.9&B&-6.045&-3.381&-0.119&0.025&0.016&0.045&0.089&0.997\\
1.5&C&-5.799&-3.298&-0.123&0.011&0.015&0.020&0.124&0.993\\
1.7&C&-5.821&-3.298&-0.111&0.011&0.014&0.020&0.104&0.996\\
1.9&C&-5.877&-3.338&-0.097&0.022&0.016&0.041&0.090&0.997\\
\hline
FO\\
\hline
1.5&A&-6.774&-3.490&-0.174&0.049&0.045&0.056&0.138&0.987\\
1.7&A&-6.916&-3.666&-0.200&0.037&0.031&0.045&0.066&0.997\\
1.9&A&-7.030&-3.731&-0.301&0.036&0.024&0.042&0.027&0.999\\
1.5&B&-6.733&-3.610&-0.252&0.060&0.059&0.058&0.091&0.990\\
1.7&B&-6.718&-3.581&-0.255&0.092&0.081&0.092&0.100&0.988\\
1.9&B&-6.817&-3.672&-0.256&0.049&0.038&0.051&0.022&0.999\\
1.5&C&-6.574&-3.564&-0.188&0.035&0.042&0.031&0.031&0.998\\
1.7&C&-6.591&-3.566&-0.216&0.040&0.046&0.042&0.025&0.998\\
\hline
\end{longtable}
\end{ThreePartTable}

\clearpage
\begin{ThreePartTable}
\begin{longtable}{cccccccccc}
\caption{\label{pwz_johK} PWZ coefficients in the Johnson-Cousins filters (W(V,K)=a + b ($\log P$ -1) +c [Fe/H]) for F and FO CCs derived by adopting A, B, C ML relations and $\alpha_{ml}$ = 1.5, 1.7 and 1.9.}\\
\hline\hline
$\alpha_{ml}$&ML&a&b&c&$\sigma_{a}$&$\sigma_{b}$&$\sigma_{c}$&$\sigma$&$R^2$\\
\hline
F\\
\hline
\endfirsthead
\caption{continued.}\\
\hline\hline
$\alpha_{ml}$&ML&a&b&c&$\sigma_{a}$&$\sigma_{b}$&$\sigma_{c}$&$\sigma$&$R^2$\\
\hline
\endhead
\hline
\endfoot
1.5&A&-6.016&-3.382&-0.145&0.008&0.014&0.018&0.101&0.995\\
1.7&A&-6.041&-3.419&-0.138&0.009&0.014&0.019&0.081&0.997\\
1.9&A&-6.126&-3.496&-0.177&0.021&0.017&0.037&0.067&0.998\\
1.5&B&-5.856&-3.300&-0.190&0.009&0.014&0.019&0.117&0.994\\
1.7&B&-5.858&-3.316&-0.195&0.010&0.013&0.020&0.101&0.996\\
1.9&B&-5.924&-3.388&-0.200&0.025&0.016&0.044&0.087&0.997\\
1.5&C&-5.702&-3.311&-0.107&0.010&0.014&0.019&0.113&0.994\\
1.7&C&-5.712&-3.312&-0.198&0.010&0.013&0.018&0.094&0.996\\
1.9&C&-5.755&-3.340&-0.106&0.021&0.015&0.039&0.086&0.997\\
\hline
FO\\
\hline
1.5&A&-6.656&-3.482&-0.089&0.048&0.045&0.056&0.136&0.987\\
1.7&A&-6.790&-3.655&-0.114&0.038&0.031&0.046&0.068&0.997\\
1.9&A&-6.899&-3.713&-0.227&0.040&0.026&0.047&0.030&0.999\\
1.5&B&-6.614&-3.599&-0.170&0.059&0.059&0.058&0.090&0.990\\
1.7&B&-6.598&-3.573&-0.177&0.091&0.080&0.091&0.099&0.989\\
1.9&B&-6.692&-3.656&-0.193&0.064&0.049&0.066&0.029&0.998\\
1.5&C&-6.473&-3.585&-0.108&0.034&0.041&0.030&0.030&0.998\\
1.7&C&-6.491&-3.586&-0.147&0.041&0.047&0.043&0.025&0.998\\
\hline
\end{longtable}
\end{ThreePartTable}

\clearpage 

\begin{table}
\caption{\label{pwz_lit}The PWZ relations (W=a + b ($\log P$ -1) +c [Fe/H]) recently derived by various authors for different CC samples. $\sigma$ is the predicted root-mean-square deviation coefficient.}
\centering
\begin{tabular}{cccccccccc}
\hline\hline
authors & sample & bands & a & b & c & $\sigma_{a}$ & $\sigma_{b}$ & $\sigma_{c}$ & $\sigma$\\
\hline
\citet[][]{Breuval2021}& MW-LMC-SMC CC & JC V-I & -5.9893 & -3.281 & -0.251 & 0.008 & 0.022 & 0.057 \\
\citet[][]{Ripepi2021}& F-FO MW-CC & JC V-K & -6.022 & -3.174 & -0.459 & 0.022 & 0.049 & 0.107 & \\
\citet[][]{Ripepi2022}& F-FO MW-CC & G-$G_{BP}$-$G_{RP}$ & -5.988 & -3.176 & -0.520 & 0.018 & 0.044 & 0.090 & 0.014\\
\hline\hline
\end{tabular}
\end{table}

\subsection{The predicted distance of the Large Magellanic Cloud: effect of adopted ML relation}

To quantify the impact of a variation of the ML relation on theoretical distance determinations, we applied the PW relations obtained for $Z=0.008$ to the OGLE catalog of LMC Cepheids \citep[][]{Sos2017}. To evaluate the influence of a variation of the ML relation on the distance scale based on predicted relations and, in turn, on the final theoretically obtained $H_{0}$ value, we considered the derived  PW relations in the Johnson-Cousins V and I filters for $Z=0.008$ with $\alpha_{ml}=1.5$ and ML cases A, B, and C. Application of these theoretical PW relations to the LMC Cepheids provided the distance moduli reported in Table \ref{moduli_lmc}.
As expected, a variation of $\Delta\log(L$/$L_\odot)$=$~0.2 \; dex$ in the assumed ML relation (e.g. between cases A and B or between cases B and C) produces a variation in the inferred LMC distance modulus of $\sim$ 0.1 dex. This exercise shows that if a variation of the Cepheid ML relation occurred from galaxy to galaxy the assumption of a unique calibrating PW relation could produce systematic errors on the inferred distances.

\begin{table}
\caption{\label{moduli_lmc} The distance moduli of LMC F-mode CCs obtained with the theoretical PW relations in the Johnson-Cousins V and I filters for $Z=0.008$ with $\alpha_{ml}=1.5$ and ML cases A, B, and C.}
\centering
\begin{tabular}{cccc}
\hline\hline
source & $\mu$ & $\alpha_{ml}$ & ML\\
\hline
F CC LMC & 18.667 & 1.5 & A \\
F CC LMC & 18.555 & 1.5 & B \\
F CC LMC & 18.403 & 1.5 & C \\
\hline\hline
\end{tabular}
\end{table}

\section{The predicted parallaxes of Galactic Cepheids}

\begin{figure}[th]
\centering
\includegraphics[width=\textwidth]{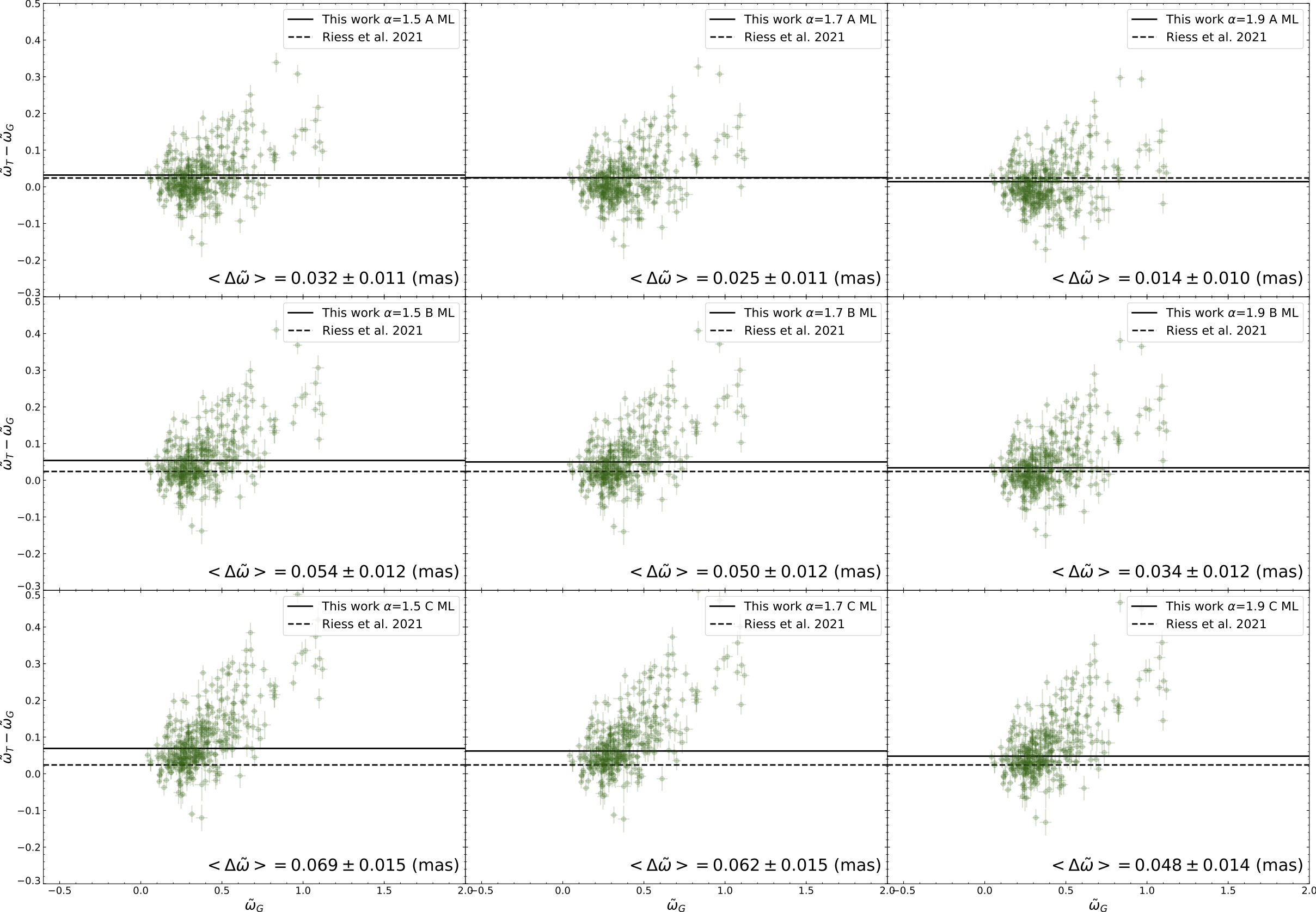}
\caption{Parallax difference between the theoretical parallax obtained by adopting the PWZ relation and Gaia parallax ($\varpi_{T}$ - $\varpi_{G}$) as a function of Gaia EDR3 parallax $\varpi_{G}$, for F-mode Galactic Cepheids.}
\label{Fig:gaia_offset_pwz_F}
\end{figure}

\begin{figure}[th]
\centering
\includegraphics[width=\textwidth]{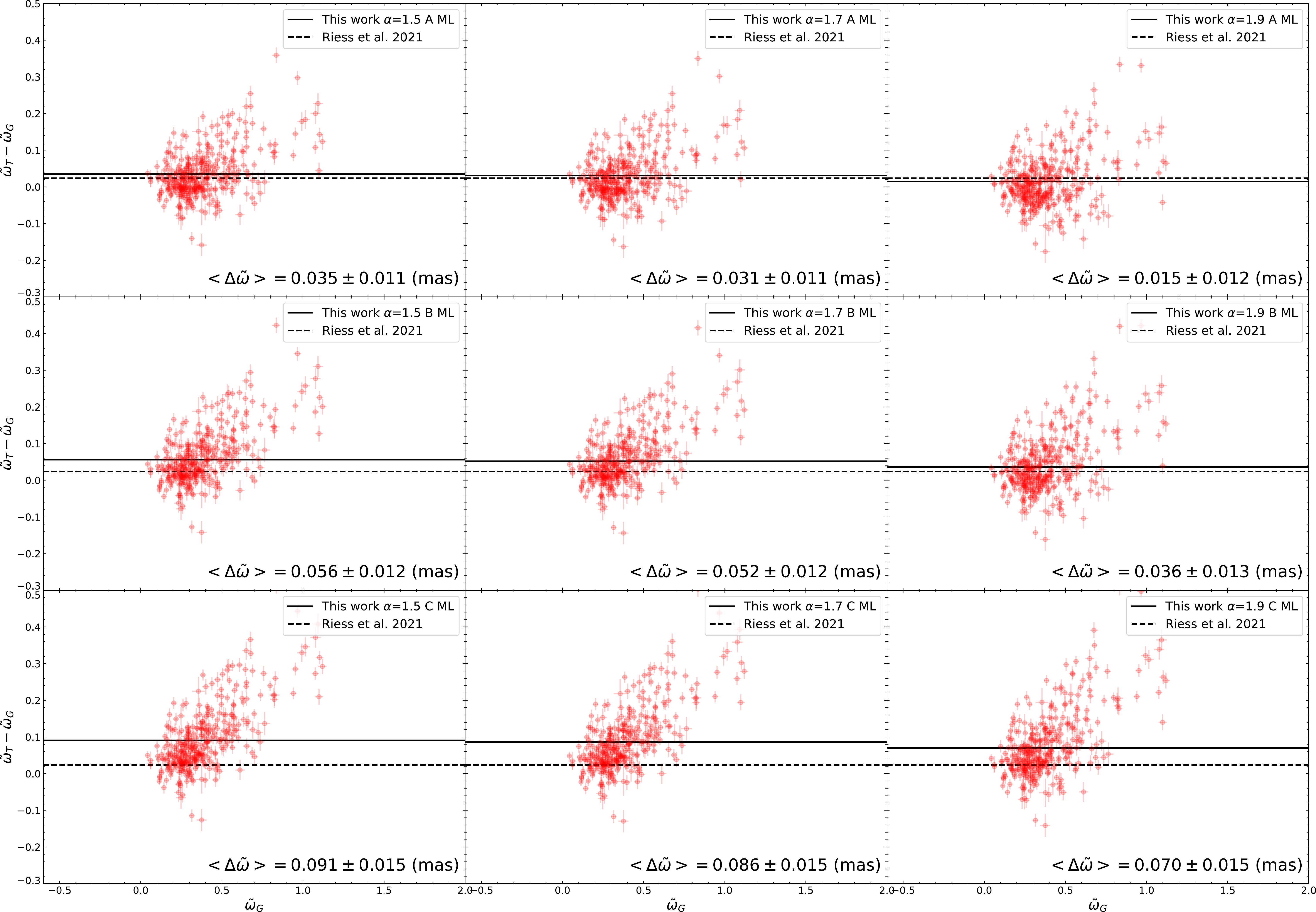}
\caption{The same as Fig.\ref{Fig:gaia_offset_pwz_F} but with the theoretical parallax obtained by applying the PW relation for $Z=0.02$.}
\label{Fig:gaia_offset_pw_F}
\end{figure}

Following the same approach as \citet[][]{Desomma2020a}, we selected a sample of Galactic Cepheids with available periods, $G$, $G_{BP}$ and $G_{RP}$ magnitudes and Gaia EDR3 parallaxes. After being ensured that the parallax zero-point offset was corrected according to \citet[][]{Lindegren2021} \citep[see][for details on this procedure]{Ripepi2021}, we derived the predicted distances and parallaxes of each Galactic Cepheid in the sample through the application of PW relations. In particular, we considered F and FO-mode Galactic Cepheids for which the G band magnitude is higher than 6. Moreover, as good astrometry is ensured by using sources for which the Renormalized Unit Weight Error (RUWE)\footnote{Section 14.1.2 of ’Gaia Data Release 2 Documentation release 1.2’;
https://gea.esac.esa.int/archive/documentation/GDR2/} coefficient is lower than 1.4 \citep[][]{Ripepi2021}, we made this assumption in the target selection. The theoretical PWZ relations obtained in Section \ref{sec_pwz} in the Gaia filters, were applied to the selected sample in order to obtain reddening free individual distances and in turn theoretical parallaxes, through inversion of the predicted distances.  The theoretical parallaxes were compared with the corresponding Gaia EDR3 values for the selected stars. The different panels of Fig. \ref{Fig:gaia_offset_pwz_F} show the difference between theoretical and Gaia parallaxes as a function of Gaia parallaxes (solid lines) compared with the value estimated for the Gaia zero-point offset published by \citet[][]{Riess2021a} \footnote{\citet[][]{Riess2021a} zero-point offset was obtained by searching for the best match between the measured Gaia EDR3 parallaxes and those predicted from their photometry, periods, and the fiducial luminosity of the HST sample of Milky Way Cepheids \citep[see][for details]{Riess2021a}.
}. The comparison is shown for each assumption about the mixing length parameter and the ML relation (see labels on the plot for details). From these plots, we can conclude that the combination of $\alpha_{ml}$ and ML that is most consistent with the value found by \citet[][]{Riess2021a} corresponds to the canonical ML relation (case A models) with $\alpha_{ml}$ =1.7. 
In order to quantify the impact of the metallicity term on the inferred offsets, we repeated the analysis using the theoretical parallaxes based on the PW relations at a fixed solar metallicity (See Fig. \ref{Fig:gaia_offset_pw_F}). Again we found that the best match with the result by \citet[][]{Riess2021a} is obtained for the canonical ML relation with $\alpha_{ml}=1.7$. Moreover, we notice that, for each assumed efficiency of super-adiabatic convection, the agreement with the offset by \citet[][]{Riess2021a} tends to worsen as the ML relation gets brighter. This occurrence holds independently of the inclusion of a metallicity term in the PW relation.

\section{Conclusions and Future Developments}

Based on an extended and detailed grid of nonlinear convective pulsation models, built by systematically varying different physical and numerical ingredients, we have presented an updated metal-dependent theoretical scenario for Classical Cepheids. By extending the model set for solar chemical abundances, published in \citet[][]{Desomma2020a}, we have computed new nonlinear convective pulsation models for three additional chemical compositions, namely $Z=0.004$ $Y=0.25$, $Z=0.008$ $Y=0.25$ and $Z=0.02$ $Y=0.28$, three different assumptions about the ML relation and three adopted values of $\alpha_{ml}$. From the integration of the nonlinear system of dynamical and convective equations, we have derived the modal stability for both F and FO-mode pulsators. As a result, we have found that:
\begin{itemize}
    \item The predicted PMR and PR relations are almost insensitive to metallicity and the efficiency of super-adiabatic convection, whereas they show a non-negligible dependence on the assumed ML relation.
    \item The predicted fundamental periods, at fixed M, L and $T_e$, show a small correlation with the assumed metal content, whereas no significant dependence is found for first overtone models. Moreover, the predicted F-mode PLMTZ relations are found to be independent of the assumed $\alpha_{ml}$ parameter.
    \item We found that the predicted instability strip, already proved in \citet[][]{Desomma2021a} and previous papers, becomes redder as the metallicity increases and also gets narrower as the super-adiabatic convection gets more efficient. These trends are confirmed for each assumption about the ML relation.
    \item The bolometric light and radial velocity curves are provided for each model set, varying simultaneously the chemical composition, the ML and $\alpha_{ml}$. As expected, at fixed M, L and $T_e$ the predicted pulsation amplitude tends to decrease as $\alpha_{ml}$ increases; whereas, the main driver of the light curve morphology at a fixed mass is the effective temperature, that is the position within the instability strip, with minor dependence on L and Z. 
    \item By adopting the transformed light curves into a variety of photometric filters, we derived the intensity weighted mean magnitudes and colors as well as the corresponding pulsation amplitudes. In agreement with previous theoretical results, we find that the pulsation amplitudes tend to decrease as the metallicity increases and, for the F-mode and at low luminosity levels, they show a linear trend with the pulsation periods. At higher luminosities, in a range where the FO-mode pulsation is no more efficient, F-mode predicted amplitudes define a bell-shape similar to the one of FO models at lower luminosity levels.
    \item The mean magnitudes and colors were used to derive PLC and PW relations for each selected chemical composition and pulsation mode, varying the ML relation, the efficiency of super-adiabatic convection and the filter combination. In particular, for a $Z$ different from the solar one, we obtained the first theoretical PLC and PW relations in the Gaia filters. Metal dependent multi-filter PW relations were also derived which pointed towards a metallicity effect on the zero point varying from $\sim$-0.1 dex to $\sim$-0.2 dex for the F-mode relations and from $\sim$-0.1 dex to $\sim$-0.3 dex for the FO-mode relations.
    \item The obtained multi-filter PWZ relations were compared with similar relations in the recent literature. The main difference between theoretical and empirical relations concerns the metallicity term coefficient, with the predicted one systematically smaller than the one obtained by other authors, in particular \citet[][]{Ripepi2021} and \citet[][]{Ripepi2022}.
    \item The PW relations in the Johnson V and I bands for $Z=0.008$ with $\alpha_{ml}$=1.5 and ML cases A, B, and C were applied to a sample of LMC Cepheids to estimate the impact of a variation in the ML relation on the theoretical calibration of the extra-galactic distance scale. This implies that possible variations between the ML relations of anchor and target Cepheid host galaxies might represent a residual source of systematic uncertainty.
    \item The PWZ relations in the Gaia filters were applied to a sample of Galactic Cepheids with Gaia EDR3 parallaxes and complementary spectroscopic metal abundances. The obtained theoretical parallaxes for different assumptions about the ML relation and $\alpha_{ml}$ were compared with Gaia EDR3 values and with the zero-point offset by \citet[][]{Riess2021a}. We found a significant dependence on the assumed ML relation and a minor sensitivity to the efficiency of super-adiabatic convection. Indeed, the best match with the \citet[][]{Riess2021a} estimate of the zero-point offset was obtained for a canonical ML and $\alpha_{ml}=1.7$.
    Similar results were obtained by replacing the PWZ relations with the relations at a fixed solar metallicity.
\end{itemize}
The obtained theoretical scenario provided sound estimates of individual and mean distances for different combinations of chemical composition, ML relation and $\alpha_{ml}$ and allowed us to provide some constraints on the physical assumptions of pulsation models. On the other hand, as we cannot exclude a variation of the efficiency of super-adiabatic convection across the instability strip \citep[see e.g.][]{Dicrisci2004,Fiorentino2007}, as well as some dispersion in the ML relation \citep[see e.g.][]{Caputo2005,Marconi2013,Marconi2017}, further theoretical tests should be performed to quantify the predicted zero-point offset. Moreover, variations in the Helium abundance are known to affect the predicted pulsation properties \citep[see e.g.][and references therein]{Marconi2005,Carini2017}. New model sets are being produced for various helium abundances to provide updated constraints on the dependence of the predicted distance scale on the helium to metal enrichment ratio and to concurrently develop theoretical tools to infer the Helium abundance from accurate astrometric distances (e.g. from Gaia and Rubin-LSST) and spectroscopic metallicities.

\appendix

\section{The instability strip boundaries} \label{sec:boundaries}

\begin{ThreePartTable}
\begin{longtable}{cccccccc}
\caption{\label{boundaries_smc} Predicted effective temperatures of the instability strip Boundaries (FOBE, FBE, FORE and FRE) for Z=0.004 Y= 0.25 pulsation models. The assumed error of the predicted boundaries of the instability strip is $T_{eff} = \pm\; 50\;K$, based on our effective temperature step assumption in the building of the pulsation model grid.}\\
\hline\hline
\msun & \lsun & $\alpha_{ml}$ & ML & FOBE & FBE & FORE & FRE \\
\endfirsthead
\caption{continued.}\\
\hline\hline
\msun & \lsun & $\alpha_{ml}$ & ML & FOBE & FBE & FORE & FRE \\
\hline\hline
\endhead
\hline
3.0 & 2.49 & 1.5 & A & 6750 & 6050 & 6050 & 5850 \\
3.0 & 2.49 & 1.7 & A & 6850 & 6150 & 6150 & 5950 \\
3.0 & 2.49 & 1.9 & A & 6750 & 6250 & 6250 & 6050 \\
3.0 & 2.69 & 1.5 & B & 6650 & 6050 & 6050 & 5650 \\
3.0 & 2.69 & 1.7 & B & 6650 & 6150 & 6150 & 5750 \\
3.0 & 2.69 & 1.9 & B & 6550 & 6250 & 6150 & 5850 \\
3.0 & 2.89 & 1.5 & C & 6450 & 6150 & 6150 & 5450 \\
3.0 & 2.89 & 1.7 & C & 6450 & 6250 & 6150 & 5550 \\
3.0 & 2.89 & 1.9 & C & 6350 & 6350 & 6250 & 5650 \\
4.0 & 2.91 & 1.5 & A & 6550 & 5950 & 5950 & 5550 \\
4.0 & 2.91 & 1.7 & A & 6550 & 6050 & 6050 & 5650 \\
4.0 & 2.91 & 1.9 & A & 6450 & 6150 & 6150 & 5750 \\
4.0 & 3.11 & 1.5 & B & 6350 & 6050 & 5550 & 5350 \\
4.0 & 3.11 & 1.7 & B & 6350 & 6150 & 5650 & 5450 \\
4.0 & 3.11 & 1.9 & B & 6250 & 6250 & 6150 & 5550 \\
4.0 & 3.31 & 1.5 & C & 6150 & 6050 & 5850 & 5050 \\
4.0 & 3.31 & 1.7 & C & 6050 & 6150 & 5950 & 5250 \\
4.0 & 3.31 & 1.9 & C &  & 6050 &  & 5350 \\
5.0 & 3.24 & 1.5 & A & 6350 & 6050 & 5950 & 5250 \\
5.0 & 3.24 & 1.7 & A & 6250 & 6050 & 6050 & 5450 \\
5.0 & 3.24 & 1.9 & A & 6250 & 6150 & 6050 & 5550 \\
5.0 & 3.44 & 1.5 & B & 6150 & 5950 & 5650 & 5050 \\
5.0 & 3.44 & 1.7 & B & 6050 & 6050 & 5750 & 5150 \\
5.0 & 3.44 & 1.9 & B & 5950 & 6050 & 5850 & 5350 \\
5.0 & 3.64 & 1.5 & C &  & 5950 &  & 4750 \\
5.0 & 3.64 & 1.7 & C &  & 5950 &  & 4950 \\
5.0 & 3.64 & 1.9 & C &  & 5950 &  & 5150 \\
6.0 & 3.5 & 1.5 & A & 6150 & 5850 & 5550 & 5150 \\
6.0 & 3.5 & 1.7 & A & 6050 & 5950 & 5650 & 5250 \\
6.0 & 3.5 & 1.9 & A & 5950 & 6050 & 5750 & 5350 \\
6.0 & 3.7 & 1.5 & B &  & 5950 &  & 4850 \\
6.0 & 3.7 & 1.7 & B &  & 5950 &  & 5050 \\
6.0 & 3.7 & 1.9 & B &  & 5950 &  & 5150 \\
6.0 & 3.9 & 1.5 & C &  & 5850 &  & 4450 \\
6.0 & 3.9 & 1.7 & C &  & 5850 &  & 4750 \\
6.0 & 3.9 & 1.9 & C &  & 5750 &  & 4950 \\
7.0 & 3.73 & 1.5 & A & 5950 & 5950 & 5750 & 4850 \\
7.0 & 3.73 & 1.7 & A &  & 5950 &  & 5050 \\
7.0 & 3.73 & 1.9 & A &  & 5950 &  & 5250 \\
7.0 & 3.93 & 1.5 & B &  & 5850 &  & 4550 \\
7.0 & 3.93 & 1.7 & B &  & 5850 &  & 4850 \\
7.0 & 3.93 & 1.9 & B &  & 5750 &  & 5050 \\
7.0 & 4.13 & 1.5 & C &  & 5750 &  & 4350 \\
7.0 & 4.13 & 1.7 & C &  & 5650 &  & 4450 \\
7.0 & 4.13 & 1.9 & C &  & 5650 &  & 4650 \\
8.0 & 3.92 & 1.5 & A &  & 5850 &  & 4750 \\
8.0 & 3.92 & 1.7 & A &  & 5850 &  & 4950 \\
8.0 & 3.92 & 1.9 & A &  & 5850 &  & 5150 \\
8.0 & 4.12 & 1.5 & B &  & 5750 &  & 4250 \\
8.0 & 4.12 & 1.7 & B &  & 5750 &  & 4550 \\
8.0 & 4.12 & 1.9 & B &  & 5650 &  & 4850 \\
8.0 & 4.32 & 1.5 & C &  & 5550 &  & 4450 \\
8.0 & 4.32 & 1.7 & C &  & 5550 &  & 4250 \\
8.0 & 4.32 & 1.9 & C &  & 5550 &  & 4350 \\
9.0 & 4.09 & 1.5 & A &  & 5750 &  & 4550 \\
9.0 & 4.09 & 1.7 & A &  & 5750 &  & 4750 \\
9.0 & 4.09 & 1.9 & A &  & 5650 &  & 4950 \\
9.0 & 4.29 & 1.5 & B &  & 5650 &  & 4250 \\
9.0 & 4.29 & 1.7 & B &  & 5650 &  & 4450 \\
9.0 & 4.29 & 1.9 & B &  & 5650 &  & 4650 \\
9.0 & 4.49 & 1.5 & C &  & 5550 &  & 4650 \\
9.0 & 4.49 & 1.7 & C &  & 5450 &  & 4550 \\
9.0 & 4.49 & 1.9 & C &  & 5450 &  & 4450 \\
10.0 & 4.25 & 1.5 & A &  & 5650 &  & 4350 \\
10.0 & 4.25 & 1.7 & A &  & 5650 &  & 4650 \\
10.0 & 4.25 & 1.9 & A &  & 5650 &  & 4850 \\
10.0 & 4.45 & 1.5 & B &  & 5550 &  & 4250 \\
10.0 & 4.45 & 1.7 & B &  & 5550 &  & 4350 \\
10.0 & 4.45 & 1.9 & B &  & 5550 &  & 4550 \\
10.0 & 4.65 & 1.5 & C &  & 5450 &  & 4850 \\
10.0 & 4.65 & 1.7 & C &  & 5450 &  & 4850 \\
10.0 & 4.65 & 1.9 & C &  & 5350 &  & 4850 \\
11.0 & 4.39 & 1.5 & A &  & 5650 &  & 4250 \\
11.0 & 4.39 & 1.7 & A &  & 5650 &  & 4450 \\
11.0 & 4.39 & 1.9 & A &  & 5550 &  & 4750 \\
11.0 & 4.59 & 1.5 & B &  & 5550 &  & 4450 \\
11.0 & 4.59 & 1.7 & B &  & 5450 &  & 4350 \\
11.0 & 4.59 & 1.9 & B &  & 5450 &  & 4350 \\
11.0 & 4.79 & 1.5 & C &  & 5350 &  & 5150 \\
11.0 & 4.79 & 1.7 & C &  & 5350 &  & 5050 \\
11.0 & 4.79 & 1.9 & C &  & 5250 &  & 5050 \\
\hline\hline
\end{longtable}
\end{ThreePartTable}

\clearpage

\begin{ThreePartTable}
\begin{longtable}{cccccccccc}
\caption{\label{boundaries_lmc} Predicted effective temperatures of the instability strip boundaries (FOBE, FBE, FORE and FRE) for Z=0.008 Y= 0.25 pulsation models. The assumed error of the predicted boundaries of the instability strip is $T_{eff} = \pm\; 50\;K$, based on our effective temperature step assumption in the building of the pulsation model grid.}\\
\hline\hline
\msun & \lsun & $\alpha_{ml}$ & ML & FOBE & FBE & FORE & FRE \\
\endfirsthead
\caption{continued.}\\
\hline\hline
\msun & \lsun & $\alpha_{ml}$ & ML & FOBE & FBE & FORE & FRE \\
\hline\hline
\endhead
\hline
3.0 & 2.39 & 1.5 & A & 6650 & 6050 & 6050 & 5950 \\
3.0 & 2.39 & 1.7 & A & 6750 &  & 6150 &  \\
3.0 & 2.39 & 1.9 & A & 6750 &  & 6350 &  \\
3.0 & 2.59 & 1.5 & B & 6750 & 6050 & 5950 & 5650 \\
3.0 & 2.59 & 1.7 & B & 6650 & 6150 & 6050 & 5750 \\
3.0 & 2.59 & 1.9 & B & 6650 & 6250 & 6150 & 5950 \\
3.0 & 2.79 & 1.5 & C & 6550 & 6150 & 6050 & 5450 \\
3.0 & 2.79 & 1.7 & C & 6550 & 6150 & 6150 & 5650 \\
3.0 & 2.79 & 1.9 & C & 6450 & 6250 & 6150 & 5750 \\
4.0 & 2.81 & 1.5 & A & 6550 & 5950 & 5850 & 5550 \\
4.0 & 2.81 & 1.7 & A & 6550 & 6050 & 5950 & 5750 \\
4.0 & 2.81 & 1.9 & A & 6550 & 6150 & 6050 & 5850 \\
4.0 & 3.01 & 1.5 & B & 6450 & 6050 & 5950 & 5350 \\
4.0 & 3.01 & 1.7 & B & 6350 & 6050 & 6050 & 5450 \\
4.0 & 3.01 & 1.9 & B & 6250 & 6150 & 6050 & 5650 \\
4.0 & 3.21 & 1.5 & C & 6250 & 5950 & 5650 & 5150 \\
4.0 & 3.21 & 1.7 & C & 6150 & 6150 & 5750 & 5250 \\
4.0 & 3.21 & 1.9 & C &  & 6150 &  & 5450 \\
5.0 & 3.14 & 1.5 & A & 6350 & 5950 & 5850 & 5250 \\
5.0 & 3.14 & 1.7 & A & 6350 & 5950 & 5950 & 5450 \\
5.0 & 3.14 & 1.9 & A & 6250 & 6050 & 6050 & 5650 \\
5.0 & 3.34 & 1.5 & B & 6150 & 5850 & 5550 & 5150 \\
5.0 & 3.34 & 1.7 & B & 6050 & 6050 & 5650 & 5350 \\
5.0 & 3.34 & 1.9 & B &  & 6050 &  & 5450 \\
5.0 & 3.54 & 1.5 & C &  & 5850 &  & 4750 \\
5.0 & 3.54 & 1.7 & C &  & 5850 &  & 4950 \\
5.0 & 3.54 & 1.9 & C &  & 5850 &  & 5150 \\
6.0 & 3.4 & 1.5 & A & 6250 & 5950 & 5450 & 5150 \\
6.0 & 3.4 & 1.7 & A & 6150 & 5950 & 5550 & 5250 \\
6.0 & 3.4 & 1.9 & A &  & 6050 &  & 5450 \\
6.0 & 3.6 & 1.5 & B &  & 5950 &  & 4750 \\
6.0 & 3.6 & 1.7 & B &  & 5950 &  & 5050 \\
6.0 & 3.6 & 1.9 & B &  & 5850 &  & 5150 \\
6.0 & 3.8 & 1.5 & C &  & 5750 &  & 4450 \\
6.0 & 3.8 & 1.7 & C &  & 5750 &  & 4750 \\
6.0 & 3.8 & 1.9 & C &  & 5650 &  & 4950 \\
7.0 & 3.63 & 1.5 & A & 5950 & 5850 & 5550 & 4850 \\
7.0 & 3.63 & 1.7 & A & 5850 & 5850 & 5650 & 5050 \\
7.0 & 3.63 & 1.9 & A &  & 5850 &  & 5250 \\
7.0 & 3.83 & 1.5 & B &  & 5850 &  & 4550 \\
7.0 & 3.83 & 1.7 & B &  & 5750 &  & 4850 \\
7.0 & 3.83 & 1.9 & B &  & 5650 &  & 5050 \\
7.0 & 4.03 & 1.5 & C &  & 5650 &  & 4250 \\
7.0 & 4.03 & 1.7 & C &  & 5550 &  & 4450 \\
7.0 & 4.03 & 1.9 & C &  & 5450 &  & 4750 \\
8.0 & 3.82 & 1.5 & A &  & 5750 &  & 4650 \\
8.0 & 3.82 & 1.7 & A &  & 5750 &  & 4950 \\
8.0 & 3.82 & 1.9 & A &  & 5750 &  & 5150 \\
8.0 & 4.02 & 1.5 & B &  & 5650 &  & 4250 \\
8.0 & 4.02 & 1.7 & B &  & 5650 &  & 4550 \\
8.0 & 4.02 & 1.9 & B &  & 5550 &  & 4850 \\
8.0 & 4.22 & 1.5 & C &  & 5550 &  & 4150 \\
8.0 & 4.22 & 1.7 & C &  & 5450 &  & 4150 \\
8.0 & 4.22 & 1.9 & C &  & 5250 &  & 4450 \\
9.0 & 3.99 & 1.5 & A &  & 5750 &  & 4450 \\
9.0 & 3.99 & 1.7 & A &  & 5650 &  & 4750 \\
9.0 & 3.99 & 1.9 & A &  & 5550 &  & 4950 \\
9.0 & 4.19 & 1.5 & B &  & 5550 &  & 4050 \\
9.0 & 4.19 & 1.7 & B &  & 5450 &  & 4350 \\
9.0 & 4.19 & 1.9 & B &  & 5350 &  & 4750 \\
9.0 & 4.39 & 1.5 & C &  & 5350 &  & 4250 \\
9.0 & 4.39 & 1.7 & C &  & 5350 &  & 4350 \\
9.0 & 4.39 & 1.9 & C &  & 5150 &  & 4250 \\
10.0 & 4.14 & 1.5 & A &  & 5550 &  & 4350 \\
10.0 & 4.14 & 1.7 & A &  & 5450 &  & 4650 \\
10.0 & 4.14 & 1.9 & A &  & 5350 &  & 4950 \\
10.0 & 4.34 & 1.5 & B &  & 5450 &  & 4050 \\
10.0 & 4.34 & 1.7 & B &  & 5250 &  & 4150 \\
10.0 & 4.34 & 1.9 & B &  & 5150 &  & 4550 \\
10.0 & 4.54 & 1.5 & C &  & 5250 &  & 4450 \\
10.0 & 4.54 & 1.7 & C &  & 5250 &  & 4450 \\
10.0 & 4.54 & 1.9 & C &  & 5050 &  & 4350 \\
11.0 & 4.28 & 1.5 & A &  & 5550 &  & 4150 \\
11.0 & 4.28 & 1.7 & A &  & 5350 &  & 4450 \\
11.0 & 4.28 & 1.9 & A &  & 5250 &  & 4850 \\
11.0 & 4.48 & 1.5 & B &  & 5350 &  & 4050 \\
11.0 & 4.48 & 1.7 & B &  & 5150 &  & 4050 \\
11.0 & 4.48 & 1.9 & B &  & 5050 &  & 4350 \\
11.0 & 4.68 & 1.5 & C &  & 5150 &  & 4650 \\
11.0 & 4.68 & 1.7 & C &  & 5150 &  & 4650 \\
11.0 & 4.68 & 1.9 & C &  & 4950 &  & 4650 \\
\hline\hline
\end{longtable}
\end{ThreePartTable}

\begin{ThreePartTable}
\begin{longtable}{cccccccccc}
\caption{\label{boundaries_m31} Predicted effective temperatures of the instability strip boundaries (FOBE, FBE, FORE and FRE) for Z=0.03 Y= 0.28 pulsation models. The assumed error of the predicted boundaries of the instability strip is $T_{eff} = \pm\; 50\;K$, based on our effective temperature step assumption in the building of the pulsation model grid.}\\
\hline\hline
\msun & \lsun & $\alpha_{ml}$ & ML & FOBE & FBE & FORE & FRE \\
\endfirsthead
\caption{continued.}\\
\hline\hline
\msun & \lsun & $\alpha_{ml}$ & ML & FOBE & FBE & FORE & FRE \\
\hline\hline
\endhead
\hline
4.0 & 2.68 & 1.5 & A & 6350.0 & 6150 & 5950.0 & 5350 \\
4.0 & 2.68 & 1.7 & A &  & 6150 &  & 5650 \\
4.0 & 2.88 & 1.5 & B &  & 5950 &  & 5250 \\
4.0 & 2.88 & 1.7 & B &  & 5850 &  & 5450 \\
4.0 & 3.08 & 1.5 & C &  & 5750 &  & 4850 \\
4.0 & 3.08 & 1.7 & C &  & 5650 &  & 5150 \\
5.0 & 3.01 & 1.5 & A &  & 5850 &  & 5250 \\
5.0 & 3.01 & 1.7 & A &  & 5850 &  & 5450 \\
5.0 & 3.21 & 1.5 & B &  & 5650 &  & 4950 \\
5.0 & 3.21 & 1.7 & B &  & 5550 &  & 5250 \\
5.0 & 3.41 & 1.5 & C &  & 5450 &  & 4550 \\
5.0 & 3.41 & 1.7 & C &  & 5250 &  & 4850 \\
6.0 & 3.27 & 1.5 & A &  & 5650 &  & 5050 \\
6.0 & 3.27 & 1.7 & A &  & 5450 &  & 5250 \\
6.0 & 3.47 & 1.5 & B &  & 5450 &  & 4650 \\
6.0 & 3.47 & 1.7 & B &  & 5250 &  & 4950 \\
6.0 & 3.67 & 1.5 & C &  & 5250 &  & 4250 \\
6.0 & 3.67 & 1.7 & C &  & 5050 &  & 4550 \\
7.0 & 3.5 & 1.5 & A &  & 5450 &  & 4750 \\
7.0 & 3.5 & 1.7 & A &  & 5250 &  & 5050 \\
7.0 & 3.7 & 1.5 & B &  & 5250 &  & 4350 \\
7.0 & 3.7 & 1.7 & B &  & 5050 &  & 4750 \\
7.0 & 3.9 & 1.5 & C &  & 5050 &  & 3950 \\
7.0 & 3.9 & 1.7 & C &  & 4850 &  & 4250 \\
8.0 & 3.69 & 1.5 & A &  & 5250 &  & 4550 \\
8.0 & 3.69 & 1.7 & A &  & 5150 &  & 4850 \\
8.0 & 3.89 & 1.5 & B &  & 5050 &  & 4150 \\
8.0 & 3.89 & 1.7 & B &  & 4850 &  & 4450 \\
8.0 & 4.09 & 1.5 & C &  & 4850 &  & 3750 \\
8.0 & 4.09 & 1.7 & C &  & 4750 &  & 4050 \\
9.0 & 3.86 & 1.5 & A &  & 5150 &  & 4350 \\
9.0 & 3.86 & 1.7 & A &  & 4950 &  & 4650 \\
9.0 & 4.06 & 1.5 & B &  & 4950 &  & 3950 \\
9.0 & 4.06 & 1.7 & B &  & 4750 &  & 4250 \\
9.0 & 4.26 & 1.5 & C &  & 4750 &  & 3650 \\
9.0 & 4.26 & 1.7 & C &  & 4650 &  & 3850 \\
10.0 & 4.02 & 1.5 & A &  & 4950 &  & 4150 \\
10.0 & 4.02 & 1.7 & A &  & 4650 &  & 4550 \\
10.0 & 4.22 & 1.5 & B &  & 4750 &  & 3750 \\
10.0 & 4.22 & 1.7 & B &  & 4650 &  & 4050 \\
10.0 & 4.42 & 1.5 & C &  & 4750 &  & 4150 \\
10.0 & 4.42 & 1.7 & C &  & 4550 &  & 3950 \\
11.0 & 4.15 & 1.5 & A &  & 4850 &  & 4050 \\
11.0 & 4.15 & 1.7 & A &  & 4650 &  & 4350 \\
11.0 & 4.35 & 1.5 & B &  & 4750 &  & 3650 \\
11.0 & 4.35 & 1.7 & B &  & 4550 &  & 3850 \\
11.0 & 4.55 & 1.5 & C &  & 4550 &  & 4150 \\
11.0 & 4.55 & 1.7 & C &  & 4450 &  & 4150 \\
\hline\hline
\end{longtable}
\end{ThreePartTable}

\section{The theoretical pulsational amplitudes} \label{sec:amplitudes}

\begin{figure}[th]
\centering
\includegraphics[width=0.8\textwidth]{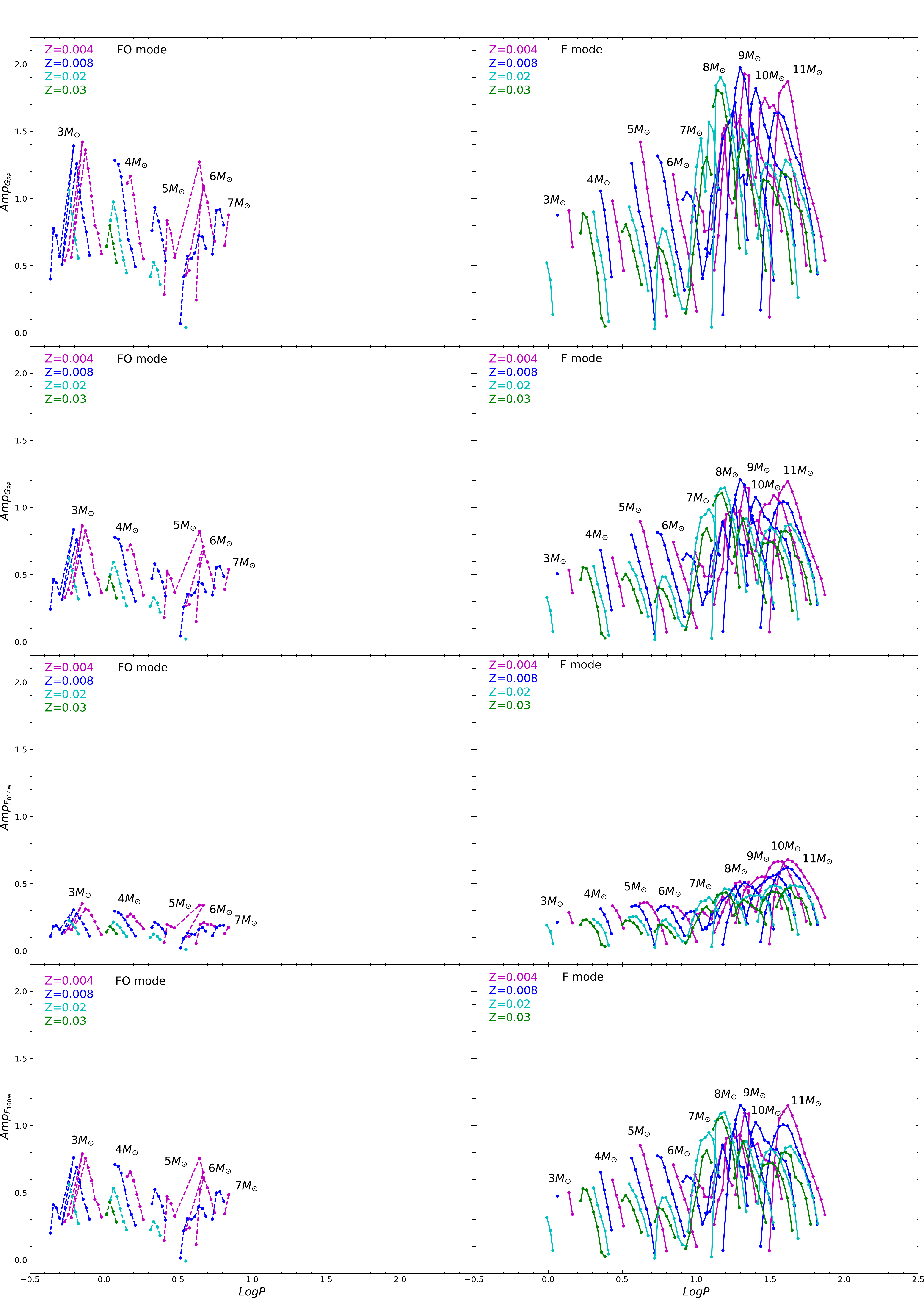}
\caption{From top to bottom, the left panels show the FO-mode theoretical amplitudes for the $G_{BP}$ and $G_{RP}$ Gaia bands and the F814W and F160W HST-WFC3 bands for Z = 0.004 Y = 0.25 (dashed magenta line), Z = 0.008 Y = 0.25 (dashed blue line), Z = 0.02 Y = 0.28 (dashed cyan line) and Z = 0.03 Y = 0.28 (dashed green line) at  fixed mixing length parameter $\alpha_{ml}=1.5$ and ML relation case A. The right panels show the same plots but for F-mode models (solid lines with the same color-code as in the left panels).}
\label{Fig:ampl_F_FO_Z}
\end{figure}

\begin{figure}[th]
\centering
\includegraphics[width=0.8\textwidth]{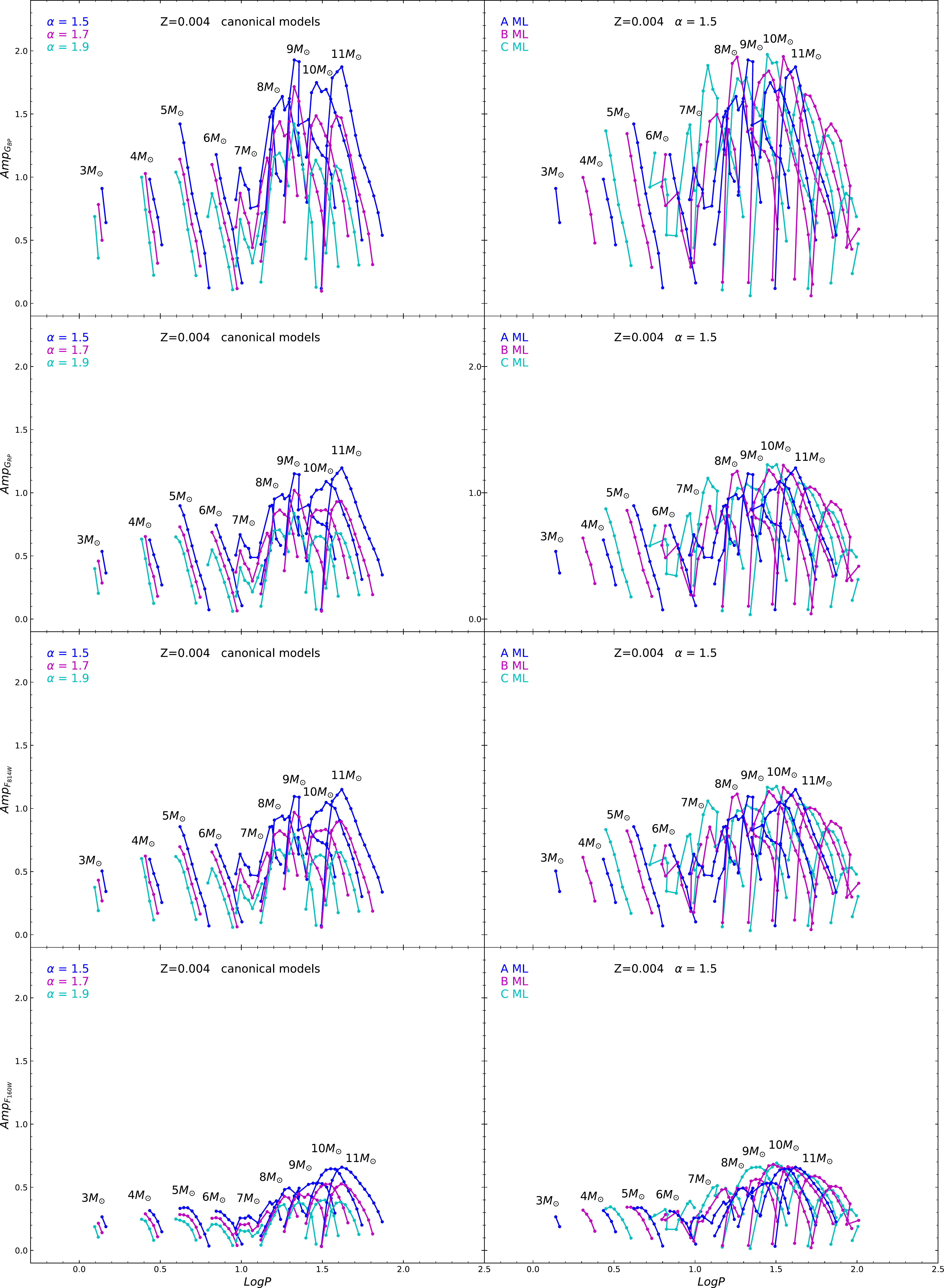}
\caption{From top to bottom, the left panels show the F-mode canonical theoretical amplitudes (solid lines) in the $G_{BP}$ and $G_{RP}$ Gaia bands and the F814W and F160W HST-WFC3 bands for Z = 0.004 Y = 0.25, for various assumptions of the $\alpha_{ml}$ parameters, while the right panels show the same F-mode theoretical amplitudes for various assumptions about the ML relation at a fixed mixing length parameter $\alpha_{ml}=1.5$.}
\label{Fig:ampl_F_SMC}
\end{figure}

\begin{figure}[th]
\centering
\includegraphics[width=0.8\textwidth]{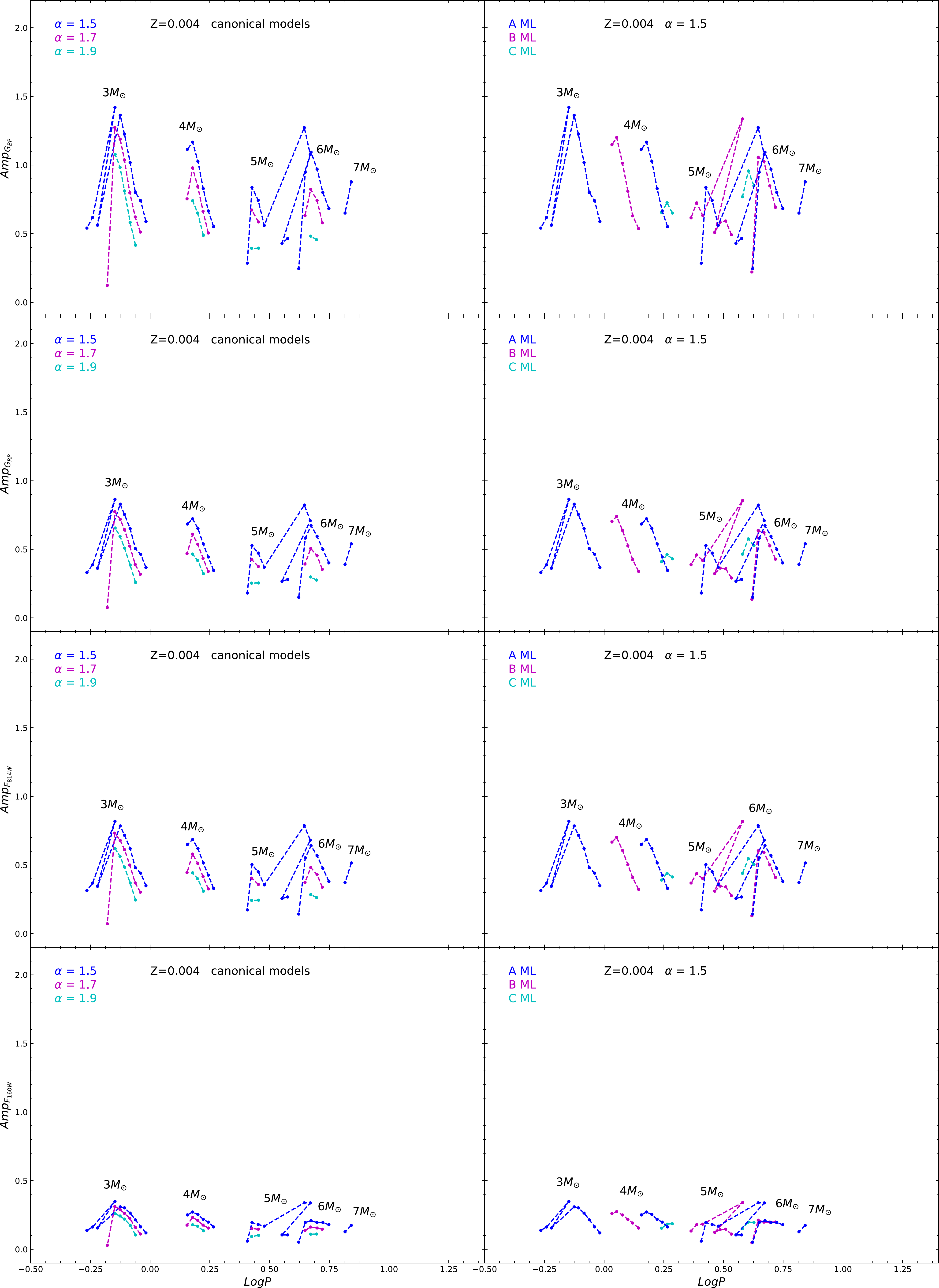}
\caption{The same as Fig.\ref{Fig:ampl_F_SMC} but for the FO-mode models.}
\label{Fig:ampl_FO_SMC}
\end{figure}

\begin{figure}[th]
\centering
\includegraphics[width=0.8\textwidth]{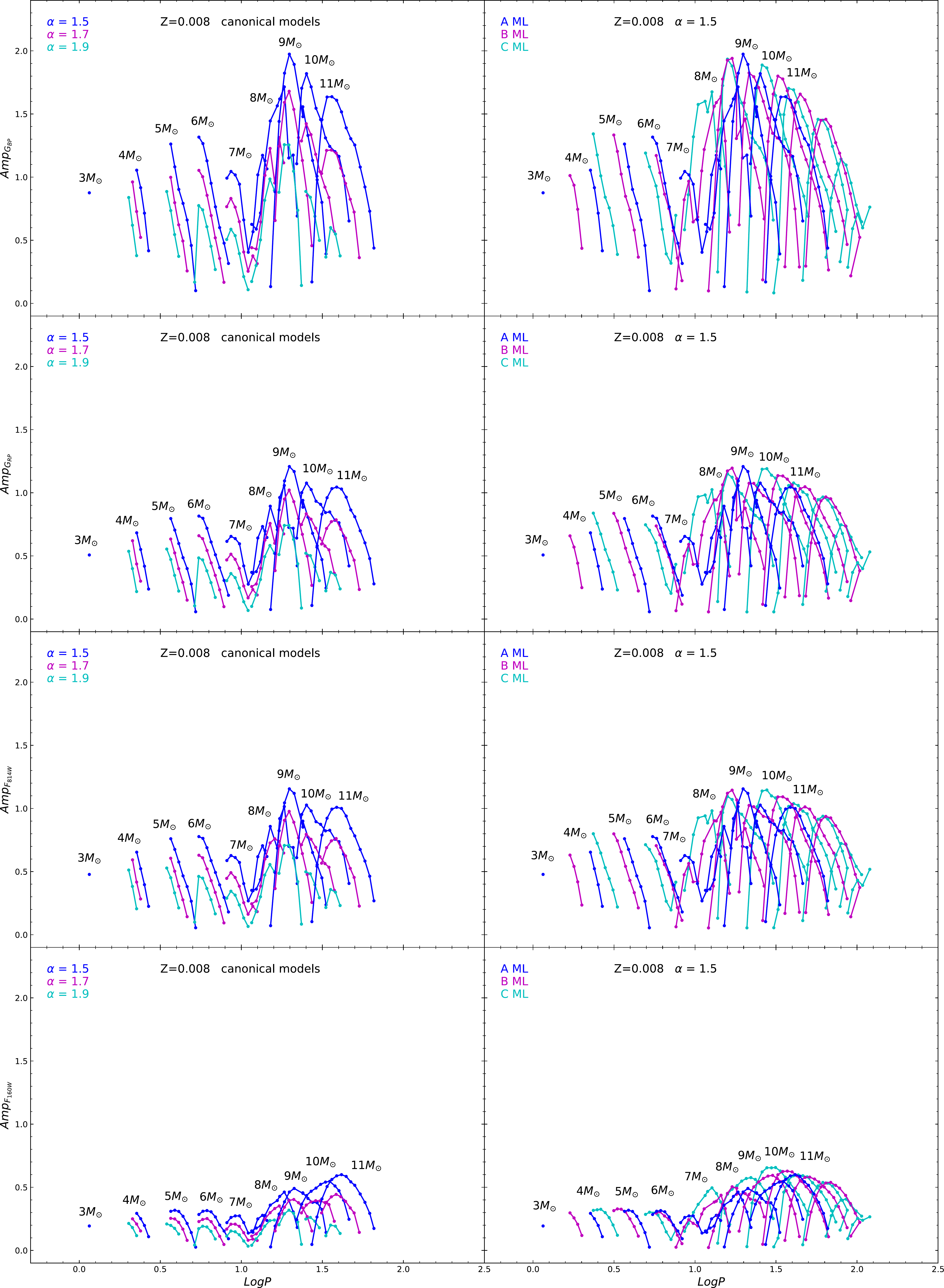}
\caption{The same as Fig.\ref{Fig:ampl_F_SMC} but for Z=0.008 and Y=0.25.}
\label{Fig:ampl_F_LMC}
\end{figure}

\begin{figure}[th]
\centering
\includegraphics[width=0.8\textwidth]{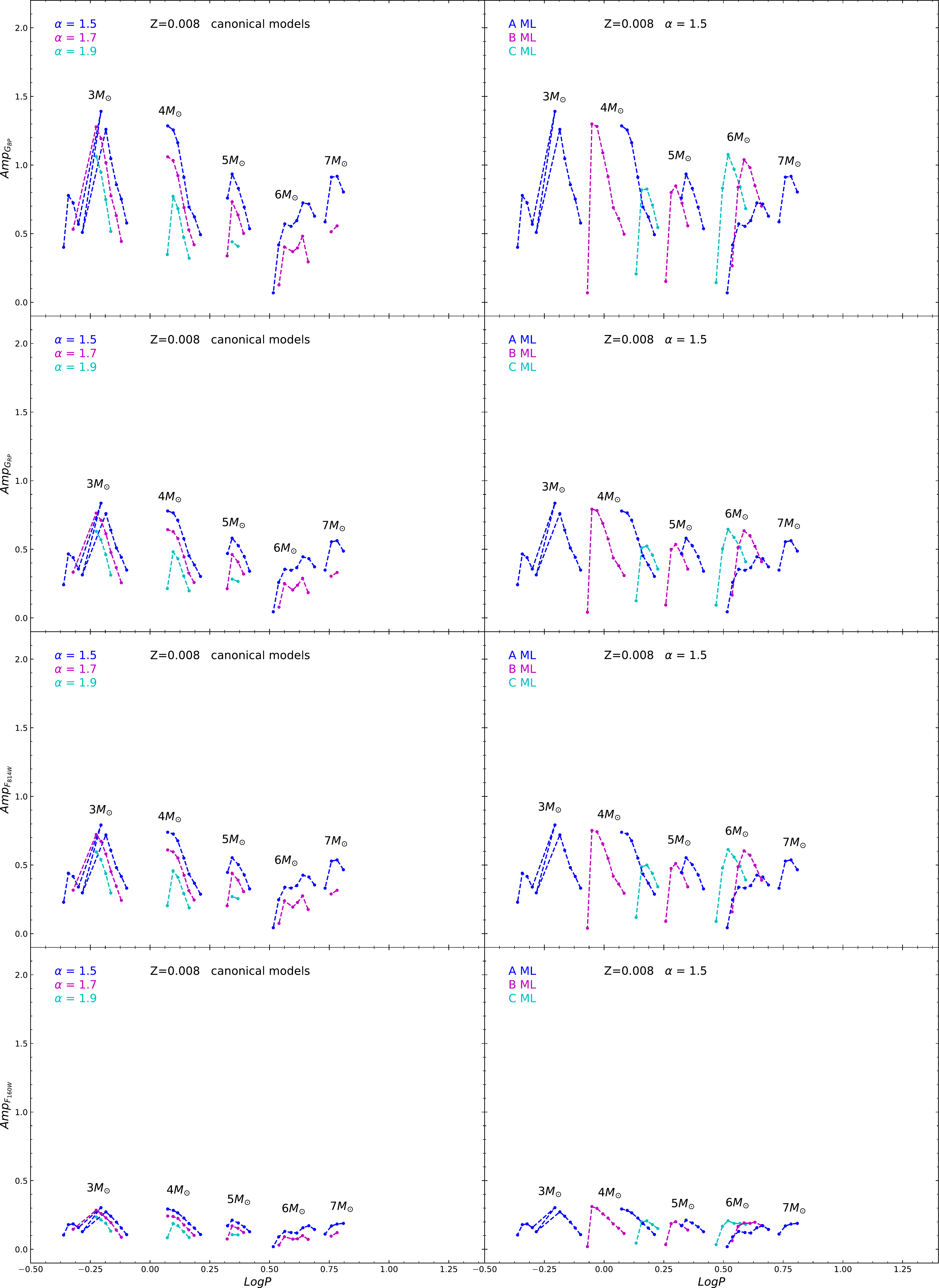}
\caption{The same as Fig.\ref{Fig:ampl_FO_SMC} but for Z=0.008 and Y=0.25.}
\label{Fig:ampl_FO_LMC}
\end{figure}

\begin{figure}[th]
\centering
\includegraphics[width=0.8\textwidth]{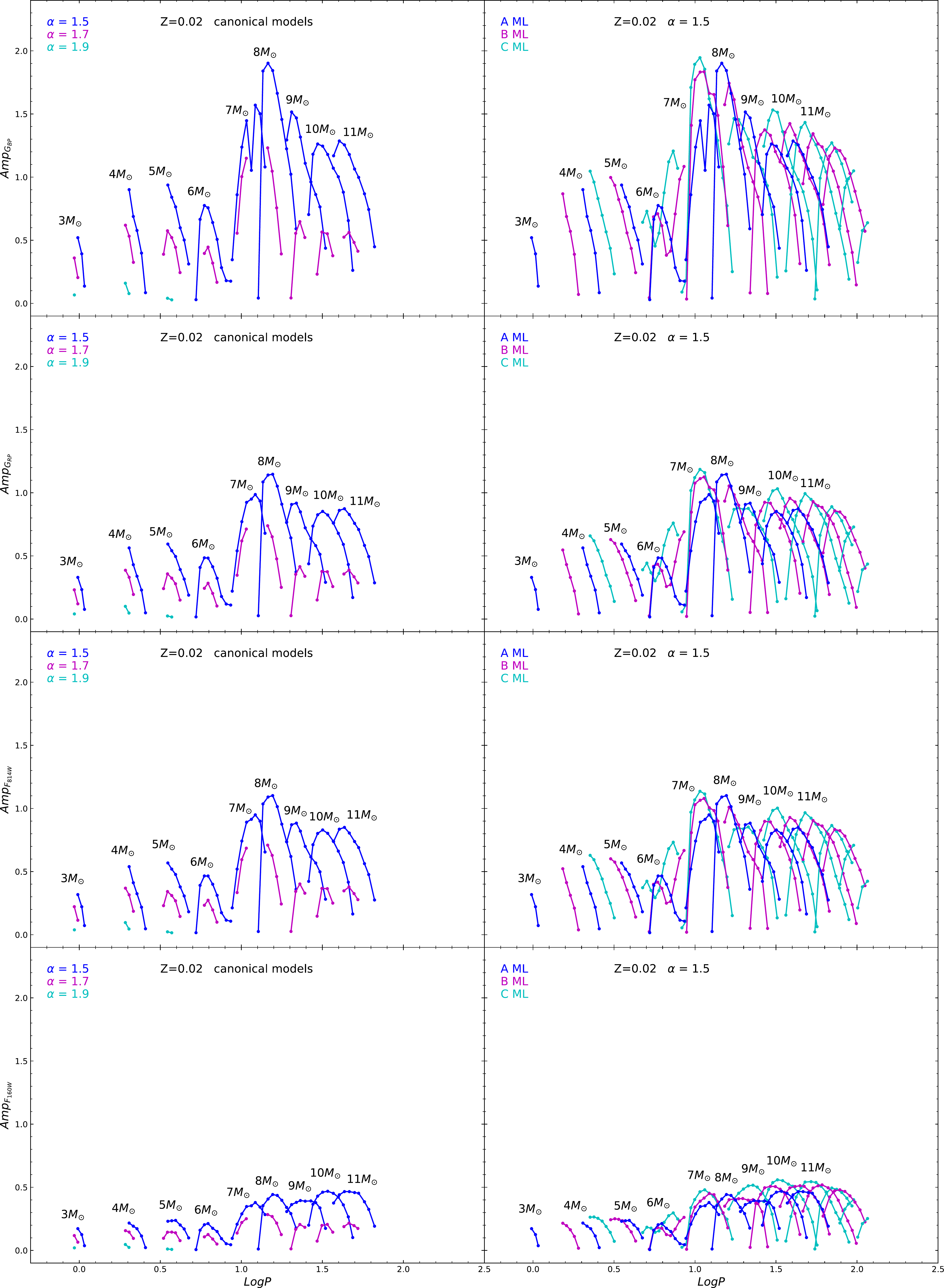}
\caption{The same as Fig.\ref{Fig:ampl_F_SMC} but for Z=0.02 and Y=0.28.}
\label{Fig:ampl_F_MW}
\end{figure}

\begin{figure}[th]
\centering
\includegraphics[width=0.8\textwidth]{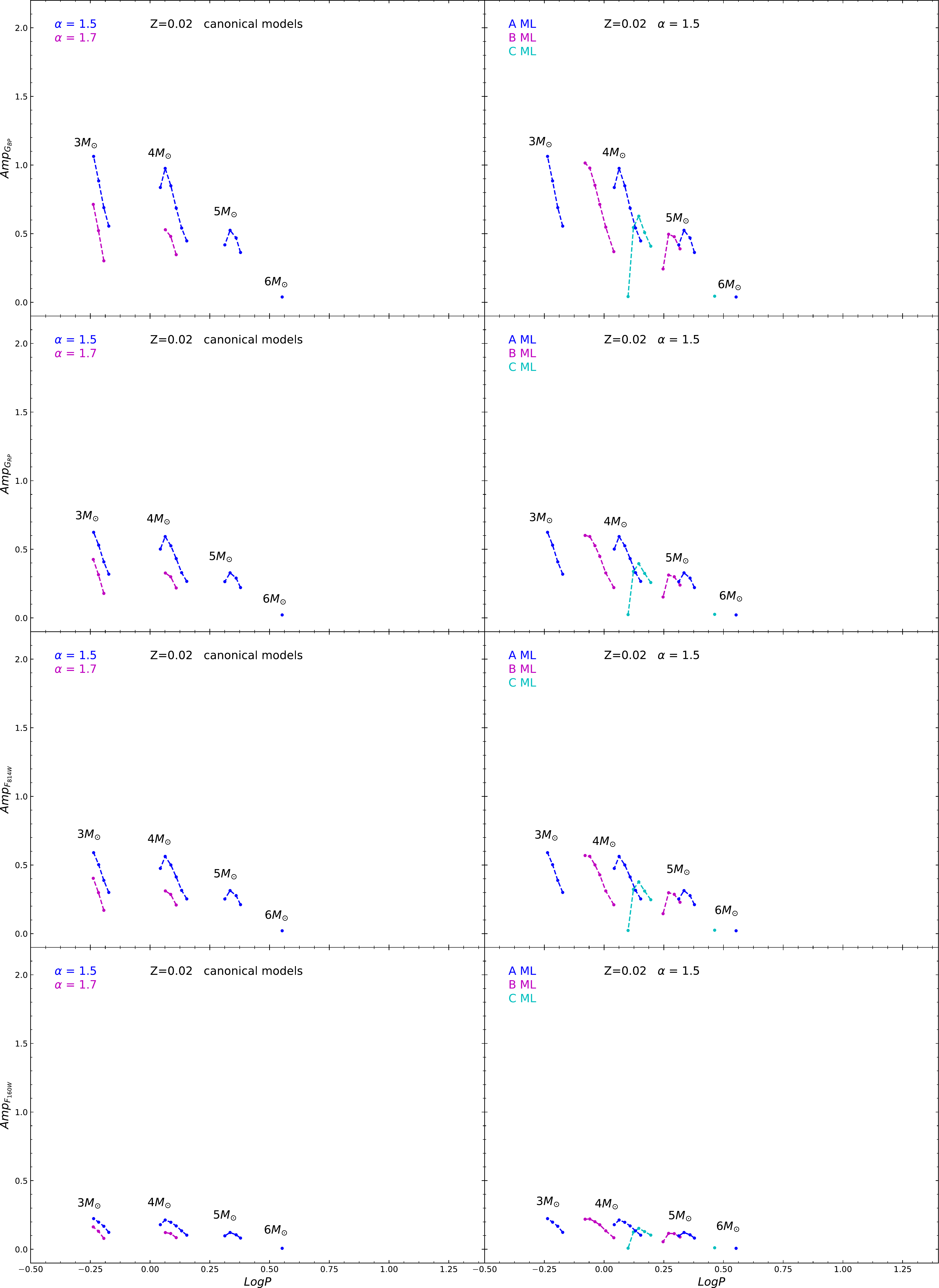}
\caption{The same as Fig.\ref{Fig:ampl_FO_SMC} but for Z=0.02 and Y=0.28.}
\label{Fig:ampl_FO_MW}
\end{figure}

\begin{figure}[th]
\centering
\includegraphics[width=0.8\textwidth]{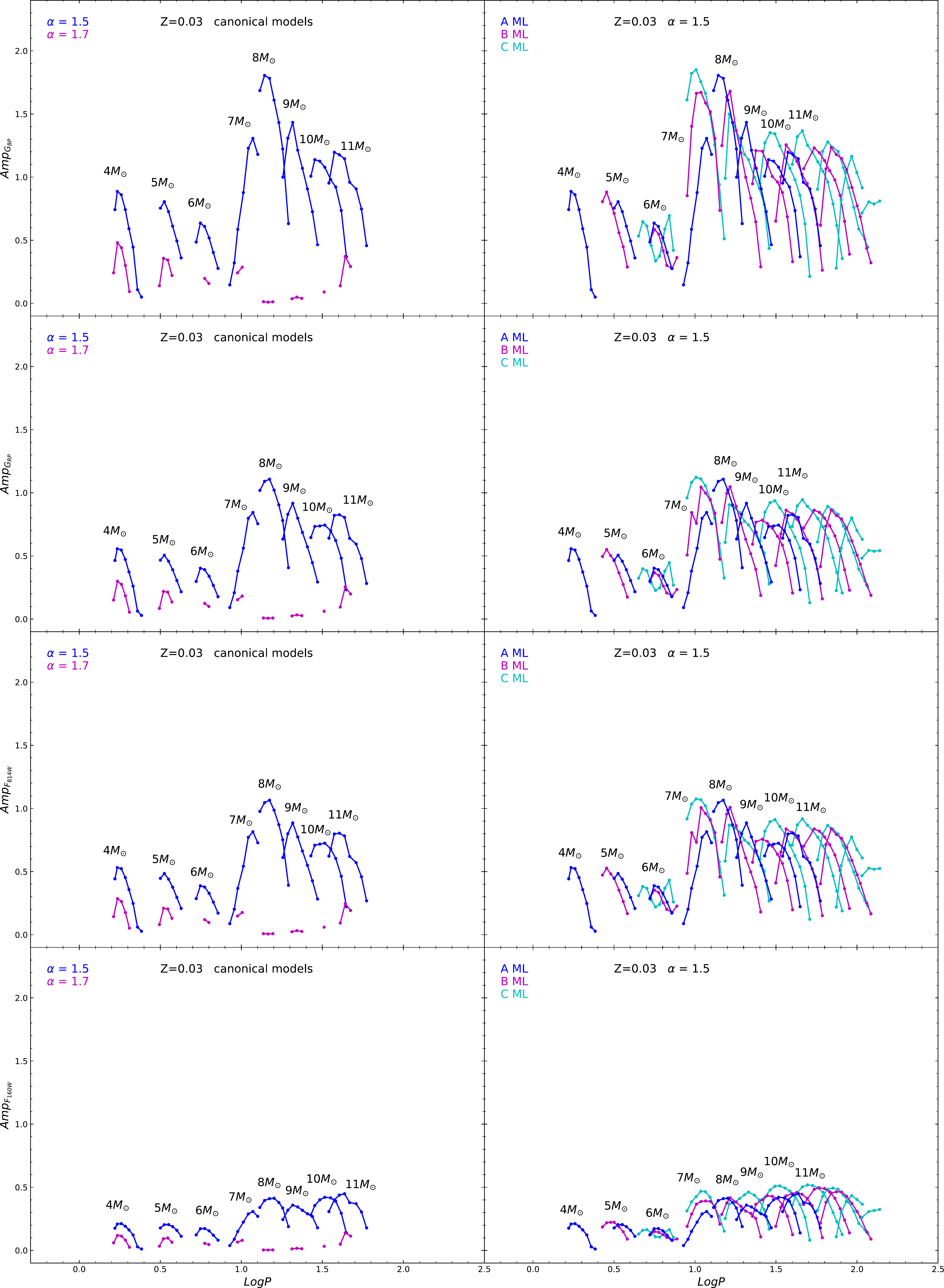}
\caption{The same as Fig.\ref{Fig:ampl_F_SMC} but for Z=0.03 and Y=0.28.}
\label{Fig:ampl_F_M31}
\end{figure}

\subsection{The comparison between theoretical and observed pulsational amplitudes for Z=0.004 and Z=0.008 in the Gaia DR2 database} 

\begin{figure}[th]
\centering
\includegraphics[width=0.5\textwidth]{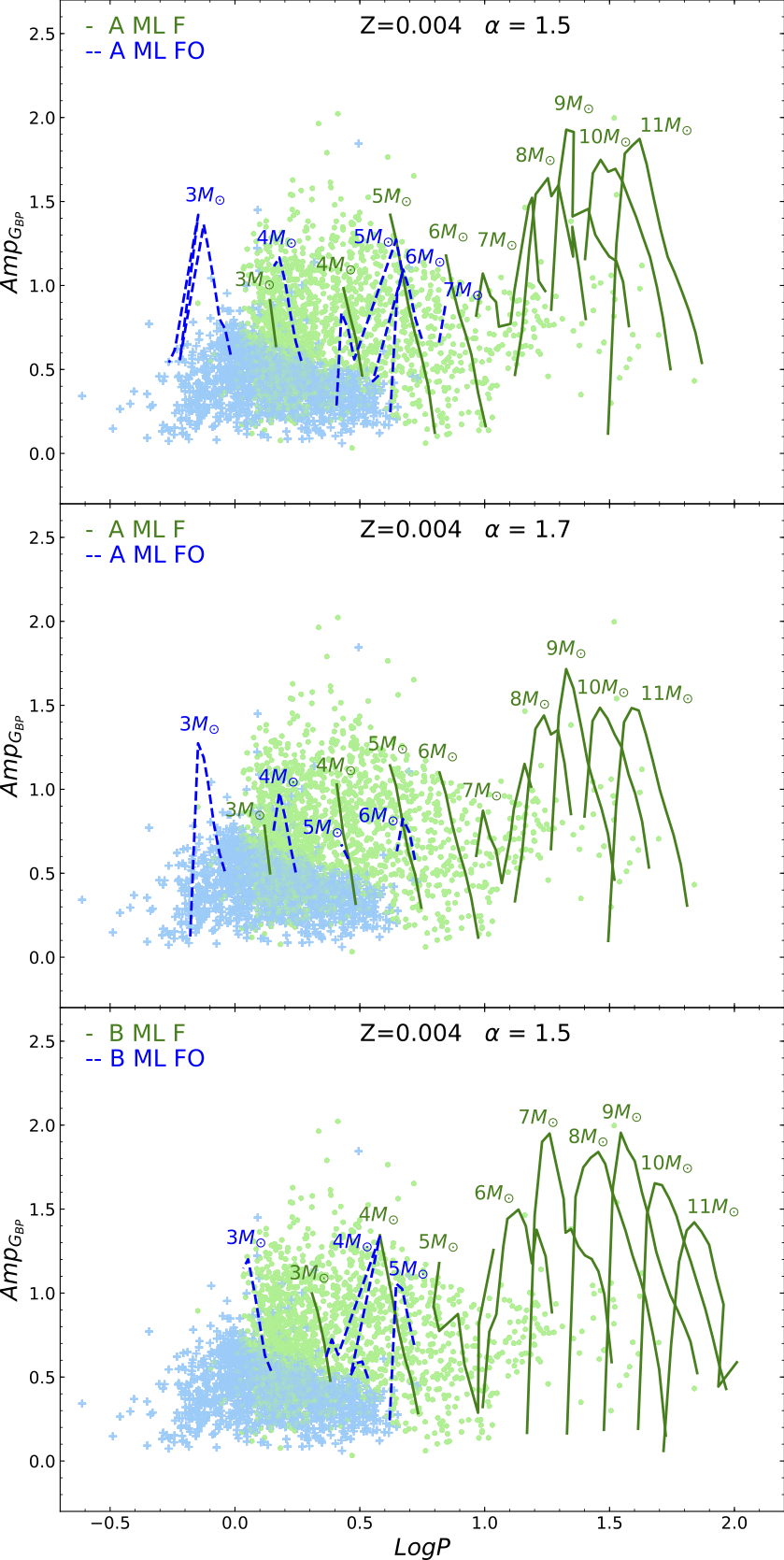}
\caption{The comparison between theoretical F and FO-mode amplitudes (solid green lines and dashed blue lines, respectively) and Z=0.004 F and FO-mode data (green points and blue plus, respectively.) The upper panel and middle panels show the comparison for canonical ML relation with $\alpha_{ml}= 1.5$ and $\alpha_{ml}= 1.7$, respectively while the bottom panel shows the comparison for noncanonical ML relation (case B) with $\alpha_{ml}= 1.5$.}
\label{Fig:amp_comp_gbp_smc}
\end{figure}

\begin{figure}[th]
\centering
\includegraphics[width=0.5\textwidth]{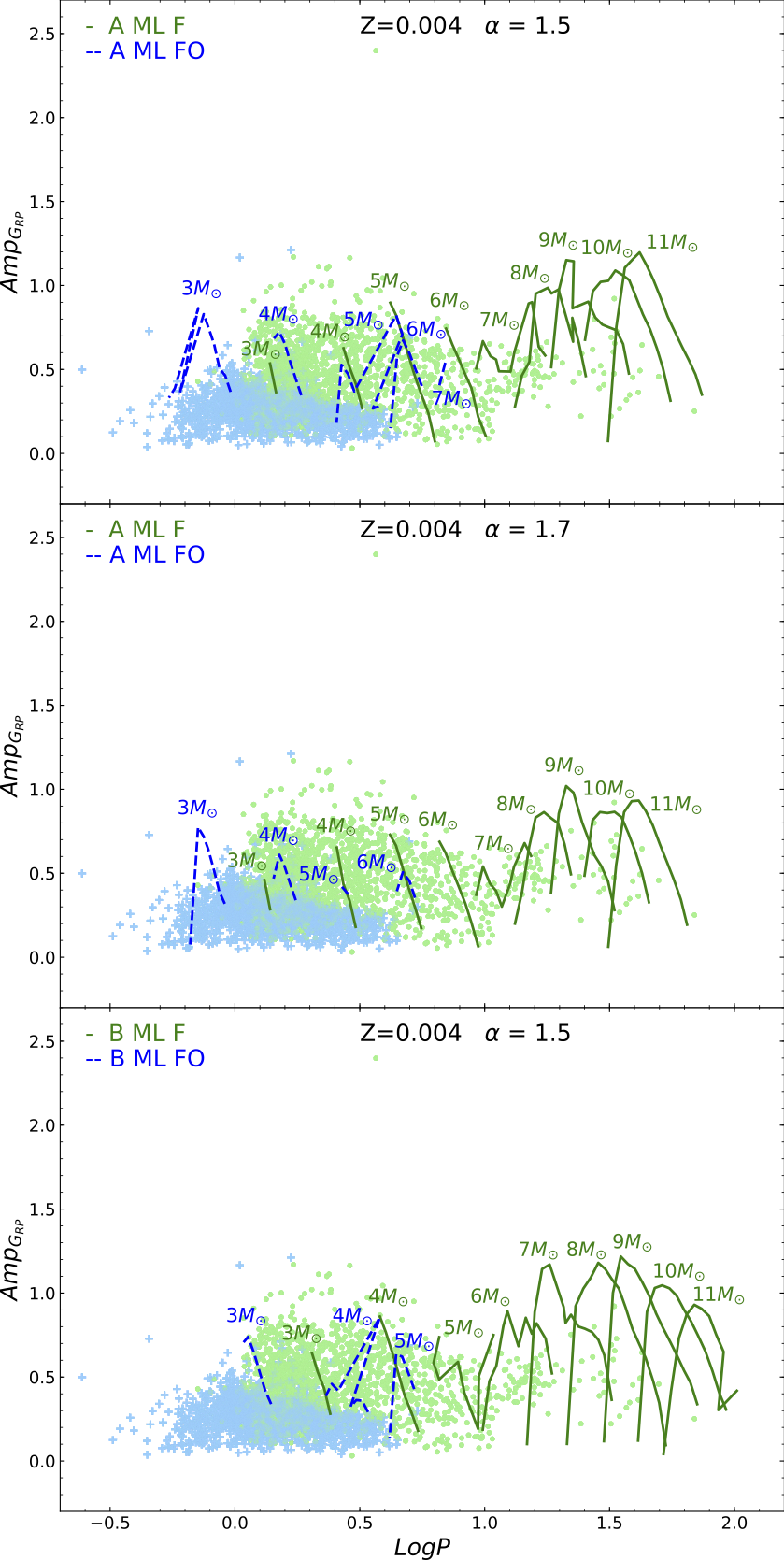}
\caption{The same as Fig.\ref{Fig:amp_comp_gbp_smc} but for the $G_{RP}$ Gaia band.}
\label{Fig:amp_comp_grp_smc}
\end{figure}

\begin{figure}[th]
\centering
\includegraphics[width=0.5\textwidth]{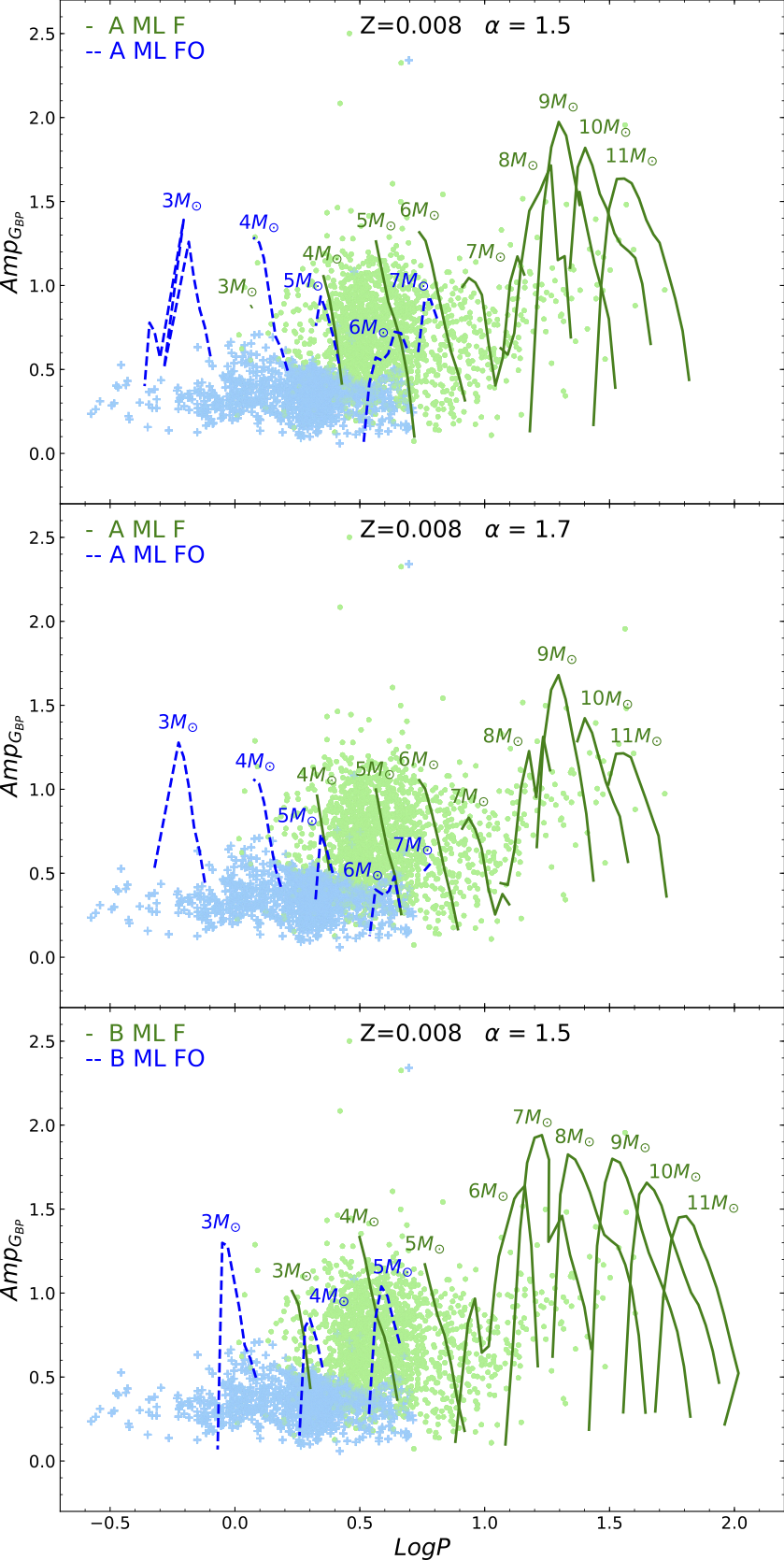}
\caption{The same as Fig.\ref{Fig:amp_comp_gbp_smc} but for Z=0.008.}
\label{Fig:amp_comp_gbp_lmc}
\end{figure}

\begin{figure}[th]
\centering
\includegraphics[width=0.5\textwidth]{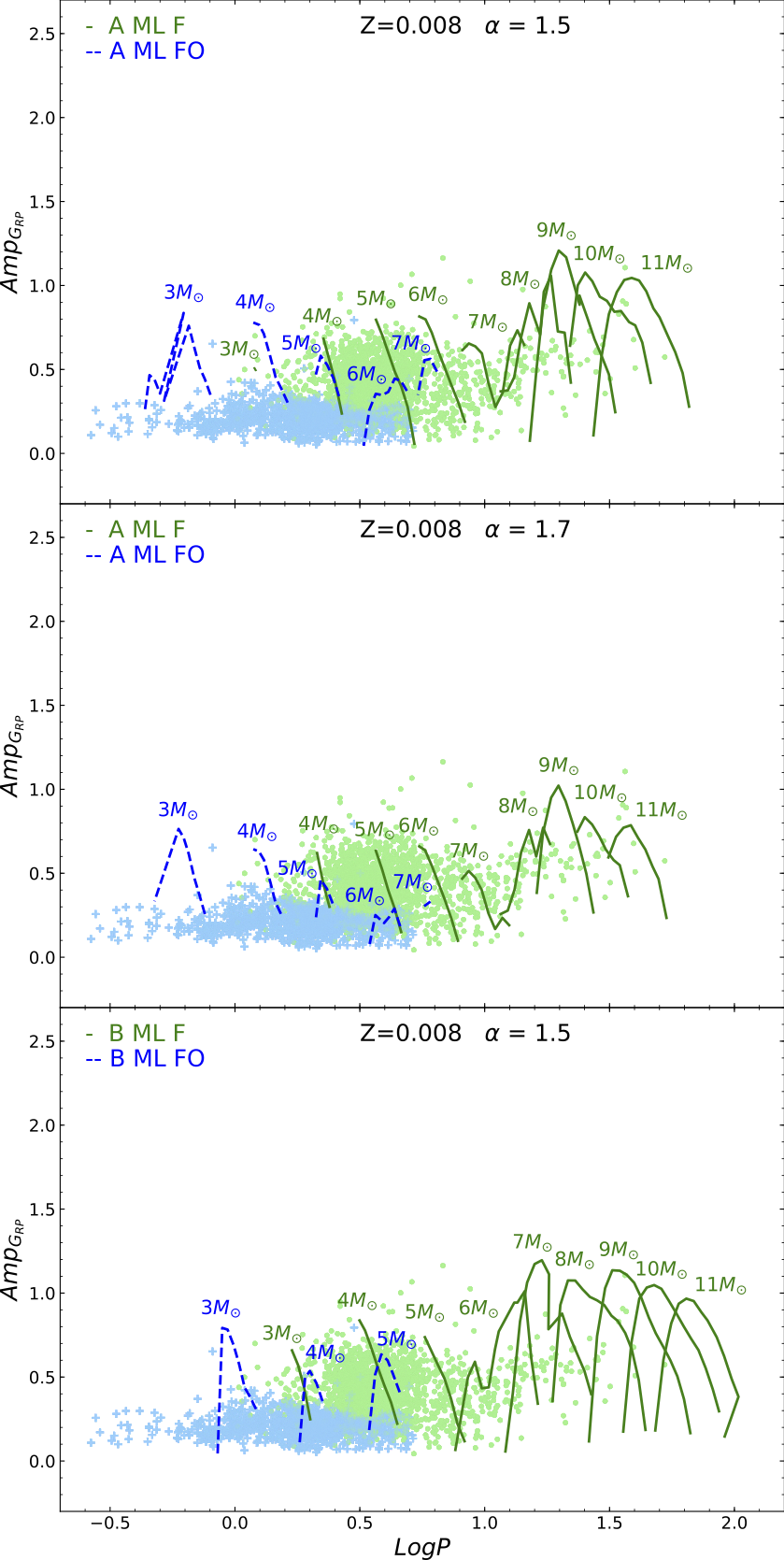}
\caption{The same as Fig.\ref{Fig:amp_comp_gbp_lmc} but for the $G_{RP}$ Gaia band.}
\label{Fig:amp_comp_grp_lmc}
\end{figure}

\clearpage

\section{PLC coefficients in the Gaia EDR3, HST-WFC3 and Johnson-Cousins filters} \label{sec:plc_coeff}

\begin{ThreePartTable}
\begin{longtable}{cccccccccccc}
\caption{\label{plc_gaia} PLC coefficients in the Gaia EDR3 filters ($<G>$=a+b$\log P$ +c($<G_{BP}>$ - $<G_{RP}>$) for F and FO CCs derived by adopting A, B, C ML relations and $\alpha_{ml}$ = 1.5, 1.7 and 1.9.}\\
\hline\hline
Z&Y&$\alpha_{ml}$&ML&a&b&c&$\sigma_{a}$&$\sigma_{b}$&$\sigma_{c}$&$\sigma$&$R^2$\\
\hline
F\\
\hline
\endfirsthead
\caption{continued.}\\
\hline\hline
Z&Y&$\alpha_{ml}$&ML&a&b&c&$\sigma_{a}$&$\sigma_{b}$&$\sigma_{c}$&$\sigma$&$R^2$\\
\hline
\endhead
\hline
\endfoot
0.004&0.25&1.5&A&-3.143&-3.741&3.107&0.019&0.013&0.033&0.029&0.999\\
0.004&0.25&1.7&A&-3.142&-3.757&3.121&0.019&0.012&0.035&0.023&1.000\\
0.004&0.25&1.9&A&-3.131&-3.781&3.132&0.022&0.013&0.043&0.019&1.000\\
0.004&0.25&1.5&B&-3.028&-3.709&3.112&0.024&0.017&0.041&0.043&0.999\\
0.004&0.25&1.7&B&-3.032&-3.731&3.153&0.017&0.012&0.032&0.026&1.000\\
0.004&0.25&1.9&B&-3.051&-3.756&3.211&0.019&0.013&0.039&0.023&1.000\\
0.004&0.25&1.5&C&-2.894&-3.683&3.100&0.030&0.017&0.047&0.046&0.999\\
0.004&0.25&1.7&C&-2.876&-3.689&3.098&0.021&0.013&0.037&0.031&0.999\\
0.004&0.25&1.9&C&-2.881&-3.710&3.136&0.021&0.014&0.041&0.026&1.000\\
0.008&0.25&1.5&A&-3.233&-3.785&3.082&0.020&0.016&0.035&0.031&0.999\\
0.008&0.25&1.7&A&-3.195&-3.785&3.021&0.023&0.018&0.045&0.023&0.999\\
0.008&0.25&1.9&A&-3.152&-3.830&3.009&0.030&0.022&0.060&0.017&1.000\\
0.008&0.25&1.5&B&-3.141&-3.765&3.136&0.023&0.018&0.040&0.043&0.999\\
0.008&0.25&1.7&B&-3.185&-3.812&3.246&0.025&0.021&0.050&0.037&0.999\\
0.008&0.25&1.9&B&-3.165&-3.849&3.261&0.039&0.031&0.081&0.032&0.999\\
0.008&0.25&1.5&C&-2.942&-3.718&3.033&0.024&0.016&0.039&0.039&0.999\\
0.008&0.25&1.7&C&-2.985&-3.735&3.099&0.023&0.016&0.042&0.033&0.999\\
0.008&0.25&1.9&C&-3.012&-3.765&3.169&0.039&0.032&0.082&0.039&0.999\\
0.02&0.28&1.5&A&-3.451&-3.854&3.233&0.037&0.028&0.062&0.039&0.999\\
0.02&0.28&1.7&A&-3.546&-4.007&3.475&0.084&0.062&0.151&0.033&0.999\\
0.02&0.28&1.9&A&-3.158&-3.850&2.860&0.058&0.025&0.097&0.002&1.000\\
0.02&0.28&1.5&B&-3.387&-3.816&3.298&0.037&0.029&0.062&0.049&0.998\\
0.02&0.28&1.7&B&-3.564&-3.961&3.630&0.088&0.070&0.159&0.060&0.998\\
0.02&0.28&1.9&B&-2.990&-3.831&2.828&0.040&0.020&0.067&0.002&1.000\\
0.02&0.28&1.5&C&-3.196&-3.744&3.182&0.034&0.024&0.053&0.048&0.998\\
0.02&0.28&1.7&C&-3.099&-3.712&3.041&0.069&0.051&0.121&0.055&0.998\\
0.02&0.28&1.9&C&-2.909&-3.920&2.980&0.038&0.030&0.073&0.004&1.000\\
0.03&0.28&1.5&A&-3.557&-3.853&3.215&0.050&0.038&0.082&0.040&0.999\\
0.03&0.28&1.7&A&-3.634&-3.978&3.388&0.221&0.166&0.376&0.045&0.999\\
\hline
FO\\
\hline
0.004&0.25&1.5&A&-3.691&-3.680&3.163&0.300&0.169&0.603&0.165&0.975\\
0.004&0.25&1.7&A&-3.549&-3.894&3.015&0.069&0.038&0.142&0.022&0.999\\
0.004&0.25&1.9&A&-3.509&-3.898&2.937&0.060&0.030&0.123&0.011&1.000\\
0.004&0.25&1.5&B&-3.173&-3.653&2.523&0.286&0.214&0.585&0.127&0.968\\
0.004&0.25&1.7&B&-3.105&-3.534&2.325&0.437&0.316&0.919&0.137&0.959\\
0.004&0.25&1.9&B&-3.396&-3.866&2.968&0.035&0.018&0.073&0.003&1.000\\
0.004&0.25&1.5&C&-3.199&-3.788&2.834&0.038&0.025&0.085&0.003&1.000\\
0.004&0.25&1.7&C&-3.489&-3.959&3.476&0.008&0.005&0.018&0.000&1.000\\
0.008&0.25&1.5&A&-3.818&-3.754&3.123&0.198&0.108&0.376&0.113&0.990\\
0.008&0.25&1.7&A&-3.622&-3.844&2.881&0.187&0.101&0.360&0.076&0.995\\
0.008&0.25&1.9&A&-3.599&-3.951&2.879&0.066&0.034&0.130&0.011&1.000\\
0.008&0.25&1.5&B&-3.475&-3.898&2.921&0.044&0.031&0.088&0.018&0.999\\
0.008&0.25&1.7&B&-3.432&-3.882&2.836&0.040&0.025&0.078&0.010&1.000\\
0.008&0.25&1.9&B&-3.450&-3.864&2.853&0.058&0.038&0.114&0.007&1.000\\
0.008&0.25&1.5&C&-3.355&-3.830&2.917&0.044&0.039&0.093&0.011&1.000\\
0.008&0.25&1.7&C&-3.411&-3.882&3.040&0.055&0.046&0.119&0.007&1.000\\
0.02&0.28&1.5&A&-3.610&-3.974&2.677&0.029&0.015&0.049&0.007&1.000\\
0.02&0.28&1.7&A&-3.661&-3.981&2.735&0.044&0.020&0.076&0.003&1.000\\
\hline
\hline
\end{longtable}
\end{ThreePartTable}

\clearpage

\begin{ThreePartTable}
\begin{longtable}{cccccccccccc}
\caption{\label{plc_hst} PLC coefficients in the HST-WFC3 filters ($<F_{160W}>$=a+b$\log P$ +c($<F_{555W}>$ - $<F_{814W}>$) for F and FO CCs derived by adopting A, B, C ML relations and $\alpha_{ml}$ = 1.5, 1.7 and  1.9.}\\
\hline\hline
Z&Y&$\alpha_{ml}$&ML&a&b&c&$\sigma_{a}$&$\sigma_{b}$&$\sigma_{c}$&$\sigma$&$R^2$\\
\hline
F\\
\hline
\endfirsthead
\caption{continued.}\\
\hline\hline
Z&Y&$\alpha_{ml}$&ML&a&b&c&$\sigma_{a}$&$\sigma_{b}$&$\sigma_{c}$&$\sigma$&$R^2$\\
\hline
\endhead
\hline
\endfoot
0.004&0.25&1.5&A&-3.190&-3.744&1.549&0.018&0.013&0.032&0.028&1.000\\
0.004&0.25&1.7&A&-3.187&-3.753&1.563&0.019&0.012&0.035&0.022&1.000\\
0.004&0.25&1.9&A&-3.175&-3.773&1.579&0.024&0.014&0.047&0.020&1.000\\
0.004&0.25&1.5&B&-3.083&-3.686&1.533&0.024&0.016&0.042&0.042&0.999\\
0.004&0.25&1.7&B&-3.088&-3.709&1.589&0.019&0.013&0.037&0.029&1.000\\
0.004&0.25&1.9&B&-3.106&-3.732&1.653&0.025&0.016&0.051&0.030&1.000\\
0.004&0.25&1.5&C&-2.948&-3.639&1.492&0.031&0.017&0.049&0.046&0.999\\
0.004&0.25&1.7&C&-2.926&-3.647&1.494&0.025&0.015&0.045&0.036&0.999\\
0.004&0.25&1.9&C&-2.932&-3.671&1.547&0.027&0.018&0.054&0.034&1.000\\
0.008&0.25&1.5&A&-3.224&-3.766&1.504&0.015&0.012&0.028&0.024&1.000\\
0.008&0.25&1.7&A&-3.214&-3.767&1.489&0.019&0.015&0.038&0.019&1.000\\
0.008&0.25&1.9&A&-3.184&-3.796&1.486&0.029&0.020&0.061&0.017&1.000\\
0.008&0.25&1.5&B&-3.119&-3.727&1.521&0.021&0.017&0.037&0.040&0.999\\
0.008&0.25&1.7&B&-3.173&-3.778&1.663&0.021&0.018&0.044&0.032&1.000\\
0.008&0.25&1.9&B&-3.175&-3.811&1.715&0.036&0.028&0.078&0.030&1.000\\
0.008&0.25&1.5&C&-2.930&-3.658&1.398&0.025&0.016&0.041&0.041&0.999\\
0.008&0.25&1.7&C&-2.967&-3.669&1.460&0.023&0.017&0.044&0.034&0.999\\
0.008&0.25&1.9&C&-2.998&-3.706&1.557&0.037&0.030&0.081&0.037&0.999\\
0.02&0.28&1.5&A&-3.305&-3.813&1.583&0.024&0.019&0.043&0.028&1.000\\
0.02&0.28&1.7&A&-3.416&-3.944&1.851&0.055&0.042&0.104&0.023&1.000\\
0.02&0.28&1.9&A&-3.088&-3.844&1.321&0.086&0.038&0.152&0.003&1.000\\
0.02&0.28&1.5&B&-3.214&-3.761&1.604&0.026&0.022&0.047&0.039&0.999\\
0.02&0.28&1.7&B&-3.353&-3.883&1.898&0.055&0.047&0.107&0.044&0.999\\
0.02&0.28&1.9&B&-2.912&-3.817&1.268&0.025&0.012&0.045&0.001&1.000\\
0.02&0.28&1.5&C&-3.020&-3.666&1.455&0.026&0.020&0.043&0.041&0.999\\
0.02&0.28&1.7&C&-2.963&-3.659&1.388&0.050&0.040&0.094&0.044&0.999\\
0.02&0.28&1.9&C&-2.811&-3.829&1.351&0.053&0.041&0.107&0.006&1.000\\
0.03&0.28&1.5&A&-3.353&-3.808&1.538&0.030&0.025&0.053&0.027&1.000\\
0.03&0.28&1.7&A&-3.469&-3.935&1.784&0.129&0.103&0.238&0.030&1.000\\
\hline
FO\\
\hline
0.004&0.25&1.5&A&-3.698&-3.690&1.570&0.283&0.159&0.581&0.157&0.982\\
0.004&0.25&1.7&A&-3.524&-3.888&1.358&0.044&0.024&0.092&0.014&1.000\\
0.004&0.25&1.9&A&-3.503&-3.899&1.328&0.043&0.022&0.090&0.008&1.000\\
0.004&0.25&1.5&B&-3.167&-3.659&0.889&0.275&0.205&0.573&0.122&0.977\\
0.004&0.25&1.7&B&-3.090&-3.542&0.682&0.419&0.303&0.899&0.133&0.971\\
0.004&0.25&1.9&B&-3.376&-3.868&1.335&0.010&0.005&0.020&0.001&1.000\\
0.004&0.25&1.5&C&-3.143&-3.800&1.132&0.030&0.020&0.069&0.003&1.000\\
0.004&0.25&1.7&C&-3.388&-3.950&1.699&0.049&0.031&0.111&0.002&1.000\\
0.008&0.25&1.5&A&-3.799&-3.754&1.562&0.179&0.097&0.354&0.104&0.993\\
0.008&0.25&1.7&A&-3.612&-3.840&1.332&0.172&0.093&0.344&0.071&0.996\\
0.008&0.25&1.9&A&-3.547&-3.928&1.256&0.039&0.021&0.081&0.007&1.000\\
0.008&0.25&1.5&B&-3.413&-3.871&1.261&0.018&0.012&0.036&0.007&1.000\\
0.008&0.25&1.7&B&-3.407&-3.872&1.253&0.020&0.012&0.040&0.005&1.000\\
0.008&0.25&1.9&B&-3.426&-3.852&1.287&0.046&0.031&0.094&0.006&1.000\\
0.008&0.25&1.5&C&-3.255&-3.817&1.203&0.018&0.016&0.039&0.004&1.000\\
0.008&0.25&1.7&C&-3.290&-3.858&1.295&0.028&0.024&0.064&0.004&1.000\\
0.02&0.28&1.5&A&-3.564&-3.949&1.155&0.025&0.013&0.046&0.006&1.000\\
0.02&0.28&1.7&A&-3.609&-3.979&1.224&0.029&0.013&0.053&0.002&1.000\\
\hline
\hline
\end{longtable}
\end{ThreePartTable}

\clearpage

\begin{ThreePartTable}
\begin{longtable}{cccccccccccc}
\caption{\label{plc_johI} PLC coefficients in the Johnson-Cousins filters ($<V>$=a+b$\log P$ +c($<V>$ - $<I>$) for F and FO CCs derived by adopting A, B, C ML relations and $\alpha_{ml}$ = 1.5, 1.7 and 1.9.}\\
\hline\hline
Z&Y&$\alpha_{ml}$&ML&a&b&c&$\sigma_{a}$&$\sigma_{b}$&$\sigma_{c}$&$\sigma$&$R^2$\\
\hline
F\\
\hline
\endfirsthead
\caption{continued.}\\
\hline\hline
Z&Y&$\alpha_{ml}$&ML&a&b&c&$\sigma_{a}$&$\sigma_{b}$&$\sigma_{c}$&$\sigma$&$R^2$\\
\hline
\endhead
\hline
\endfoot
0.004&0.25&1.5&A&-3.248&-3.727&3.830&0.019&0.013&0.037&0.029&0.999\\
0.004&0.25&1.7&A&-3.248&-3.743&3.851&0.020&0.013&0.042&0.024&1.000\\
0.004&0.25&1.9&A&-3.224&-3.767&3.849&0.026&0.015&0.056&0.022&1.000\\
0.004&0.25&1.5&B&-3.138&-3.690&3.839&0.024&0.016&0.045&0.042&0.999\\
0.004&0.25&1.7&B&-3.142&-3.713&3.889&0.019&0.012&0.039&0.028&0.999\\
0.004&0.25&1.9&B&-3.158&-3.739&3.955&0.023&0.015&0.052&0.028&1.000\\
0.004&0.25&1.5&C&-3.002&-3.660&3.816&0.030&0.017&0.053&0.045&0.999\\
0.004&0.25&1.7&C&-2.980&-3.667&3.814&0.023&0.014&0.045&0.033&0.999\\
0.004&0.25&1.9&C&-2.985&-3.690&3.861&0.025&0.016&0.054&0.031&0.999\\
0.008&0.25&1.5&A&-3.289&-3.763&3.781&0.018&0.014&0.036&0.028&0.999\\
0.008&0.25&1.7&A&-3.272&-3.763&3.746&0.023&0.017&0.049&0.023&0.999\\
0.008&0.25&1.9&A&-3.233&-3.805&3.740&0.033&0.023&0.076&0.019&1.000\\
0.008&0.25&1.5&B&-3.178&-3.733&3.805&0.021&0.017&0.041&0.041&0.999\\
0.008&0.25&1.7&B&-3.235&-3.783&3.956&0.021&0.018&0.048&0.032&0.999\\
0.008&0.25&1.9&B&-3.236&-3.822&4.011&0.038&0.030&0.090&0.032&0.999\\
0.008&0.25&1.5&C&-2.992&-3.682&3.701&0.024&0.016&0.043&0.039&0.999\\
0.008&0.25&1.7&C&-3.036&-3.697&3.780&0.022&0.015&0.044&0.032&0.999\\
0.008&0.25&1.9&C&-3.068&-3.729&3.872&0.037&0.030&0.088&0.037&0.999\\
0.02&0.28&1.5&A&-3.365&-3.809&3.811&0.024&0.019&0.047&0.028&0.999\\
0.02&0.28&1.7&A&-3.496&-3.961&4.141&0.059&0.045&0.124&0.026&1.000\\
0.02&0.28&1.9&A&-3.125&-3.827&3.468&0.071&0.031&0.137&0.002&1.000\\
0.02&0.28&1.5&B&-3.245&-3.755&3.797&0.022&0.019&0.043&0.034&0.999\\
0.02&0.28&1.7&B&-3.401&-3.880&4.132&0.051&0.045&0.110&0.043&0.999\\
0.02&0.28&1.9&B&-2.955&-3.805&3.426&0.030&0.015&0.059&0.002&1.000\\
0.02&0.28&1.5&C&-3.067&-3.681&3.679&0.023&0.018&0.040&0.037&0.999\\
0.02&0.28&1.7&C&-3.040&-3.679&3.642&0.050&0.040&0.102&0.045&0.999\\
0.02&0.28&1.9&C&-2.871&-3.861&3.574&0.051&0.039&0.111&0.005&1.000\\
0.03&0.28&1.5&A&-3.396&-3.796&3.710&0.030&0.025&0.057&0.028&0.999\\
0.03&0.28&1.7&A&-3.595&-3.987&4.133&0.142&0.115&0.288&0.034&0.999\\
\hline
FO\\
\hline
0.004&0.25&1.5&A&-3.700&-3.690&3.770&0.285&0.163&0.636&0.161&0.976\\
0.004&0.25&1.7&A&-3.502&-3.878&3.480&0.058&0.033&0.132&0.019&1.000\\
0.004&0.25&1.9&A&-3.462&-3.883&3.396&0.066&0.034&0.151&0.013&1.000\\
0.004&0.25&1.5&B&-3.169&-3.652&3.027&0.278&0.210&0.629&0.125&0.968\\
0.004&0.25&1.7&B&-3.096&-3.536&2.801&0.420&0.310&0.982&0.136&0.958\\
0.004&0.25&1.9&B&-3.331&-3.842&3.387&0.046&0.024&0.106&0.004&1.000\\
0.004&0.25&1.5&C&-3.166&-3.780&3.311&0.028&0.019&0.069&0.002&1.000\\
0.004&0.25&1.7&C&-3.398&-3.924&3.887&0.019&0.012&0.046&0.001&1.000\\
0.008&0.25&1.5&A&-3.812&-3.754&3.757&0.181&0.099&0.388&0.107&0.991\\
0.008&0.25&1.7&A&-3.608&-3.835&3.462&0.175&0.095&0.379&0.073&0.995\\
0.008&0.25&1.9&A&-3.526&-3.929&3.332&0.058&0.031&0.129&0.011&1.000\\
0.008&0.25&1.5&B&-3.430&-3.870&3.434&0.036&0.026&0.082&0.015&1.000\\
0.008&0.25&1.7&B&-3.388&-3.855&3.342&0.040&0.025&0.090&0.010&1.000\\
0.008&0.25&1.9&B&-3.375&-3.837&3.296&0.055&0.038&0.124&0.007&1.000\\
0.008&0.25&1.5&C&-3.312&-3.810&3.441&0.032&0.029&0.078&0.008&1.000\\
0.008&0.25&1.7&C&-3.355&-3.862&3.555&0.037&0.032&0.092&0.005&1.000\\
0.02&0.28&1.5&A&-3.546&-3.943&3.207&0.028&0.015&0.056&0.007&1.000\\
0.02&0.28&1.7&A&-3.572&-3.956&3.227&0.045&0.021&0.090&0.003&1.000\\
\hline
\hline
\end{longtable}
\end{ThreePartTable}

\clearpage

\begin{ThreePartTable}
\begin{longtable}{cccccccccccc}
\caption{\label{plc_johK} PLC coefficients in the Johnson-Cousins filters ($<V>$=a+b$\log P$ +c($<V>$ - $<K>$) for F and FO CCs derived by adopting A, B, C ML relations and $\alpha_{ml}$ = 1.5, 1.7 and 1.9.}\\
\hline\hline
Z&Y&$\alpha_{ml}$&ML&a&b&c&$\sigma_{a}$&$\sigma_{b}$&$\sigma_{c}$&$\sigma$&$R^2$\\
\hline
F\\
\hline
\endfirsthead
\caption{continued.}\\
\hline\hline
Z&Y&$\alpha_{ml}$&ML&a&b&c&$\sigma_{a}$&$\sigma_{b}$&$\sigma_{c}$&$\sigma$&$R^2$\\
\hline
\endhead
\hline
\endfoot
0.004&0.25&1.5&A&-3.099&-3.772&1.667&0.018&0.014&0.016&0.029&0.999\\
0.004&0.25&1.7&A&-3.099&-3.774&1.675&0.019&0.013&0.018&0.024&1.000\\
0.004&0.25&1.9&A&-3.094&-3.790&1.689&0.024&0.015&0.024&0.022&1.000\\
0.004&0.25&1.5&B&-2.997&-3.686&1.644&0.026&0.018&0.022&0.047&0.998\\
0.004&0.25&1.7&B&-3.006&-3.710&1.678&0.022&0.016&0.022&0.036&0.999\\
0.004&0.25&1.9&B&-3.021&-3.728&1.708&0.028&0.020&0.029&0.036&0.999\\
0.004&0.25&1.5&C&-2.866&-3.622&1.614&0.033&0.019&0.025&0.052&0.998\\
0.004&0.25&1.7&C&-2.844&-3.628&1.617&0.029&0.018&0.025&0.044&0.999\\
0.004&0.25&1.9&C&-2.849&-3.653&1.646&0.031&0.021&0.031&0.041&0.999\\
0.008&0.25&1.5&A&-3.101&-3.777&1.626&0.013&0.011&0.012&0.021&1.000\\
0.008&0.25&1.7&A&-3.122&-3.783&1.645&0.017&0.014&0.017&0.018&1.000\\
0.008&0.25&1.9&A&-3.105&-3.795&1.647&0.029&0.022&0.031&0.018&1.000\\
0.008&0.25&1.5&B&-2.984&-3.720&1.615&0.020&0.017&0.018&0.042&0.999\\
0.008&0.25&1.7&B&-3.044&-3.771&1.693&0.019&0.017&0.020&0.031&0.999\\
0.008&0.25&1.9&B&-3.069&-3.798&1.735&0.033&0.028&0.038&0.031&0.999\\
0.008&0.25&1.5&C&-2.808&-3.627&1.547&0.026&0.018&0.021&0.047&0.999\\
0.008&0.25&1.7&C&-2.845&-3.633&1.577&0.025&0.018&0.023&0.039&0.999\\
0.008&0.25&1.9&C&-2.877&-3.668&1.628&0.034&0.030&0.038&0.039&0.999\\
0.02&0.28&1.5&A&-3.071&-3.798&1.605&0.017&0.015&0.016&0.023&1.000\\
0.02&0.28&1.7&A&-3.164&-3.879&1.712&0.033&0.029&0.034&0.017&1.000\\
0.02&0.28&1.9&A&-3.010&-3.863&1.595&0.097&0.047&0.091&0.003&1.000\\
0.02&0.28&1.5&B&-2.963&-3.734&1.598&0.020&0.019&0.019&0.035&0.999\\
0.02&0.28&1.7&B&-3.039&-3.808&1.691&0.032&0.032&0.034&0.032&0.999\\
0.02&0.28&1.9&B&-2.836&-3.834&1.568&0.019&0.010&0.018&0.001&1.000\\
0.02&0.28&1.5&C&-2.778&-3.617&1.515&0.023&0.019&0.019&0.041&0.999\\
0.02&0.28&1.7&C&-2.742&-3.618&1.502&0.035&0.031&0.034&0.037&0.999\\
0.02&0.28&1.9&C&-2.707&-3.805&1.576&0.047&0.038&0.050&0.005&1.000\\
0.03&0.28&1.5&A&-3.090&-3.794&1.577&0.018&0.018&0.017&0.020&1.000\\
0.03&0.28&1.7&A&-3.176&-3.864&1.668&0.041&0.039&0.042&0.013&1.000\\
\hline
FO\\
\hline
0.004&0.25&1.5&A&-3.739&-3.748&1.798&0.266&0.158&0.277&0.152&0.978\\
0.004&0.25&1.7&A&-3.514&-3.914&1.640&0.031&0.018&0.033&0.010&1.000\\
0.004&0.25&1.9&A&-3.511&-3.929&1.646&0.030&0.016&0.032&0.006&1.000\\
0.004&0.25&1.5&B&-3.179&-3.698&1.424&0.262&0.206&0.277&0.120&0.970\\
0.004&0.25&1.7&B&-3.127&-3.594&1.350&0.400&0.307&0.438&0.132&0.961\\
0.004&0.25&1.9&B&-3.375&-3.902&1.645&0.021&0.011&0.023&0.002&1.000\\
0.004&0.25&1.5&C&-3.096&-3.824&1.491&0.041&0.030&0.048&0.004&1.000\\
0.004&0.25&1.7&C&-3.328&-3.987&1.775&0.080&0.057&0.094&0.003&1.000\\
0.008&0.25&1.5&A&-3.792&-3.788&1.756&0.166&0.095&0.168&0.101&0.992\\
0.008&0.25&1.7&A&-3.613&-3.868&1.648&0.163&0.092&0.168&0.069&0.996\\
0.008&0.25&1.9&A&-3.534&-3.939&1.600&0.028&0.015&0.029&0.005&1.000\\
0.008&0.25&1.5&B&-3.365&-3.880&1.559&0.012&0.009&0.013&0.005&1.000\\
0.008&0.25&1.7&B&-3.389&-3.894&1.589&0.018&0.011&0.019&0.005&1.000\\
0.008&0.25&1.9&B&-3.437&-3.873&1.640&0.041&0.027&0.044&0.005&1.000\\
0.008&0.25&1.5&C&-3.178&-3.828&1.505&0.021&0.021&0.024&0.006&1.000\\
0.008&0.25&1.7&C&-3.203&-3.858&1.542&0.033&0.031&0.039&0.005&1.000\\
0.02&0.28&1.5&A&-3.543&-3.959&1.549&0.028&0.015&0.027&0.007&1.000\\
0.02&0.28&1.7&A&-3.604&-4.008&1.610&0.031&0.015&0.030&0.002&1.000\\
\hline
\hline
\end{longtable}
\end{ThreePartTable}

\clearpage

\section{PW coefficients in the Gaia EDR3, HST-WFC3 and Johnson-Cousins filters} \label{sec:pw_coeff}

\begin{ThreePartTable}
\begin{longtable}{cccccccccc}
\caption{\label{pw_gaia} PW coefficients in the Gaia EDR3 filters ($<W>=<G> -1.9 <G_{BP}-G_{RP}> = a+b\log P$) for F and FO CCs derived by adopting A, B, C ML relations and $\alpha_{ml}$ = 1.5, 1.7 and 1.9.}\\
\hline\hline
Z&Y&$\alpha_{ml}$&ML&a&b&$\sigma_{a}$&$\sigma_{b}$&$\sigma$&$R^2$\\
\hline
F\\
\hline
\endfirsthead
\caption{continued.}\\
\hline\hline
Z&Y&$\alpha_{ml}$&ML&a&b&$\sigma_{a}$&$\sigma_{b}$&$\sigma$&$R^2$\\
\hline
\endhead
\hline
\endfoot
0.004&0.25&1.5&A&-2.580&-3.350&0.045&0.034&0.123&0.992\\
0.004&0.25&1.7&A&-2.559&-3.403&0.039&0.031&0.102&0.995\\
0.004&0.25&1.9&A&-2.540&-3.453&0.030&0.026&0.078&0.997\\
0.004&0.25&1.5&B&-2.471&-3.318&0.048&0.033&0.137&0.991\\
0.004&0.25&1.7&B&-2.474&-3.336&0.039&0.028&0.114&0.994\\
0.004&0.25&1.9&B&-2.472&-3.360&0.034&0.024&0.098&0.996\\
0.004&0.25&1.5&C&-2.272&-3.365&0.051&0.035&0.136&0.991\\
0.004&0.25&1.7&C&-2.319&-3.340&0.043&0.029&0.117&0.994\\
0.004&0.25&1.9&C&-2.333&-3.340&0.039&0.026&0.100&0.996\\
0.008&0.25&1.5&A&-2.670&-3.341&0.044&0.036&0.122&0.991\\
0.008&0.25&1.7&A&-2.668&-3.384&0.037&0.031&0.082&0.996\\
0.008&0.25&1.9&A&-2.634&-3.460&0.027&0.025&0.055&0.998\\
0.008&0.25&1.5&B&-2.581&-3.281&0.046&0.032&0.143&0.991\\
0.008&0.25&1.7&B&-2.596&-3.287&0.039&0.028&0.120&0.994\\
0.008&0.25&1.9&B&-2.555&-3.358&0.031&0.024&0.082&0.997\\
0.008&0.25&1.5&C&-2.365&-3.339&0.044&0.030&0.127&0.993\\
0.008&0.25&1.7&C&-2.419&-3.321&0.041&0.027&0.112&0.995\\
0.008&0.25&1.9&C&-2.460&-3.296&0.033&0.023&0.088&0.997\\
0.02&0.28&1.5&A&-2.698&-3.300&0.035&0.029&0.110&0.995\\
0.02&0.28&1.7&A&-2.677&-3.370&0.028&0.025&0.070&0.998\\
0.02&0.28&1.9&A&-2.582&-3.605&0.014&0.034&0.013&1.000\\
0.02&0.28&1.5&B&-2.636&-3.205&0.041&0.029&0.132&0.993\\
0.02&0.28&1.7&B&-2.641&-3.218&0.039&0.028&0.110&0.996\\
0.02&0.28&1.9&B&-2.439&-3.585&0.028&0.069&0.020&0.999\\
0.02&0.28&1.5&C&-2.476&-3.225&0.049&0.034&0.142&0.992\\
0.02&0.28&1.7&C&-2.479&-3.249&0.036&0.026&0.092&0.997\\
0.02&0.28&1.9&C&-2.342&-3.493&0.028&0.045&0.026&0.999\\
0.03&0.28&1.5&A&-2.779&-3.264&0.032&0.028&0.095&0.996\\
0.03&0.28&1.7&A&-2.764&-3.326&0.028&0.027&0.061&0.999\\
\hline
FO\\
\hline
0.004&0.25&1.5&A&-3.069&-3.378&0.043&0.092&0.177&0.978\\
0.004&0.25&1.7&A&-3.011&-3.629&0.015&0.041&0.050&0.998\\
0.004&0.25&1.9&A&-3.005&-3.665&0.013&0.035&0.033&0.999\\
0.004&0.25&1.5&B&-2.876&-3.477&0.063&0.136&0.131&0.975\\
0.004&0.25&1.7&B&-2.906&-3.411&0.073&0.167&0.138&0.970\\
0.004&0.25&1.9&B&-2.878&-3.631&0.020&0.061&0.028&0.999\\
0.004&0.25&1.5&C&-2.783&-3.536&0.027&0.059&0.020&0.999\\
0.004&0.25&1.7&C&-2.778&-3.545&0.049&0.130&0.027&0.997\\
0.008&0.25&1.5&A&-3.178&-3.451&0.027&0.061&0.131&0.990\\
0.008&0.25&1.7&A&-3.114&-3.600&0.022&0.054&0.088&0.995\\
0.008&0.25&1.9&A&-3.104&-3.740&0.011&0.053&0.032&0.998\\
0.008&0.25&1.5&B&-2.966&-3.603&0.020&0.053&0.056&0.996\\
0.008&0.25&1.7&B&-2.959&-3.627&0.017&0.045&0.039&0.998\\
0.008&0.25&1.9&B&-2.971&-3.621&0.016&0.095&0.032&0.997\\
0.008&0.25&1.5&C&-2.887&-3.474&0.032&0.078&0.042&0.995\\
0.008&0.25&1.7&C&-2.892&-3.478&0.027&0.069&0.031&0.998\\
0.02&0.28&1.5&A&-3.155&-3.775&0.010&0.037&0.031&0.999\\
0.02&0.28&1.7&A&-3.179&-3.808&0.011&0.066&0.020&0.999\\
\hline
\end{longtable}
\end{ThreePartTable}

\clearpage

\begin{ThreePartTable}
\begin{longtable}{cccccccccc}
\caption{\label{pw_hst} PW coefficients in the HST-WFC3 filters ($<W>=<F_{160W}> - 0.386 <F_{555W}-F_{814W}> = a+b\log P$) for F and FO CCs derived by adopting A, B, C ML relations and $\alpha_{ml}$ = 1.5, 1.7 and 1.9.}\\
\hline\hline
Z&Y&$\alpha_{ml}$&ML&a&b&$\sigma_{a}$&$\sigma_{b}$&$\sigma$&$R^2$\\
\hline
F\\
\hline
\endfirsthead
\caption{continued.}\\
\hline\hline
Z&Y&$\alpha_{ml}$&ML&a&b&$\sigma_{a}$&$\sigma_{b}$&$\sigma$&$R^2$\\
\hline
\endhead
\hline
\endfoot
0.004&0.25&1.5&A&-2.652&-3.385&0.042&0.032&0.115&0.993\\
0.004&0.25&1.7&A&-2.629&-3.431&0.036&0.029&0.095&0.995\\
0.004&0.25&1.9&A&-2.607&-3.475&0.028&0.024&0.073&0.997\\
0.004&0.25&1.5&B&-2.561&-3.334&0.044&0.031&0.127&0.992\\
0.004&0.25&1.7&B&-2.556&-3.351&0.037&0.026&0.108&0.995\\
0.004&0.25&1.9&B&-2.550&-3.371&0.033&0.023&0.094&0.997\\
0.004&0.25&1.5&C&-2.381&-3.364&0.047&0.032&0.124&0.992\\
0.004&0.25&1.7&C&-2.415&-3.342&0.040&0.027&0.108&0.995\\
0.004&0.25&1.9&C&-2.420&-3.344&0.036&0.025&0.094&0.996\\
0.008&0.25&1.5&A&-2.712&-3.364&0.041&0.033&0.112&0.993\\
0.008&0.25&1.7&A&-2.713&-3.395&0.035&0.029&0.078&0.996\\
0.008&0.25&1.9&A&-2.682&-3.456&0.026&0.024&0.052&0.998\\
0.008&0.25&1.5&B&-2.628&-3.299&0.042&0.030&0.131&0.992\\
0.008&0.25&1.7&B&-2.637&-3.302&0.037&0.027&0.112&0.995\\
0.008&0.25&1.9&B&-2.599&-3.359&0.030&0.023&0.079&0.998\\
0.008&0.25&1.5&C&-2.435&-3.336&0.040&0.027&0.114&0.994\\
0.008&0.25&1.7&C&-2.478&-3.317&0.036&0.025&0.101&0.996\\
0.008&0.25&1.9&C&-2.507&-3.295&0.031&0.021&0.081&0.998\\
0.02&0.28&1.5&A&-2.679&-3.323&0.032&0.026&0.100&0.996\\
0.02&0.28&1.7&A&-2.661&-3.372&0.026&0.023&0.064&0.999\\
0.02&0.28&1.9&A&-2.559&-3.621&0.013&0.033&0.012&1.000\\
0.02&0.28&1.5&B&-2.619&-3.227&0.037&0.026&0.119&0.995\\
0.02&0.28&1.7&B&-2.616&-3.236&0.035&0.025&0.098&0.997\\
0.02&0.28&1.9&B&-2.417&-3.599&0.026&0.063&0.018&0.999\\
0.02&0.28&1.5&C&-2.471&-3.234&0.043&0.029&0.124&0.994\\
0.02&0.28&1.7&C&-2.462&-3.257&0.031&0.022&0.080&0.998\\
0.02&0.28&1.9&C&-2.332&-3.471&0.025&0.040&0.023&0.999\\
0.03&0.28&1.5&A&-2.730&-3.290&0.029&0.025&0.085&0.997\\
0.03&0.28&1.7&A&-2.716&-3.332&0.023&0.022&0.049&0.999\\
\hline
FO\\
\hline
0.004&0.25&1.5&A&-3.127&-3.415&0.041&0.087&0.168&0.981\\
0.004&0.25&1.7&A&-3.064&-3.662&0.013&0.034&0.042&0.999\\
0.004&0.25&1.9&A&-3.056&-3.693&0.011&0.031&0.029&0.999\\
0.004&0.25&1.5&B&-2.932&-3.521&0.060&0.130&0.125&0.977\\
0.004&0.25&1.7&B&-2.954&-3.459&0.071&0.162&0.134&0.972\\
0.004&0.25&1.9&B&-2.925&-3.665&0.018&0.054&0.024&0.999\\
0.004&0.25&1.5&C&-2.818&-3.605&0.022&0.047&0.016&0.999\\
0.004&0.25&1.7&C&-2.809&-3.613&0.042&0.110&0.023&0.998\\
0.008&0.25&1.5&A&-3.207&-3.475&0.025&0.056&0.121&0.991\\
0.008&0.25&1.7&A&-3.141&-3.617&0.021&0.051&0.082&0.996\\
0.008&0.25&1.9&A&-3.123&-3.747&0.009&0.045&0.028&0.999\\
0.008&0.25&1.5&B&-2.993&-3.630&0.016&0.043&0.045&0.998\\
0.008&0.25&1.7&B&-2.986&-3.647&0.015&0.040&0.034&0.999\\
0.008&0.25&1.9&B&-2.991&-3.631&0.015&0.089&0.030&0.997\\
0.008&0.25&1.5&C&-2.894&-3.543&0.025&0.060&0.032&0.997\\
0.008&0.25&1.7&C&-2.894&-3.547&0.021&0.053&0.024&0.999\\
0.02&0.28&1.5&A&-3.140&-3.764&0.009&0.035&0.030&0.999\\
0.02&0.28&1.7&A&-3.154&-3.815&0.011&0.064&0.020&0.999\\
\hline
\end{longtable}
\end{ThreePartTable}

\clearpage

\begin{ThreePartTable}
\begin{longtable}{cccccccccc}
\caption{\label{pw_joh_I} PW coefficients in the Johnson-Cousins filters ($<W>=<I> -1.55 < V - I > = a+b\log P$) for F and FO CCs derived by adopting A, B, C ML relations and $\alpha_{ml}$ = 1.5, 1.7 and 1.9.}\\
\hline\hline
Z&Y&$\alpha_{ml}$&ML&a&b&$\sigma_{a}$&$\sigma_{b}$&$\sigma$&$R^2$\\
\hline
F\\
\hline
\endfirsthead
\caption{continued.}\\
\hline\hline
Z&Y&$\alpha_{ml}$&ML&a&b&$\sigma_{a}$&$\sigma_{b}$&$\sigma$&$R^2$\\
\hline
\endhead
\hline
\endfoot
0.004&0.25&1.5&A&-2.700&-3.364&0.042&0.032&0.117&0.993\\
0.004&0.25&1.7&A&-2.676&-3.416&0.037&0.029&0.097&0.995\\
0.004&0.25&1.9&A&-2.652&-3.467&0.029&0.024&0.074&0.997\\
0.004&0.25&1.5&B&-2.593&-3.328&0.046&0.031&0.131&0.992\\
0.004&0.25&1.7&B&-2.591&-3.348&0.038&0.026&0.109&0.995\\
0.004&0.25&1.9&B&-2.585&-3.372&0.033&0.024&0.094&0.997\\
0.004&0.25&1.5&C&-2.398&-3.371&0.049&0.034&0.129&0.992\\
0.004&0.25&1.7&C&-2.437&-3.349&0.041&0.028&0.111&0.995\\
0.004&0.25&1.9&C&-2.447&-3.352&0.037&0.025&0.096&0.996\\
0.008&0.25&1.5&A&-2.768&-3.356&0.042&0.033&0.115&0.992\\
0.008&0.25&1.7&A&-2.769&-3.393&0.035&0.029&0.078&0.996\\
0.008&0.25&1.9&A&-2.732&-3.467&0.026&0.024&0.053&0.998\\
0.008&0.25&1.5&B&-2.676&-3.298&0.043&0.030&0.134&0.992\\
0.008&0.25&1.7&B&-2.690&-3.302&0.037&0.027&0.113&0.995\\
0.008&0.25&1.9&B&-2.650&-3.367&0.030&0.023&0.080&0.998\\
0.008&0.25&1.5&C&-2.469&-3.347&0.041&0.028&0.119&0.994\\
0.008&0.25&1.7&C&-2.518&-3.329&0.038&0.026&0.105&0.995\\
0.008&0.25&1.9&C&-2.554&-3.306&0.032&0.022&0.084&0.998\\
0.02&0.28&1.5&A&-2.763&-3.332&0.031&0.026&0.098&0.996\\
0.02&0.28&1.7&A&-2.745&-3.391&0.027&0.024&0.066&0.998\\
0.02&0.28&1.9&A&-2.649&-3.626&0.012&0.030&0.011&1.000\\
0.02&0.28&1.5&B&-2.690&-3.247&0.035&0.025&0.115&0.995\\
0.02&0.28&1.7&B&-2.697&-3.254&0.035&0.024&0.097&0.997\\
0.02&0.28&1.9&B&-2.506&-3.605&0.024&0.059&0.017&0.999\\
0.02&0.28&1.5&C&-2.537&-3.258&0.042&0.029&0.123&0.994\\
0.02&0.28&1.7&C&-2.540&-3.276&0.032&0.023&0.082&0.998\\
0.02&0.28&1.9&C&-2.406&-3.512&0.025&0.039&0.023&0.999\\
0.03&0.28&1.5&A&-2.827&-3.311&0.028&0.024&0.082&0.997\\
0.03&0.28&1.7&A&-2.818&-3.359&0.025&0.024&0.054&0.999\\
\hline
FO\\
\hline
0.004&0.25&1.5&A&-3.158&-3.424&0.042&0.089&0.170&0.980\\
0.004&0.25&1.7&A&-3.097&-3.676&0.012&0.032&0.040&0.999\\
0.004&0.25&1.9&A&-3.093&-3.710&0.010&0.029&0.027&0.999\\
0.004&0.25&1.5&B&-2.964&-3.530&0.061&0.132&0.127&0.977\\
0.004&0.25&1.7&B&-2.990&-3.470&0.072&0.165&0.137&0.971\\
0.004&0.25&1.9&B&-2.965&-3.674&0.015&0.046&0.021&0.999\\
0.004&0.25&1.5&C&-2.861&-3.592&0.021&0.044&0.015&0.999\\
0.004&0.25&1.7&C&-2.857&-3.601&0.040&0.105&0.022&0.998\\
0.008&0.25&1.5&A&-3.255&-3.488&0.025&0.057&0.122&0.991\\
0.008&0.25&1.7&A&-3.191&-3.635&0.021&0.051&0.082&0.996\\
0.008&0.25&1.9&A&-3.177&-3.778&0.008&0.041&0.025&0.999\\
0.008&0.25&1.5&B&-3.040&-3.644&0.016&0.042&0.045&0.998\\
0.008&0.25&1.7&B&-3.035&-3.664&0.013&0.036&0.031&0.999\\
0.008&0.25&1.9&B&-3.044&-3.666&0.012&0.071&0.024&0.998\\
0.008&0.25&1.5&C&-2.951&-3.531&0.026&0.063&0.034&0.997\\
0.008&0.25&1.7&C&-2.954&-3.541&0.022&0.056&0.025&0.999\\
0.02&0.28&1.5&A&-3.215&-3.799&0.008&0.029&0.024&0.999\\
0.02&0.28&1.7&A&-3.237&-3.834&0.008&0.049&0.015&0.999\\
\hline
\end{longtable}
\end{ThreePartTable}

\clearpage

\begin{ThreePartTable}
\begin{longtable}{cccccccccc}
\caption{\label{pw_joh_K} PW coefficients in the Johnson-Cousins filters ($<W>=<K> - 0.13 < V - K > = a+b\log P$) for F and FO CCs derived by adopting A, B, C ML relations and $\alpha_{ml}$ = 1.5, 1.7 and 1.9.}\\
\hline\hline
Z&Y&$\alpha_{ml}$&ML&a&b&$\sigma_{a}$&$\sigma_{b}$&$\sigma$&$R^2$\\
\hline
F\\
\hline
\endfirsthead
\caption{continued.}\\
\hline\hline
Z&Y&$\alpha_{ml}$&ML&a&b&$\sigma_{a}$&$\sigma_{b}$&$\sigma$&$R^2$\\
\hline
\endhead
\hline
\endfoot
0.004&0.25&1.5&A&-2.618&-3.407&0.041&0.031&0.113&0.993\\
0.004&0.25&1.7&A&-2.597&-3.449&0.036&0.028&0.094&0.996\\
0.004&0.25&1.9&A&-2.576&-3.489&0.028&0.024&0.072&0.998\\
0.004&0.25&1.5&B&-2.533&-3.350&0.043&0.030&0.124&0.993\\
0.004&0.25&1.7&B&-2.528&-3.365&0.037&0.026&0.106&0.995\\
0.004&0.25&1.9&B&-2.522&-3.382&0.033&0.023&0.093&0.997\\
0.004&0.25&1.5&C&-2.361&-3.373&0.046&0.031&0.121&0.993\\
0.004&0.25&1.7&C&-2.392&-3.351&0.039&0.026&0.106&0.995\\
0.004&0.25&1.9&C&-2.395&-3.352&0.036&0.024&0.093&0.996\\
0.008&0.25&1.5&A&-2.670&-3.392&0.039&0.031&0.107&0.994\\
0.008&0.25&1.7&A&-2.676&-3.414&0.034&0.028&0.076&0.996\\
0.008&0.25&1.9&A&-2.651&-3.465&0.026&0.023&0.052&0.998\\
0.008&0.25&1.5&B&-2.585&-3.327&0.040&0.028&0.123&0.993\\
0.008&0.25&1.7&B&-2.597&-3.325&0.035&0.025&0.107&0.996\\
0.008&0.25&1.9&B&-2.566&-3.370&0.029&0.022&0.077&0.998\\
0.008&0.25&1.5&C&-2.405&-3.352&0.037&0.025&0.107&0.995\\
0.008&0.25&1.7&C&-2.448&-3.330&0.035&0.023&0.095&0.996\\
0.008&0.25&1.9&C&-2.475&-3.307&0.030&0.020&0.077&0.998\\
0.02&0.28&1.5&A&-2.619&-3.375&0.028&0.023&0.087&0.997\\
0.02&0.28&1.7&A&-2.612&-3.402&0.023&0.020&0.057&0.999\\
0.02&0.28&1.9&A&-2.519&-3.630&0.013&0.033&0.013&1.000\\
0.02&0.28&1.5&B&-2.551&-3.287&0.032&0.023&0.104&0.996\\
0.02&0.28&1.7&B&-2.549&-3.289&0.029&0.021&0.083&0.998\\
0.02&0.28&1.9&B&-2.378&-3.608&0.026&0.064&0.018&0.999\\
0.02&0.28&1.5&C&-2.412&-3.283&0.036&0.025&0.105&0.996\\
0.02&0.28&1.7&C&-2.403&-3.300&0.027&0.019&0.068&0.998\\
0.02&0.28&1.9&C&-2.294&-3.476&0.024&0.039&0.023&0.999\\
0.03&0.28&1.5&A&-2.661&-3.356&0.025&0.021&0.073&0.998\\
0.03&0.28&1.7&A&-2.657&-3.375&0.018&0.018&0.039&0.999\\
\hline
FO\\
\hline
0.004&0.25&1.5&A&-3.103&-3.422&0.041&0.087&0.167&0.981\\
0.004&0.25&1.7&A&-3.040&-3.668&0.013&0.034&0.043&0.999\\
0.004&0.25&1.9&A&-3.031&-3.698&0.012&0.033&0.031&0.999\\
0.004&0.25&1.5&B&-2.908&-3.528&0.060&0.129&0.124&0.978\\
0.004&0.25&1.7&B&-2.930&-3.465&0.070&0.161&0.133&0.973\\
0.004&0.25&1.9&B&-2.898&-3.671&0.019&0.057&0.026&0.999\\
0.004&0.25&1.5&C&-2.792&-3.616&0.022&0.047&0.016&0.999\\
0.004&0.25&1.7&C&-2.782&-3.624&0.043&0.112&0.023&0.998\\
0.008&0.25&1.5&A&-3.180&-3.480&0.025&0.056&0.121&0.991\\
0.008&0.25&1.7&A&-3.113&-3.619&0.021&0.051&0.082&0.995\\
0.008&0.25&1.9&A&-3.094&-3.747&0.010&0.047&0.029&0.999\\
0.008&0.25&1.5&B&-2.966&-3.636&0.016&0.043&0.045&0.998\\
0.008&0.25&1.7&B&-2.958&-3.651&0.015&0.042&0.036&0.999\\
0.008&0.25&1.9&B&-2.962&-3.627&0.016&0.094&0.031&0.997\\
0.008&0.25&1.5&C&-2.864&-3.555&0.025&0.060&0.032&0.997\\
0.008&0.25&1.7&C&-2.864&-3.556&0.021&0.053&0.024&0.999\\
0.02&0.28&1.5&A&-3.107&-3.764&0.010&0.037&0.032&0.999\\
0.02&0.28&1.7&A&-3.118&-3.819&0.011&0.069&0.021&0.999\\
\hline
\end{longtable}
\end{ThreePartTable}

\acknowledgments
We thank the anonymous Referee for her/his comments that significantly improved the content and readability of the manuscript.
We acknowledge the financial support from ASI-Gaia (“Missione Gaia Partecipazione italiana al DPAC – Operazioni e Attività di Analisi dati”).
GDS thanks the support from Istituto Nazionale di Fisica Nucleare (INFN), Naples section-Specific Initiative Moonlight2.

\bibliography{desomma_main_apjs}{}

\begin{thebibliography}{}
\expandafter\ifx\csname natexlab\endcsname\relax\def\natexlab#1{#1}\fi
\providecommand{\url}[1]{\href{#1}{#1}}
\providecommand{\dodoi}[1]{doi:~\href{http://doi.org/#1}{\nolinkurl{#1}}}
\providecommand{\doeprint}[1]{\href{http://ascl.net/#1}{\nolinkurl{http://ascl.net/#1}}}
\providecommand{\doarXiv}[1]{\href{https://arxiv.org/abs/#1}{\nolinkurl{https://arxiv.org/abs/#1}}}

\bibitem[{{Anderson} {et~al.}(2016){Anderson}, {Saio}, {Ekstr{\"o}m}, {Georgy},
  \& {Meynet}}]{Anderson2016}
{Anderson}, R.~I., {Saio}, H., {Ekstr{\"o}m}, S., {Georgy}, C., \& {Meynet}, G.
  2016, \aap, 591, A8, \dodoi{10.1051/0004-6361/201528031}

\bibitem[{{Bono} {et~al.}(1997){Bono}, {Caputo}, {Cassisi}, {Castellani}, \&
  {Marconi}}]{Bono1997}
{Bono}, G., {Caputo}, F., {Cassisi}, S., {Castellani}, V., \& {Marconi}, M.
  1997, apj, 489, 822, \dodoi{10.1086/304807}

\bibitem[{{Bono} {et~al.}(1999){Bono}, {Caputo}, {Castellani}, \&
  {Marconi}}]{Bono1999}
{Bono}, G., {Caputo}, F., {Castellani}, V., \& {Marconi}, M. 1999, \apj, 512,
  711, \dodoi{10.1086/306815}

\bibitem[{{Bono} {et~al.}(1998){Bono}, {Caputo}, \& {Marconi}}]{Bono1998pr}
{Bono}, G., {Caputo}, F., \& {Marconi}, M. 1998, \apjl, 497, L43,
  \dodoi{10.1086/311270}

\bibitem[{{Bono} {et~al.}(2010){Bono}, {Caputo}, {Marconi}, \&
  {Musella}}]{Bono2010}
{Bono}, G., {Caputo}, F., {Marconi}, M., \& {Musella}, I. 2010, \apj, 715, 277,
  \dodoi{10.1088/0004-637X/715/1/277}

\bibitem[{{Bono} {et~al.}(2000{\natexlab{a}}){Bono}, {Castellani}, \&
  {Marconi}}]{BonoCM2000}
{Bono}, G., {Castellani}, V., \& {Marconi}, M. 2000{\natexlab{a}}, \apj, 529,
  293, \dodoi{10.1086/308263}

\bibitem[{{Bono} {et~al.}(2000{\natexlab{b}}){Bono}, {Marconi}, \&
  {Stellingwerf}}]{BonoHP2000}
{Bono}, G., {Marconi}, M., \& {Stellingwerf}, R.~F. 2000{\natexlab{b}}, \aap,
  360, 245.
\newblock \doarXiv{astro-ph/0006229}

\bibitem[{{Breuval} {et~al.}(2021){Breuval}, {Kervella}, {Wielg{\'o}rski},
  {Gieren}, {Graczyk}, {Trahin}, {Pietrzy{\'n}ski}, {Arenou}, {Javanmardi}, \&
  {Zgirski}}]{Breuval2021}
{Breuval}, L., {Kervella}, P., {Wielg{\'o}rski}, P., {et~al.} 2021, \apj, 913,
  38, \dodoi{10.3847/1538-4357/abf0ae}

\bibitem[{{Caputo} {et~al.}(2005){Caputo}, {Bono}, {Fiorentino}, {Marconi}, \&
  {Musella}}]{Caputo2005}
{Caputo}, F., {Bono}, G., {Fiorentino}, G., {Marconi}, M., \& {Musella}, I.
  2005, \apj, 629, 1021, \dodoi{10.1086/431641}

\bibitem[{{Carini} {et~al.}(2017){Carini}, {Brocato}, {Raimondo}, \&
  {Marconi}}]{Carini2017}
{Carini}, R., {Brocato}, E., {Raimondo}, G., \& {Marconi}, M. 2017, \mnras,
  469, 1532, \dodoi{10.1093/mnras/stx927}

\bibitem[{{Chen} {et~al.}(2019){Chen}, {Girardi}, {Fu}, {Bressan}, {Aringer},
  {Dal Tio}, {Pastorelli}, {Marigo}, {Costa}, \& {Zhang}}]{Chen2019}
{Chen}, Y., {Girardi}, L., {Fu}, X., {et~al.} 2019, \aap, 632, A105,
  \dodoi{10.1051/0004-6361/201936612}

\bibitem[{{De Somma} {et~al.}(2020{\natexlab{a}}){De Somma}, {Marconi},
  {Cassisi}, {Ripepi}, {Leccia}, {Molinaro}, \& {Musella}}]{Desomma2020b}
{De Somma}, G., {Marconi}, M., {Cassisi}, S., {et~al.} 2020{\natexlab{a}},
  \mnras, 496, 5039, \dodoi{10.1093/mnras/staa1834}

\bibitem[{{De Somma} {et~al.}(2021{\natexlab{a}}){De Somma}, {Marconi},
  {Cassisi}, {Ripepi}, {Pietrinferni}, {Molinaro}, {Leccia}, \&
  {Musella}}]{Desomma2021a}
---. 2021{\natexlab{a}}, \mnras, 508, 1473, \dodoi{10.1093/mnras/stab2611}

\bibitem[{{De Somma} {et~al.}(2020{\natexlab{b}}){De Somma}, {Marconi},
  {Molinaro}, {Cignoni}, {Musella}, \& {Ripepi}}]{Desomma2020a}
{De Somma}, G., {Marconi}, M., {Molinaro}, R., {et~al.} 2020{\natexlab{b}},
  \apjs, 247, 30, \dodoi{10.3847/1538-4365/ab7204}

\bibitem[{{De Somma} {et~al.}(2021{\natexlab{b}}){De Somma}, {Marconi},
  {Molinaro}, {Cignoni}, {Musella}, \& {Ripepi}}]{DeSomma2021proc}
{De Somma}, G., {Marconi}, M., {Molinaro}, R., {et~al.} 2021{\natexlab{b}}, in
  Astronomical Society of the Pacific Conference Series, Vol. 529, RR
  Lyrae/Cepheid 2019: Frontiers of Classical Pulsators, ed. K.~{Kinemuchi},
  C.~{Lovekin}, H.~{Neilson}, \& K.~{Vivas}, 27

\bibitem[{{Di Criscienzo} {et~al.}(2004){Di Criscienzo}, {Marconi}, \&
  {Caputo}}]{Dicrisci2004}
{Di Criscienzo}, M., {Marconi}, M., \& {Caputo}, F. 2004, \apj, 612, 1092,
  \dodoi{10.1086/422742}

\bibitem[{{Fausnaugh} {et~al.}(2015){Fausnaugh}, {Kochanek}, {Gerke}, {Macri},
  {Riess}, \& {Stanek}}]{Fausnaugh2015}
{Fausnaugh}, M.~M., {Kochanek}, C.~S., {Gerke}, J.~R., {et~al.} 2015, \mnras,
  450, 3597, \dodoi{10.1093/mnras/stv881}

\bibitem[{{Fiorentino} {et~al.}(2007){Fiorentino}, {Marconi}, {Musella}, \&
  {Caputo}}]{Fiorentino2007}
{Fiorentino}, G., {Marconi}, M., {Musella}, I., \& {Caputo}, F. 2007, \aap,
  476, 863, \dodoi{10.1051/0004-6361:20077587}

\bibitem[{{Freedman} \& {Madore}(2011)}]{Freedman2011}
{Freedman}, W.~L., \& {Madore}, B.~F. 2011, \apj, 734, 46,
  \dodoi{10.1088/0004-637X/734/1/46}

\bibitem[{{Gaia Collaboration} {et~al.}(2018){Gaia Collaboration}, {Brown},
  {Vallenari}, {Prusti}, {de Bruijne}, {Babusiaux}, {Bailer-Jones}, {Biermann},
  {Evans}, {Eyer}, {Jansen}, {Jordi}, {Klioner}, {Lammers}, {Lindegren},
  {Luri}, {Mignard}, {Panem}, {Pourbaix}, {Randich}, {Sartoretti}, {Siddiqui},
  {Soubiran}, {van Leeuwen}, {Walton}, {Arenou}, {Bastian}, {Cropper},
  {Drimmel}, {Katz}, {Lattanzi}, {Bakker}, {Cacciari}, {Casta{\~n}eda},
  {Chaoul}, {Cheek}, {De Angeli}, {Fabricius}, {Guerra}, {Holl}, {Masana},
  {Messineo}, {Mowlavi}, {Nienartowicz}, {Panuzzo}, {Portell}, {Riello},
  {Seabroke}, {Tanga}, {Th{\'e}venin}, {Gracia-Abril}, {Comoretto},
  {Garcia-Reinaldos}, {Teyssier}, {Altmann}, {Andrae}, {Audard},
  {Bellas-Velidis}, {Benson}, {Berthier}, {Blomme}, {Burgess}, {Busso},
  {Carry}, {Cellino}, {Clementini}, {Clotet}, {Creevey}, {Davidson}, {De
  Ridder}, {Delchambre}, {Dell'Oro}, {Ducourant},
  {Fern{\'a}ndez-Hern{\'a}ndez}, {Fouesneau}, {Fr{\'e}mat}, {Galluccio},
  {Garc{\'\i}a-Torres}, {Gonz{\'a}lez-N{\'u}{\~n}ez}, {Gonz{\'a}lez-Vidal},
  {Gosset}, {Guy}, {Halbwachs}, {Hambly}, {Harrison}, {Hern{\'a}ndez},
  {Hestroffer}, {Hodgkin}, {Hutton}, {Jasniewicz}, {Jean-Antoine-Piccolo},
  {Jordan}, {Korn}, {Krone-Martins}, {Lanzafame}, {Lebzelter}, {L{\"o}ffler},
  {Manteiga}, {Marrese}, {Mart{\'\i}n-Fleitas}, {Moitinho}, {Mora}, {Muinonen},
  {Osinde}, {Pancino}, {Pauwels}, {Petit}, {Recio-Blanco}, {Richards},
  {Rimoldini}, {Robin}, {Sarro}, {Siopis}, {Smith}, {Sozzetti}, {S{\"u}veges},
  {Torra}, {van Reeven}, {Abbas}, {Abreu Aramburu}, {Accart}, {Aerts},
  {Altavilla}, {{\'A}lvarez}, {Alvarez}, {Alves}, {Anderson}, {Andrei},
  {Anglada Varela}, {Antiche}, {Antoja}, {Arcay}, {Astraatmadja}, {Bach},
  {Baker}, {Balaguer-N{\'u}{\~n}ez}, {Balm}, {Barache}, {Barata}, {Barbato},
  {Barblan}, {Barklem}, {Barrado}, {Barros}, {Barstow}, {Bartholom{\'e}
  Mu{\~n}oz}, {Bassilana}, {Becciani}, {Bellazzini}, {Berihuete}, {Bertone},
  {Bianchi}, {Bienaym{\'e}}, {Blanco-Cuaresma}, {Boch}, {Boeche}, {Bombrun},
  {Borrachero}, {Bossini}, {Bouquillon}, {Bourda}, {Bragaglia}, {Bramante},
  {Breddels}, {Bressan}, {Brouillet}, {Br{\"u}semeister}, {Brugaletta},
  {Bucciarelli}, {Burlacu}, {Busonero}, {Butkevich}, {Buzzi}, {Caffau},
  {Cancelliere}, {Cannizzaro}, {Cantat-Gaudin}, {Carballo}, {Carlucci},
  {Carrasco}, {Casamiquela}, {Castellani}, {Castro-Ginard}, {Charlot},
  {Chemin}, {Chiavassa}, {Cocozza}, {Costigan}, {Cowell}, {Crifo}, {Crosta},
  {Crowley}, {Cuypers}, {Dafonte}, {Damerdji}, {Dapergolas}, {David}, {David},
  {de Laverny}, {De Luise}, {De March}, {de Martino}, {de Souza}, {de Torres},
  {Debosscher}, {del Pozo}, {Delbo}, {Delgado}, {Delgado}, {Di Matteo},
  {Diakite}, {Diener}, {Distefano}, {Dolding}, {Drazinos}, {Dur{\'a}n},
  {Edvardsson}, {Enke}, {Eriksson}, {Esquej}, {Eynard Bontemps}, {Fabre},
  {Fabrizio}, {Faigler}, {Falc{\~a}o}, {Farr{\`a}s Casas}, {Federici},
  {Fedorets}, {Fernique}, {Figueras}, {Filippi}, {Findeisen}, {Fonti},
  {Fraile}, {Fraser}, {Fr{\'e}zouls}, {Gai}, {Galleti}, {Garabato},
  {Garc{\'\i}a-Sedano}, {Garofalo}, {Garralda}, {Gavel}, {Gavras}, {Gerssen},
  {Geyer}, {Giacobbe}, {Gilmore}, {Girona}, {Giuffrida}, {Glass}, {Gomes},
  {Granvik}, {Gueguen}, {Guerrier}, {Guiraud}, {Guti{\'e}rrez-S{\'a}nchez},
  {Haigron}, {Hatzidimitriou}, {Hauser}, {Haywood}, {Heiter}, {Helmi}, {Heu},
  {Hilger}, {Hobbs}, {Hofmann}, {Holland}, {Huckle}, {Hypki}, {Icardi},
  {Jan{\ss}en}, {Jevardat de Fombelle}, {Jonker}, {Juh{\'a}sz}, {Julbe},
  {Karampelas}, {Kewley}, {Klar}, {Kochoska}, {Kohley}, {Kolenberg},
  {Kontizas}, {Kontizas}, {Koposov}, {Kordopatis}, {Kostrzewa-Rutkowska},
  {Koubsky}, {Lambert}, {Lanza}, {Lasne}, {Lavigne}, {Le Fustec}, {Le
  Poncin-Lafitte}, {Lebreton}, {Leccia}, {Leclerc}, {Lecoeur-Taibi},
  {Lenhardt}, {Leroux}, {Liao}, {Licata}, {Lindstr{\o}m}, {Lister}, {Livanou},
  {Lobel}, {L{\'o}pez}, {Managau}, {Mann}, {Mantelet}, {Marchal}, {Marchant},
  {Marconi}, {Marinoni}, {Marschalk{\'o}}, {Marshall}, {Martino}, {Marton},
  {Mary}, {Massari}, {Matijevi{\v{c}}}, {Mazeh}, {McMillan}, {Messina},
  {Michalik}, {Millar}, {Molina}, {Molinaro}, {Moln{\'a}r}, {Montegriffo},
  {Mor}, {Morbidelli}, {Morel}, {Morris}, {Mulone}, {Muraveva}, {Musella},
  {Nelemans}, {Nicastro}, {Noval}, {O'Mullane}, {Ord{\'e}novic},
  {Ord{\'o}{\~n}ez-Blanco}, {Osborne}, {Pagani}, {Pagano}, {Pailler},
  {Palacin}, {Palaversa}, {Panahi}, {Pawlak}, {Piersimoni}, {Pineau}, {Plachy},
  {Plum}, {Poggio}, {Poujoulet}, {Pr{\v{s}}a}, {Pulone}, {Racero}, {Ragaini},
  {Rambaux}, {Ramos-Lerate}, {Regibo}, {Reyl{\'e}}, {Riclet}, {Ripepi}, {Riva},
  {Rivard}, {Rixon}, {Roegiers}, {Roelens}, {Romero-G{\'o}mez}, {Rowell},
  {Royer}, {Ruiz-Dern}, {Sadowski}, {Sagrist{\`a} Sell{\'e}s}, {Sahlmann},
  {Salgado}, {Salguero}, {Sanna}, {Santana-Ros}, {Sarasso}, {Savietto},
  {Schultheis}, {Sciacca}, {Segol}, {Segovia}, {S{\'e}gransan}, {Shih},
  {Siltala}, {Silva}, {Smart}, {Smith}, {Solano}, {Solitro}, {Sordo}, {Soria
  Nieto}, {Souchay}, {Spagna}, {Spoto}, {Stampa}, {Steele},
  {Steidelm{\"u}ller}, {Stephenson}, {Stoev}, {Suess}, {Surdej}, {Szabados},
  {Szegedi-Elek}, {Tapiador}, {Taris}, {Tauran}, {Taylor}, {Teixeira},
  {Terrett}, {Teyssand ier}, {Thuillot}, {Titarenko}, {Torra Clotet}, {Turon},
  {Ulla}, {Utrilla}, {Uzzi}, {Vaillant}, {Valentini}, {Valette}, {van Elteren},
  {Van Hemelryck}, {van Leeuwen}, {Vaschetto}, {Vecchiato}, {Veljanoski},
  {Viala}, {Vicente}, {Vogt}, {von Essen}, {Voss}, {Votruba}, {Voutsinas},
  {Walmsley}, {Weiler}, {Wertz}, {Wevers}, {Wyrzykowski}, {Yoldas},
  {{\v{Z}}erjal}, {Ziaeepour}, {Zorec}, {Zschocke}, {Zucker}, {Zurbach}, \&
  {Zwitter}}]{Brown2018}
{Gaia Collaboration}, {Brown}, A.~G.~A., {Vallenari}, A., {et~al.} 2018, \aap,
  616, A1, \dodoi{10.1051/0004-6361/201833051}

\bibitem[{{Gaia Collaboration} {et~al.}(2021){Gaia Collaboration}, {Brown},
  {Vallenari}, {Prusti}, {de Bruijne}, {Babusiaux}, {Biermann}, {Creevey},
  {Evans}, {Eyer}, {Hutton}, {Jansen}, {Jordi}, {Klioner}, {Lammers},
  {Lindegren}, {Luri}, {Mignard}, {Panem}, {Pourbaix}, {Randich}, {Sartoretti},
  {Soubiran}, {Walton}, {Arenou}, {Bailer-Jones}, {Bastian}, {Cropper},
  {Drimmel}, {Katz}, {Lattanzi}, {van Leeuwen}, {Bakker}, {Cacciari},
  {Casta{\~n}eda}, {De Angeli}, {Ducourant}, {Fabricius}, {Fouesneau},
  {Fr{\'e}mat}, {Guerra}, {Guerrier}, {Guiraud}, {Jean-Antoine Piccolo},
  {Masana}, {Messineo}, {Mowlavi}, {Nicolas}, {Nienartowicz}, {Pailler},
  {Panuzzo}, {Riclet}, {Roux}, {Seabroke}, {Sordo}, {Tanga}, {Th{\'e}venin},
  {Gracia-Abril}, {Portell}, {Teyssier}, {Altmann}, {Andrae}, {Bellas-Velidis},
  {Benson}, {Berthier}, {Blomme}, {Brugaletta}, {Burgess}, {Busso}, {Carry},
  {Cellino}, {Cheek}, {Clementini}, {Damerdji}, {Davidson}, {Delchambre},
  {Dell'Oro}, {Fern{\'a}ndez-Hern{\'a}ndez}, {Galluccio}, {Garc{\'\i}a-Lario},
  {Garcia-Reinaldos}, {Gonz{\'a}lez-N{\'u}{\~n}ez}, {Gosset}, {Haigron},
  {Halbwachs}, {Hambly}, {Harrison}, {Hatzidimitriou}, {Heiter},
  {Hern{\'a}ndez}, {Hestroffer}, {Hodgkin}, {Holl}, {Jan{\ss}en}, {Jevardat de
  Fombelle}, {Jordan}, {Krone-Martins}, {Lanzafame}, {L{\"o}ffler}, {Lorca},
  {Manteiga}, {Marchal}, {Marrese}, {Moitinho}, {Mora}, {Muinonen}, {Osborne},
  {Pancino}, {Pauwels}, {Petit}, {Recio-Blanco}, {Richards}, {Riello},
  {Rimoldini}, {Robin}, {Roegiers}, {Rybizki}, {Sarro}, {Siopis}, {Smith},
  {Sozzetti}, {Ulla}, {Utrilla}, {van Leeuwen}, {van Reeven}, {Abbas}, {Abreu
  Aramburu}, {Accart}, {Aerts}, {Aguado}, {Ajaj}, {Altavilla}, {{\'A}lvarez},
  {{\'A}lvarez Cid-Fuentes}, {Alves}, {Anderson}, {Anglada Varela}, {Antoja},
  {Audard}, {Baines}, {Baker}, {Balaguer-N{\'u}{\~n}ez}, {Balbinot}, {Balog},
  {Barache}, {Barbato}, {Barros}, {Barstow}, {Bartolom{\'e}}, {Bassilana},
  {Bauchet}, {Baudesson-Stella}, {Becciani}, {Bellazzini}, {Bernet}, {Bertone},
  {Bianchi}, {Blanco-Cuaresma}, {Boch}, {Bombrun}, {Bossini}, {Bouquillon},
  {Bragaglia}, {Bramante}, {Breedt}, {Bressan}, {Brouillet}, {Bucciarelli},
  {Burlacu}, {Busonero}, {Butkevich}, {Buzzi}, {Caffau}, {Cancelliere},
  {C{\'a}novas}, {Cantat-Gaudin}, {Carballo}, {Carlucci}, {Carnerero},
  {Carrasco}, {Casamiquela}, {Castellani}, {Castro-Ginard}, {Castro Sampol},
  {Chaoul}, {Charlot}, {Chemin}, {Chiavassa}, {Cioni}, {Comoretto}, {Cooper},
  {Cornez}, {Cowell}, {Crifo}, {Crosta}, {Crowley}, {Dafonte}, {Dapergolas},
  {David}, {David}, {de Laverny}, {De Luise}, {De March}, {De Ridder}, {de
  Souza}, {de Teodoro}, {de Torres}, {del Peloso}, {del Pozo}, {Delbo},
  {Delgado}, {Delgado}, {Delisle}, {Di Matteo}, {Diakite}, {Diener},
  {Distefano}, {Dolding}, {Eappachen}, {Edvardsson}, {Enke}, {Esquej}, {Fabre},
  {Fabrizio}, {Faigler}, {Fedorets}, {Fernique}, {Fienga}, {Figueras},
  {Fouron}, {Fragkoudi}, {Fraile}, {Franke}, {Gai}, {Garabato},
  {Garcia-Gutierrez}, {Garc{\'\i}a-Torres}, {Garofalo}, {Gavras}, {Gerlach},
  {Geyer}, {Giacobbe}, {Gilmore}, {Girona}, {Giuffrida}, {Gomel}, {Gomez},
  {Gonzalez-Santamaria}, {Gonz{\'a}lez-Vidal}, {Granvik},
  {Guti{\'e}rrez-S{\'a}nchez}, {Guy}, {Hauser}, {Haywood}, {Helmi}, {Hidalgo},
  {Hilger}, {H{\l}adczuk}, {Hobbs}, {Holland}, {Huckle}, {Jasniewicz},
  {Jonker}, {Juaristi Campillo}, {Julbe}, {Karbevska}, {Kervella}, {Khanna},
  {Kochoska}, {Kontizas}, {Kordopatis}, {Korn}, {Kostrzewa-Rutkowska},
  {Kruszy{\'n}ska}, {Lambert}, {Lanza}, {Lasne}, {Le Campion}, {Le Fustec},
  {Lebreton}, {Lebzelter}, {Leccia}, {Leclerc}, {Lecoeur-Taibi}, {Liao},
  {Licata}, {Lindstr{\o}m}, {Lister}, {Livanou}, {Lobel}, {Madrero Pardo},
  {Managau}, {Mann}, {Marchant}, {Marconi}, {Marcos Santos}, {Marinoni},
  {Marocco}, {Marshall}, {Martin Polo}, {Mart{\'\i}n-Fleitas}, {Masip},
  {Massari}, {Mastrobuono-Battisti}, {Mazeh}, {McMillan}, {Messina},
  {Michalik}, {Millar}, {Mints}, {Molina}, {Molinaro}, {Moln{\'a}r},
  {Montegriffo}, {Mor}, {Morbidelli}, {Morel}, {Morris}, {Mulone}, {Munoz},
  {Muraveva}, {Murphy}, {Musella}, {Noval}, {Ord{\'e}novic}, {Orr{\`u}},
  {Osinde}, {Pagani}, {Pagano}, {Palaversa}, {Palicio}, {Panahi}, {Pawlak},
  {Pe{\~n}alosa Esteller}, {Penttil{\"a}}, {Piersimoni}, {Pineau}, {Plachy},
  {Plum}, {Poggio}, {Poretti}, {Poujoulet}, {Pr{\v{s}}a}, {Pulone}, {Racero},
  {Ragaini}, {Rainer}, {Raiteri}, {Rambaux}, {Ramos}, {Ramos-Lerate}, {Re
  Fiorentin}, {Regibo}, {Reyl{\'e}}, {Ripepi}, {Riva}, {Rixon}, {Robichon},
  {Robin}, {Roelens}, {Rohrbasser}, {Romero-G{\'o}mez}, {Rowell}, {Royer},
  {Rybicki}, {Sadowski}, {Sagrist{\`a} Sell{\'e}s}, {Sahlmann}, {Salgado},
  {Salguero}, {Samaras}, {Sanchez Gimenez}, {Sanna}, {Santove{\~n}a},
  {Sarasso}, {Schultheis}, {Sciacca}, {Segol}, {Segovia}, {S{\'e}gransan},
  {Semeux}, {Shahaf}, {Siddiqui}, {Siebert}, {Siltala}, {Slezak}, {Smart},
  {Solano}, {Solitro}, {Souami}, {Souchay}, {Spagna}, {Spoto}, {Steele},
  {Steidelm{\"u}ller}, {Stephenson}, {S{\"u}veges}, {Szabados}, {Szegedi-Elek},
  {Taris}, {Tauran}, {Taylor}, {Teixeira}, {Thuillot}, {Tonello}, {Torra},
  {Torra}, {Turon}, {Unger}, {Vaillant}, {van Dillen}, {Vanel}, {Vecchiato},
  {Viala}, {Vicente}, {Voutsinas}, {Weiler}, {Wevers}, {Wyrzykowski}, {Yoldas},
  {Yvard}, {Zhao}, {Zorec}, {Zucker}, {Zurbach}, \& {Zwitter}}]{Brown2021}
---. 2021, \aap, 650, C3, \dodoi{10.1051/0004-6361/202039657e}

\bibitem[{{Graczyk} {et~al.}(2020){Graczyk}, {Pietrzy{\'n}ski}, {Thompson},
  {Gieren}, {Zgirski}, {Villanova}, {G{\'o}rski}, {Wielg{\'o}rski},
  {Karczmarek}, {Narloch}, {Pilecki}, {Taormina}, {Smolec}, {Suchomska},
  {Gallenne}, {Nardetto}, {Storm}, {Kudritzki}, {Ka{\l}uszy{\'n}ski}, \&
  {Pych}}]{Grac_smc_2020}
{Graczyk}, D., {Pietrzy{\'n}ski}, G., {Thompson}, I.~B., {et~al.} 2020, \apj,
  904, 13, \dodoi{10.3847/1538-4357/abbb2b}

\bibitem[{{Groenewegen}(2013)}]{Groenewegen2013}
{Groenewegen}, M.~A.~T. 2013, \aap, 550, A70,
  \dodoi{10.1051/0004-6361/201220446}

\bibitem[{{Hertzsprung}(1926)}]{Hertzprung1926}
{Hertzsprung}, E. 1926, \bain, 3, 115

\bibitem[{{Hidalgo} {et~al.}(2018){Hidalgo}, {Pietrinferni}, {Cassisi},
  {Salaris}, {Mucciarelli}, {Savino}, {Aparicio}, {Silva Aguirre}, \&
  {Verma}}]{Hidalgo2018}
{Hidalgo}, S.~L., {Pietrinferni}, A., {Cassisi}, S., {et~al.} 2018, \apj, 856,
  125, \dodoi{10.3847/1538-4357/aab158}

\bibitem[{{Kennicutt} {et~al.}(1998){Kennicutt}, {Stetson}, {Saha}, {Kelson},
  {Rawson}, {Sakai}, {Madore}, {Mould}, {Freedman}, {Bresolin}, {Ferrarese},
  {Ford}, {Gibson}, {Graham}, {Han}, {Harding}, {Hoessel}, {Huchra}, {Hughes},
  {Illingworth}, {Macri}, {Phelps}, {Silbermann}, {Turner}, \&
  {Wood}}]{Kennicutt1998}
{Kennicutt}, Robert~C., J., {Stetson}, P.~B., {Saha}, A., {et~al.} 1998, \apj,
  498, 181, \dodoi{10.1086/305538}

\bibitem[{{Kodric} {et~al.}(2013){Kodric}, {Riffeser}, {Hopp}, {Seitz},
  {Koppenhoefer}, {Bender}, {Goessl}, {Snigula}, {Lee}, {Ngeow}, {Chambers},
  {Magnier}, {Price}, {Burgett}, {Hodapp}, {Kaiser}, \&
  {Kudritzki}}]{Kodric2013}
{Kodric}, M., {Riffeser}, A., {Hopp}, U., {et~al.} 2013, \aj, 145, 106,
  \dodoi{10.1088/0004-6256/145/4/106}

\bibitem[{{Lindegren} {et~al.}(2021){Lindegren}, {Bastian}, {Biermann},
  {Bombrun}, {de Torres}, {Gerlach}, {Geyer}, {Hern{\'a}ndez}, {Hilger},
  {Hobbs}, {Klioner}, {Lammers}, {McMillan}, {Ramos-Lerate},
  {Steidelm{\"u}ller}, {Stephenson}, \& {van Leeuwen}}]{Lindegren2021}
{Lindegren}, L., {Bastian}, U., {Biermann}, M., {et~al.} 2021, \aap, 649, A4,
  \dodoi{10.1051/0004-6361/202039653}

\bibitem[{{Macri} {et~al.}(2006){Macri}, {Stanek}, {Bersier}, {Greenhill}, \&
  {Reid}}]{Macri2006}
{Macri}, L.~M., {Stanek}, K.~Z., {Bersier}, D., {Greenhill}, L.~J., \& {Reid},
  M.~J. 2006, \apj, 652, 1133, \dodoi{10.1086/508530}

\bibitem[{{Marconi} {et~al.}(2020){Marconi}, {De Somma}, {Ripepi}, {Molinaro},
  {Musella}, {Leccia}, \& {Moretti}}]{Marconi2020}
{Marconi}, M., {De Somma}, G., {Ripepi}, V., {et~al.} 2020, \apjl, 898, L7,
  \dodoi{10.3847/2041-8213/aba12b}

\bibitem[{{Marconi} {et~al.}(2005){Marconi}, {Musella}, \&
  {Fiorentino}}]{Marconi2005}
{Marconi}, M., {Musella}, I., \& {Fiorentino}, G. 2005, \apj, 632, 590,
  \dodoi{10.1086/432790}

\bibitem[{{Marconi} {et~al.}(2010){Marconi}, {Musella}, {Fiorentino},
  {Clementini}, {Aloisi}, {Annibali}, {Contreras Ramos}, {Saha}, {Tosi}, \&
  {van der Marel}}]{Marconi2010}
{Marconi}, M., {Musella}, I., {Fiorentino}, G., {et~al.} 2010, \apj, 713, 615,
  \dodoi{10.1088/0004-637X/713/1/615}

\bibitem[{{Marconi} {et~al.}(2013){Marconi}, {Molinaro}, {Bono},
  {Pietrzy{\'n}ski}, {Gieren}, {Pilecki}, {Stellingwerf}, {Graczyk}, {Smolec},
  {Konorski}, {Suchomska}, {G{\'o}rski}, \& {Karczmarek}}]{Marconi2013}
{Marconi}, M., {Molinaro}, R., {Bono}, G., {et~al.} 2013, \apjl, 768, L6,
  \dodoi{10.1088/2041-8205/768/1/L6}

\bibitem[{{Marconi} {et~al.}(2017){Marconi}, {Molinaro}, {Ripepi}, {Cioni},
  {Clementini}, {Moretti}, {Ragosta}, {de Grijs}, {Groenewegen}, \&
  {Ivanov}}]{Marconi2017}
{Marconi}, M., {Molinaro}, R., {Ripepi}, V., {et~al.} 2017, \mnras, 466, 3206,
  \dodoi{10.1093/mnras/stw3289}

\bibitem[{{Pejcha} \& {Kochanek}(2012)}]{Pejcha2012}
{Pejcha}, O., \& {Kochanek}, C.~S. 2012, \apj, 748, 107,
  \dodoi{10.1088/0004-637X/748/2/107}

\bibitem[{{Pietrzy{\'n}ski} {et~al.}(2019){Pietrzy{\'n}ski}, {Graczyk},
  {Gallenne}, {Gieren}, {Thompson}, {Pilecki}, {Karczmarek}, {G{\'o}rski},
  {Suchomska}, {Taormina}, {Zgirski}, {Wielg{\'o}rski}, {Ko{\l}aczkowski},
  {Konorski}, {Villanova}, {Nardetto}, {Kervella}, {Bresolin}, {Kudritzki},
  {Storm}, {Smolec}, \& {Narloch}}]{Pietr_lmc_2019}
{Pietrzy{\'n}ski}, G., {Graczyk}, D., {Gallenne}, A., {et~al.} 2019, \nat, 567,
  200, \dodoi{10.1038/s41586-019-0999-4}

\bibitem[{{Riess} {et~al.}(2021{\natexlab{a}}){Riess}, {Casertano}, {Yuan},
  {Bowers}, {Macri}, {Zinn}, \& {Scolnic}}]{Riess2021a}
{Riess}, A.~G., {Casertano}, S., {Yuan}, W., {et~al.} 2021{\natexlab{a}},
  \apjl, 908, L6, \dodoi{10.3847/2041-8213/abdbaf}

\bibitem[{{Riess} {et~al.}(2016){Riess}, {Macri}, {Hoffmann}, {Scolnic},
  {Casertano}, {Filippenko}, {Tucker}, {Reid}, {Jones}, {Silverman},
  {Chornock}, {Challis}, {Yuan}, {Brown}, \& {Foley}}]{Riess2016}
{Riess}, A.~G., {Macri}, L.~M., {Hoffmann}, S.~L., {et~al.} 2016, \apj, 826,
  56, \dodoi{10.3847/0004-637X/826/1/56}

\bibitem[{{Riess} {et~al.}(2021{\natexlab{b}}){Riess}, {Yuan}, {Macri},
  {Scolnic}, {Brout}, {Casertano}, {Jones}, {Murakami}, {Breuval}, {Brink},
  {Filippenko}, {Hoffmann}, {Jha}, {Kenworthy}, {Mackenty}, {Stahl}, \&
  {Zheng}}]{Riess2021b}
{Riess}, A.~G., {Yuan}, W., {Macri}, L.~M., {et~al.} 2021{\natexlab{b}}, arXiv
  e-prints, arXiv:2112.04510.
\newblock \doarXiv{2112.04510}

\bibitem[{{Riess} {et~al.}(2021{\natexlab{c}}){Riess}, {Yuan}, {Macri},
  {Scolnic}, {Brout}, {Casertano}, {Jones}, {Murakami}, {Breuval}, {Brink},
  {Filippenko}, {Hoffmann}, {Jha}, {Kenworthy}, {Mackenty}, {Stahl}, \&
  {Zheng}}]{Riess2021arX}
---. 2021{\natexlab{c}}, arXiv e-prints, arXiv:2112.04510.
\newblock \doarXiv{2112.04510}

\bibitem[{{Ripepi} {et~al.}(2019){Ripepi}, {Molinaro}, {Musella}, {Marconi},
  {Leccia}, \& {Eyer}}]{Ripepi2019}
{Ripepi}, V., {Molinaro}, R., {Musella}, I., {et~al.} 2019, \aap, 625, A14,
  \dodoi{10.1051/0004-6361/201834506}

\bibitem[{{Ripepi} {et~al.}(2020){Ripepi}, {Catanzaro}, {Molinaro}, {Marconi},
  {Clementini}, {Cusano}, {De Somma}, {Leccia}, {Musella}, \&
  {Testa}}]{Ripepi2020}
{Ripepi}, V., {Catanzaro}, G., {Molinaro}, R., {et~al.} 2020, \aap, 642, A230,
  \dodoi{10.1051/0004-6361/202038714}

\bibitem[{{Ripepi} {et~al.}(2021){Ripepi}, {Catanzaro}, {Molinaro}, {Gatto},
  {De Somma}, {Marconi}, {Romaniello}, {Leccia}, {Musella}, {Trentin},
  {Clementini}, {Testa}, {Cusano}, \& {Storm}}]{Ripepi2021}
---. 2021, \mnras, 508, 4047, \dodoi{10.1093/mnras/stab2460}

\bibitem[{{Ripepi} {et~al.}(2022){Ripepi}, {Catanzaro}, {Clementini}, {De
  Somma}, {Drimmel}, {Leccia}, {Marconi}, {Molinaro}, {Musella}, \&
  {Poggio}}]{Ripepi2022}
{Ripepi}, V., {Catanzaro}, G., {Clementini}, G., {et~al.} 2022, arXiv e-prints,
  arXiv:2201.01126.
\newblock \doarXiv{2201.01126}

\bibitem[{{Romaniello} {et~al.}(2008){Romaniello}, {Primas}, {Mottini},
  {Pedicelli}, {Lemasle}, {Bono}, {Fran{\c{c}}ois}, {Groenewegen}, \&
  {Laney}}]{Romaniello2008}
{Romaniello}, M., {Primas}, F., {Mottini}, M., {et~al.} 2008, \aap, 488, 731,
  \dodoi{10.1051/0004-6361:20065661}

\bibitem[{{Romaniello} {et~al.}(2022){Romaniello}, {Riess}, {Mancino},
  {Anderson}, {Freudling}, {Kudritzki}, {Macr{\`\i}}, {Mucciarelli}, \&
  {Yuan}}]{Romaniello2022}
{Romaniello}, M., {Riess}, A., {Mancino}, S., {et~al.} 2022, \aap, 658, A29,
  \dodoi{10.1051/0004-6361/202142441}

\bibitem[{{Shappee} \& {Stanek}(2011)}]{Shappee2011}
{Shappee}, B.~J., \& {Stanek}, K.~Z. 2011, \apj, 733, 124,
  \dodoi{10.1088/0004-637X/733/2/124}

\bibitem[{{Soszy{\'n}ski} {et~al.}(2017){Soszy{\'n}ski}, {Udalski},
  {Szyma{\'n}ski}, {Wyrzykowski}, {Ulaczyk}, {Poleski}, {Pietrukowicz},
  {Koz{\l}owski}, {Skowron}, {Skowron}, {Mr{\'o}z}, \& {Pawlak}}]{Sos2017}
{Soszy{\'n}ski}, I., {Udalski}, A., {Szyma{\'n}ski}, M.~K., {et~al.} 2017,
  \actaa, 67, 103, \dodoi{10.32023/0001-5237/67.2.1}

\bibitem[{{Verde} {et~al.}(2019){Verde}, {Treu}, \& {Riess}}]{Verde2019}
{Verde}, L., {Treu}, T., \& {Riess}, A.~G. 2019, Nature Astronomy, 3, 891,
  \dodoi{10.1038/s41550-019-0902-0}

\bibitem[{{Wielg{\'o}rski} {et~al.}(2017){Wielg{\'o}rski}, {Pietrzy{\'n}ski},
  {Gieren}, {G{\'o}rski}, {Kudritzki}, {Zgirski}, {Bresolin}, {Storm},
  {Matsunaga}, {Graczyk}, \& {Soszy{\'n}ski}}]{Wiel2017}
{Wielg{\'o}rski}, P., {Pietrzy{\'n}ski}, G., {Gieren}, W., {et~al.} 2017, \apj,
  842, 116, \dodoi{10.3847/1538-4357/aa7565}

\end{thebibliography}
\bibliographystyle{aasjournal}

\end{document}